\documentclass[11pt]{article}
\usepackage{amsmath}
\usepackage{graphicx}
\usepackage{enumerate}
\usepackage{natbib}
\usepackage{url} 
\usepackage{amssymb}
\usepackage{physics}
\usepackage{bbm}
\usepackage{booktabs}
\usepackage{graphicx}
\usepackage{float}
\usepackage{caption}
\usepackage{mathtools}
\usepackage{authblk}
\usepackage{bookmark}

\allowdisplaybreaks 

\newcommand{\indep}{\perp \!\!\! \perp}
\newcommand{\normt}[1]{\norm\big{#1}_2}
\newcommand{\normp}[1]{\norm\Big{#1}_{L_2(\mathbb{P})}}

\newcommand{\varphidiff}{\varphi(\mathbf{O}; \widehat{\boldsymbol{\eta}}) - \varphi(\mathbf{O}; \boldsymbol{\eta})}
\newcommand{\varphidiffk}{\varphi(\mathbf{O}; \widehat{\boldsymbol{\eta}}^{(-k)}) - \varphi(\mathbf{O}; \boldsymbol{\eta})}

\newcommand{\E}[3]{\mathbb{E} \left#1 #3 \right#2}
\newcommand{\ED}[3]{\mathbb{E} \left#1 #3 \middle| D \right#2}
\newcommand{\EDA}[3]{\mathbb{E} \left#1 #3 \middle| D, \mathbf{A}, \mathbf{X}, N \right#2}

\newcommand{\V}[3]{\text{Var} \left#1 #3 \right#2}
\newcommand{\VD}[3]{\text{Var} \left#1 #3 \middle| D \right#2}
\newcommand{\VDA}[3]{\text{Var} \left#1 #3 \middle| D, \mathbf{A}, \mathbf{X}, N \right#2}

\newcommand{\wt}[1]{w(\mathbf{#1})}
\newcommand{\wh}[1]{\widehat{w}(\mathbf{#1})}

\newcommand{\phit}{\phi(\mathbf{a})}
\newcommand{\phih}{\widehat{\phi}(\mathbf{a})}

\newcommand{\Gt}[1]{G(\mathbf{#1})}
\newcommand{\Gh}[1]{\widehat{G}(\mathbf{#1})}

\newcommand{\Ht}[1]{H(\mathbf{#1})}
\newcommand{\Hh}[1]{\widehat{H}(\mathbf{#1})}

\newcommand{\suma}{\sum_{\mathbf{a} \in \mathcal{A}(N)}}

\newcommand{\oP}[1]{o_{\mathbb{P}}(#1)}
\newcommand{\OP}[1]{O_{\mathbb{P}}(#1)}

\newcommand{\aq}{\mathbf{a}^{(q)}}
\newcommand{\as}{\mathbf{a}^{*}}
\newcommand{\sumaq}{\frac{1}{r} \sum_{q=1}^{r}}

\newcommand{\varphidiffZ}{\varphi(\mathbf{Z}; \widehat{\boldsymbol{\eta}}) - \varphi(\mathbf{Z}; \boldsymbol{\eta})}
\newcommand{\varphidiffkZ}{\varphi(\mathbf{Z}; \widehat{\boldsymbol{\eta}}^{(-k)}) - \varphi(\mathbf{Z}; \boldsymbol{\eta})}

\newtheorem{theorem}{Theorem}
\newtheorem{lemma}{Lemma}
\newtheorem{corollary}{Corollary}

\usepackage[margin = 1in]{geometry}

\begin{document}

\def\spacingset#1{\renewcommand{\baselinestretch}%
{#1}\small\normalsize} \spacingset{1}


\title{\bf Efficient Nonparametric Estimation of Stochastic Policy Effects with Clustered Interference}

\author{Chanhwa Lee, Donglin Zeng, and Michael G. Hudgens \\
  \hspace{.2cm}\\
  \{chanhwa, mhudgens\}@email.unc.edu \\
  Department of Biostatistics, University of North Carolina at Chapel Hill \\
  \hspace{.2cm}\\
  dzeng@umich.edu \\
  Department of Biostatistics, University of Michigan Ann Arbor \\
  }
  
\maketitle

\begin{abstract}
\spacingset{1} 
Interference occurs when a unit's treatment (or exposure) affects another unit's outcome. 
In some settings, units may be grouped into clusters such that it is reasonable to assume that interference, if present, only occurs between individuals in the same cluster, i.e., there is clustered interference. 
Various causal estimands have been proposed to quantify treatment effects under clustered interference from observational data, 
but these estimands either entail treatment policies lacking real-world relevance or are based on parametric propensity score models.
Here, we propose new causal estimands based on modification of the propensity score distribution which may be more relevant in many contexts and are not based on parametric models.
Nonparametric sample splitting estimators of the new estimands are constructed, which allow for flexible data-adaptive estimation of nuisance functions and are consistent, asymptotically normal, and efficient, converging at the usual parametric rate.
Simulations show the finite sample performance of the proposed estimators. 
The proposed methods are applied to evaluate the effect of water, sanitation, and hygiene facilities on diarrhea among children in Senegal.
\end{abstract}

\noindent%

\spacingset{1.2} 
\section{Introduction}
\label{sec:intro}

A standard assumption in causal inference is no interference between units, which supposes that a unit's treatment does not affect the outcome of other units.
However, this assumption might be unrealistic in some circumstances;
for example, a family member getting the COVID-19 vaccine may protect not only themselves but also other family members from serious illness or death due to
SARS-CoV-2 infections \citep{salo22, prunas22}.
The no interference assumption is sometimes relaxed by instead assuming there is clustered (or partial) interference \citep{sobel06, hudgens08, barkley20}.
Clustered interference assumes units can be partitioned into clusters such that a unit's potential outcome may depend on the treatment of others in the same cluster but not on the others in different clusters.
Clusters may be households, classrooms, or villages; for example,
an intervention encouraging students to attend class may affect the attendance rates of siblings in the same household \citep{barrera11},
and the use of bed nets in a household may affect the malaria incidence of other households within a village \citep{kilpatrick21}.
Under clustered interference,
units may be affected by treatment either via direct receipt of treatment or due to interference. 
The goal of this paper is to develop nonparametric methods for quantifying these treatment effects using data from observational studies where there may be clustered interference.

A variety of causal estimands (and corresponding inferential methods) have been proposed to characterize the different effects of treatment in the presence of clustered interference.
These estimands typically entail contrasts in average potential outcomes under different counterfactual scenarios or policies. 
For example, we might consider 
the prevalence of COVID-19 in a city
when 50\% of citizens are vaccinated 
compared to when 30\% of citizens are vaccinated, 
or 
the risk of COVID-19
when an individual is vaccinated versus not vaccinated 
when 50\% of other individuals in the same city are vaccinated. 
\citet{tchetgen12}, \citet{perez14}, \citet{liu19}, and many others consider such estimands under a counterfactual policy that units independently select treatment with the same probability, namely type B policy. 
However, the real-world relevance of such estimands is not clear because,
in many settings, realistic counterfactual scenarios would allow for heterogeneity between units' propensity to select treatment.
For instance, the impact of a policy encouraging vaccination may result in increased uptake of a vaccine, but the propensity for individuals to get vaccinated would likely not be uniform across the population.
Alternatively, \citet{Papadogeorgou19} and \citet{barkley20} propose estimands based on a shift in the distribution of propensity scores according to an assumed parametric model. 
These estimands are appealing because they describe counterfactual settings that allow for differences between unit propensity scores; 
however, in the presence of model mis-specification, interpretation of these estimands is ambiguous.

To overcome these limitations, in this paper methods are developed for drawing inference about a general class of causal estimands that do not require parametric modeling of propensity scores, 
thus obviating the potential for model mis-specification. 
The class of estimands considered describes counterfactual scenarios where the propensity score distribution is modified, 
which we refer to as treatment allocation policies. 
The target estimands may have utility in many settings because the counterfactual scenarios considered allow for the probability of treatment selection to vary depending on a unit’s pre-treatment covariates. 
The class of estimands includes generalizations of several treatment allocation policies that have been studied in the absence of interference \citep{munoz12, kennedy19, wen23}. 
Nonparametric sample splitting estimators of the new estimands are constructed, 
which allow for flexible data-adaptive estimation of nuisance functions and are consistent, asymptotically normal, and efficient, converging at the usual parametric rate. 
To illustrate the methods, four specific treatment allocation policies are considered.

The methods are applied to the Senegal Demographic and Health Survey (DHS) data \citeyearpar[ANSD \& ICF][]{dhs20}
to assess the effect of water, sanitation, and hygiene (WASH) facilities on diarrhea incidence among children.
Prior research suggests the lack of WASH facilities in Senegal \citep{un21} may result in elevated risk of diarrhea among children \citep{thiam17, park21}.
Furthermore, it is plausible that WASH facilities may have interference effects, i.e., a WASH facility in one household may afford beneficial effects to a neighboring household \citep{benjamin2018randomized}.
In the survey, households were randomly sampled from census blocks.
Based on spatial separation between the census blocks,
it may be reasonable to assume there is clustered interference in this setting,
considering census blocks as clusters and households as treatment units. 

The outline of the rest of this paper is as follows. 
In Section 2, the general class of treatment allocations policies and corresponding causal estimands are defined, 
and identifiability is established assuming clustered interference and other standard casual assumptions. 
Section 3 presents the proposed nonparametric sample splitting estimators, and their asymptotic properties are investigated. 
Four specific treatment allocation policies are considered in Section 4.
Simulation study results are then presented in Section 5.
Finally,
the proposed methods are applied to the Senegal DHS data in Section 6,
and some concluding remarks are given in Section 7.

\section{Treatment allocation policies and causal estimands}
\label{s:methods}
\subsection{Data structure and potential outcomes}

Assume there exists a super population of clusters of units, and data from $m$ clusters are observed.
Let $N_i$ equal the number of units in cluster $i \in \{1, \dots, m\}$.
For unit $j \in \{1, \dots, N_i\}$ in cluster $i$, 
let $Y_{ij} \in \mathbb{R}$ denote the outcome of interest, 
$A_{ij} \in \{0,1\}$ denote the binary treatment status ($A_{ij} = 1$ if treated and $A_{ij} = 0$ if untreated),
and $\mathbf{X}_{ij} \in \mathbb{R}^{p}$ denote the vector of pre-treatment covariates.
Let $\mathbf{Y}_i = (Y_{i1}, \dots,  Y_{iN_i})^\top$, 
$\mathbf{A}_i = (A_{i1}, \dots, A_{iN_i})^\top$, 
$\mathbf{X}_i = (\mathbf{X}_{i1}^\top, \dots, \mathbf{X}_{iN_i}^\top)^\top$
denote the outcome, treatment, and covariates vectors for cluster $i$.
Denote the observed data for cluster $i$ by
$\mathbf{O}_i = (\mathbf{Y}_i, \mathbf{A}_i, \mathbf{X}_i, N_i)$,
and assume $(\mathbf{O}_1, \dots, \mathbf{O}_m)$ is an independent and identically distributed random sample from the super population $\mathbb{P}$.
Finally, let
$\mathcal{Y}(n_i)$, $\mathcal{A}(n_i)$ and $\mathcal{X}(n_i)$ denote the support of $\mathbf{Y}_i$, $\mathbf{A}_i$, and $\mathbf{X}_i$ given $N_i = n_i$, respectively, 
where $\mathcal{A}(n_i)$ equals the set of all length $n_i$ binary vectors.

Assuming clustered interference, the potential outcome for a unit may depend on the unit's treatment as well as on those of other units in the same cluster.
The potential outcome for unit $j$ in cluster $i$ when the cluster $i$ receives $\mathbf{a}_i \in \mathcal{A}(N_i)$ is denoted by $Y_{ij}(\mathbf{a}_i)$, and let
$\mathbf{Y}_i(\mathbf{a}_i) = (Y_{i1}(\mathbf{a}_i), \dots, Y_{iN_i}(\mathbf{a}_i))^\top$ denote the vector of potential outcomes.
For notational convenience, 
$\mathbf{a}_i$ is sometimes denoted by
$(a_{ij}, \mathbf{a}_{i(-j)})$, 
where $a_{ij}$ is the treatment status of unit $j$ in cluster $i$, 
and $\mathbf{a}_{i(-j)} = (a_{i1}, \dots, a_{i(j-1)}, a_{i(j+1)}, \dots, a_{iN_i})^\top \in \mathcal{A}(N_i-1)$ is the vector of treatment status for cluster $i$ excluding unit $j$. 
Likewise, the potential outcome $Y_{ij}(\mathbf{a}_i)$ will sometimes be denoted by 
$Y_{ij}(a_{ij}, \mathbf{a}_{i(-j)})$.
If there is no interference, 
$Y_{ij}(a_{ij}, \mathbf{a}_{i(-j)}) 
= 
Y_{ij}(a_{ij}, \mathbf{a}'_{i(-j)})$, 
i.e., potential outcome does not depend on other units' treatments, $\mathbf{a}_{i(-j)}$.

\subsection{Treatment allocation policy}

Traditional causal inference methods target estimands corresponding to deterministic policies. 
For example, the average treatment effect \citep{hernan20} compares the average potential outcome under the policy where all units receive treatment versus the policy where all units do not receive treatment. 
Such deterministic policies may not be of practical relevance because oftentimes treatment is not applied uniformly over units in the real world.
Stochastic policies, on the contrary, allow that units can be treated with some probability possibly depending on covariates.

To formally define stochastic policies, 
consider the counterfactual scenario that a cluster of size $N_i$ with cluster-level covariate $\mathbf{X}_i$ receives treatment $\mathbf{a}_i \in \mathcal{A}(N_i)$ with probability $Q(\mathbf{a}_i|\mathbf{X}_i, N_i)$.
Here $Q(\cdot|\mathbf{X}_i, N_i)$ is a probability distribution on $\mathcal{A}(N_i)$
and corresponds to a stochastic treatment allocation policy \citep{munoz12}.
The distribution for a specific policy will be denoted by
$Q_{\scriptscriptstyle <\textup{NAME}>}(\mathbf{a}_i|\mathbf{X}_i, N_i;\theta)$
for policy ${\scriptstyle <\textup{NAME}>}$ 
where $\theta$ is an optional user-specified parameter or function of the observed data (see examples below).
For example, under the type B policy described below, 
individuals receive treatment independently with the same probability $\alpha$ such that 
$Q_{\scriptscriptstyle \textup{B}}(\mathbf{a}_i|\mathbf{X}_i, N_i; \alpha) = \prod_{j=1}^{N_i} \alpha^{a_{ij}} (1-\alpha)^{1-a_{ij}}$.
Outside of the context of randomized studies, the type B policy may not be particularly realistic or relevant in many settings.
\citet{Papadogeorgou19} and \citet{barkley20} consider more realistic policies that allow the probability of treatment receipt to vary across individuals; 
however, their estimands are defined with respect to parametric propensity score models, 
and in the presence of model mis-specification, interpretation of these estimands is unclear. 
This article focuses on policies which modify the propensity score distribution but do not rely on any parametric models. 
The methods developed in Section 3 are applicable to any stochastic policy, under some mild conditions. 
Four example policies are described in the following subsections.

%


\subsubsection{Type B}
The type B policy often serves as the basis for defining target causal estimands in the presence of interference \citep[e.g.,][]{perez14} due to its simplicity and ease of interpretation.
Regarding inferential methods for type B policy estimands,
\cite{tchetgen12} suggested parametric inverse probability weighted estimators, 
\cite{liu19} proposed parametric doubly robust estimators,
and
\cite{park2022efficient} investigated nonparametric estimators.
In this paper, the type B policy is used to illustrate how the methods can be applied to a treatment policy that does not depend on the observed data distribution.

\subsubsection{Cluster incremental propensity score policy}

Unlike the type B policy, a more realistic policy might allow units to receive treatment with different probabilities based on their covariates.
To that end, a new policy is considered based on an extension of the incremental propensity score intervention \citep{kennedy19} to the clustered interference setting,
which is referred to as the cluster incremental propensity score (CIPS) policy henceforth.
The CIPS policy estimands describe average outcomes that would occur if the propensity score distribution was shifted such that receipt of treatment was more (or less) likely. 
Specifically,
the policy distribution is given as
$Q_{\scriptscriptstyle \textup{CIPS}}(\mathbf{a}_i | \mathbf{X}_i, N_i; \delta) 
    =
    \prod_{j = 1}^{N_i} 
    \allowbreak
        (\pi_{ij,\delta})^{a_{ij}}
        \allowbreak
        (1 - \pi_{ij,\delta})^{1-a_{ij}}$,
where
$\pi_{ij} \allowbreak
= \mathbb{P}(A_{ij} = 1 | \mathbf{X}_i, N_i)$
denotes the propensity score of unit $j$ in cluster $i$,
$\pi_{ij, \delta} \allowbreak
= \delta(\mathbf{X}_i, N_i)
\pi_{ij} / \allowbreak
\{ \delta(\mathbf{X}_i, N_i) \pi_{ij} + 1 - \allowbreak  \pi_{ij} \}$
denotes the shifted propensity score,
and
$\delta: \bigcup_{n \in \mathbb{N}} \{
\mathcal{X}(n) \times {n} \} 
\mapsto (0,\infty)$ 
is a user-specified known function of $\mathbf{X}_i$ and $N_i$.
The CIPS policy corresponds to shifting the propensity score distribution such that the counterfactual odds of treatment
$\pi_{ij, \delta} \allowbreak 
/ \allowbreak 
(1-\pi_{ij, \delta})$
is $\delta(\mathbf{X}_i, N_i)$ times the observed odds of being treated
$\pi_{ij} \allowbreak / \allowbreak (1-\pi_{ij})$.
Like the policies in \citet{Papadogeorgou19} and \citet{barkley20}, the CIPS policy has the appealing property that the ranking of units within clusters by the probability of treatment selection is preserved across policies, 
i.e., $\pi_{ij} < \pi_{ik}$ implies $\pi_{ij, \delta} < \pi_{ik, \delta}$. 
However, the CIPS policy has the additional advantage of not relying on any parametric models.
From the individual-level shifted propensity score $\pi_{ij, \delta}$,
the policy distribution is obtained assuming treatment between two units in the same cluster are conditionally independent given covariates (but not necessarily marginally independent).

Different choices of the function $\delta(\mathbf{X}_i, N_i)$ give rise to different versions of the CIPS policy. 
One example CIPS policy is $\delta(\mathbf{X}_i, N_i) = \delta_0$, i.e., the multiplicative change in the odds of being treated is the same across clusters.
If there is no interference, i.e., $N_i \equiv 1$ for all $i$, then this policy reduces to the incremental propensity score intervention \citep{kennedy19}.
A second example CIPS policy is $\delta(\mathbf{X}_i, N_i) = \delta_0 (1+1/N_i)$, which corresponds to larger changes in the odds of treatment for smaller clusters.
In the context of the WASH facilities example, this policy would correspond to treating a greater proportion of households in smaller census blocks.

\subsubsection{Cluster multiplicative shift policy}

Similar to the CIPS policy, 
another new policy considered is the cluster multiplicative shift (CMS) policy which is an extension of the multiplicative shift policy \citep{wen23}
to the clustered interference setting.
Earlier $\mathbf{X}_{ij}$ was defined to be the vector of covariates for individual $j$ in cluster $i$. 
Suppose one of these covariates is binary, 
say $X_{ij}^*$. 
The CMS policy describes the counterfactual scenario where the propensity for receiving treatment is increased only for individuals where $X_{ij}^* = 1$. 
For example, $X_{ij}^*$ may be an indicator someone is at high risk for an adverse outcome, 
and it may be of interest to estimate counterfactual outcomes if a greater proportion of high risk individuals were treated.
The corresponding policy distribution is
$Q_{\scriptscriptstyle \textup{CMS}}(\mathbf{a}_i | \mathbf{X}_i, N_i; \lambda) 
    =
    \prod_{j = 1}^{N_i} 
    \allowbreak
        (\pi_{ij,\lambda})^{a_{ij}}
        \allowbreak
        (1 - \pi_{ij,\lambda})^{1-a_{ij}}$,
where
$\pi_{ij, \lambda} \allowbreak
= \mathbb{P}_{\lambda}(A_{ij} = 1 | \mathbf{X}_i, N_i) \allowbreak
= (1-\lambda+\lambda\pi_{ij}) X_{ij}^* + \pi_{ij}(1 - X_{ij}^*)$
denotes the shifted propensity score,
and $\lambda \in [0,1]$ is a user-specified factor.
Note the propensity score distribution is shifted
for individual $j$ only if 
$X_{ij}^* = 1$ and not otherwise.
The policy distribution is obtained again assuming conditional independence of treatment selection between units in the same cluster given covariates.

\subsubsection{Treated proportion bound}

Different from the aforementioned policies, the treated proportion bound (TPB) policy is constructed based on the joint (cluster-level) probability of treatment rather than individual-level propensity scores.
The TPB policy corresponds to the counterfactual scenario that the proportion of treated individuals in each cluster is at least some threshold $\rho \in [0,1]$.
Specifically,
$Q_{\scriptscriptstyle \textup{TPB}}(\mathbf{a}_i | \mathbf{X}_i, N_i; \rho) 
    =
    \mathbbm{1}(\overline{\mathbf{a}}_i \ge \rho)
    {\mathbb{P}(\mathbf{a}_i | \mathbf{X}_i, N_i)}
    \big/
    \big\{\sum_{\overline{\mathbf{a}}'_i \ge \rho} \mathbb{P}(\mathbf{a}'_i | \mathbf{X}_i, N_i) \big\}$,
where
$\overline{\mathbf{a}}_{i} = N_i^{-1} \sum_{j=1}^{N_i} a_{ij}$,
and
$\mathbb{P}(\mathbf{a}_i | \mathbf{X}_i, N_i) = \mathbb{P}(\mathbf{A}_i = \mathbf{a}_i | \mathbf{X}_i, N_i)$ is
the observed joint probability of treatment vector $\mathbf{a}_i$ conditional on $\mathbf{X}_i, N_i$.
Under the TPB policy, every cluster has treatment coverage of at least $\rho$ 
because the treatment $\mathbf{a}_i$ occurs with zero probability if
$\overline{\mathbf{a}}_i < \rho$, 
while otherwise the assignment probability of $\mathbf{a}_i$ is proportional to the observed probability.
In the context of COVID-19 vaccination, the TPB policy estimands could be used to quantify the risk of COVID-19 infections when at least 50\% (for example) of individuals in a cluster are vaccinated.

\subsection{Causal estimands and assumptions}

Causal estimands considered here are defined by  the average of potential outcomes and their contrasts under stochastic policy $Q$.
First, denote
the expected value of the average potential outcome under the policy $Q$ by
$\mu(Q) 
= 
\mathbb{E} \left\{ 
    N_i^{-1}
    \sum_{j = 1}^{N_i} 
    \sum_{\mathbf{a}_i \in \mathcal{A} (N_i)} \allowbreak
    Y_{ij}(\mathbf{a}_i) \allowbreak
    Q(\mathbf{a}_i | \mathbf{X}_i, N_i) 
\right\}.$
Note the potential outcome is first averaged over all possible configurations of cluster-level treatment vectors under the policy distribution, 
then averaged over units in a cluster, 
and finally the expectation is taken over the super population of clusters to yield $\mu(Q)$.
Similarly, the expected value of the average potential outcome when treated under the policy $Q$ is defined by
$\mu_{1}(Q) 
= 
\mathbb{E} \Big\{ 
    N_i^{-1} 
    \sum_{j = 1}^{N_i} \allowbreak
    \sum_{\mathbf{a}_{i(-j)} \in \mathcal{A} (N_i-1)} \allowbreak
    Y_{ij}(1, \mathbf{a}_{i(-j)}) 
    Q(\mathbf{a}_{i(-j)} \allowbreak
    | \allowbreak
    \mathbf{X}_i, N_i) 
\Big\},$
where 
$Q(\mathbf{a}_{i(-j)} | \mathbf{X}_i, N_i) 
=  
Q(1, \mathbf{a}_{i(-j)} | \mathbf{X}_i, N_i) 
+  
Q(0, \mathbf{a}_{i(-j)} | \mathbf{X}_i, N_i)$ 
is the probability of all units in cluster $i$ other than $j$ receiving treatment $\mathbf{a}_{i(-j)}$ under the policy $Q$. 
The expected value of the average potential outcome when untreated under the policy $Q$ is defined analogously and is denoted by $\mu_{0}(Q)$.
While $\mu(Q)$ describes the overall expected outcome under the policy $Q$,
$\mu_t(Q)$ describes the expected outcome when a unit's treatment is fixed at $t \in \{0,1\}$ and other units receive treatments according to $Q$.
In the context of the COVID-19 vaccine example, 
$\mu(Q)$ quantifies the overall risk of COVID-19 under the policy $Q$ 
while
$\mu_1(Q)$ quantifies the risk of COVID-19 when an individual is vaccinated.
In the absence of interference, 
$Y_{ij}(t, \mathbf{a}_{i(-j)})$ does not depend on $\mathbf{a}_{i(-j)}$ and may be written simply as $Y_{ij}(t)$, 
in which case $\mu_{t}(Q) 
= 
\mathbb{E} \Big\{ 
    N_i^{-1} 
    \sum_{j = 1}^{N_i} \allowbreak
    Y_{ij}(t) 
\Big\}$
regardless of the choice of $Q$. 
Thus, changes in $\mu_t(Q)$ with respect to $Q$ provide a measure of the degree of interference present.

Causal effects are defined by contrasts of $\mu(Q)$, $\mu_1(Q)$, and $\mu_0(Q)$. 
For example, the direct effect for policy $Q$ is defined to be
$DE(Q) = \mu_1(Q) - \mu_0(Q)$,
which quantifies the effect of a unit receiving treatment under policy $Q$.
For policies $Q$ and $Q'$, 
the overall effect is defined to be
$OE(Q, Q') = \mu(Q) - \mu(Q')$,
which compares two policies overall.
The spillover effect when treated is defined by 
$SE_1(Q, Q') = \mu_1(Q) - \mu_1(Q')$,
which compares average potential outcomes when treated under policy $Q$ versus $Q'$.
For the COVID-19 vaccine example, $SE_1(Q, Q')$ could be used to quantify the difference between a vaccinated individual's risk of COVID-19 when 50\% versus 30\% of their neighbors are vaccinated. 
Similarly, define the spillover effect when untreated by
$SE_0(Q, Q') \allowbreak = \mu_0(Q) - \mu_0(Q')$.
Finally, the total effect is defined by
$TE(Q, Q') = \mu_1(Q) - \mu_0(Q')$,
which can be viewed as the sum of direct effect $\mu_1(Q) - \mu_0(Q)$
and spillover effect $\mu_0(Q) - \mu_0(Q')$,
measuring the direct treatment effect and the indirect effect spilled over from others due to interference.
For notational convenience, 
$\mu(Q)$ under policy $Q_{\scriptscriptstyle <\textup{NAME}>}(\mathbf{a}_i|\mathbf{X}_i, N_i;\theta)$ is sometimes denoted by
$\mu_{\scriptscriptstyle <\textup{NAME}>}(\theta)$.
Other causal estimands are denoted similarly, for example, 
$\mu_{\scriptscriptstyle <\textup{NAME}>, \scriptstyle 1}(\theta)$ instead of $\mu_1(Q)$ and
$OE_{\scriptscriptstyle <\textup{NAME}>}(\theta, \theta')$ instead of $OE(Q, Q')$.
For more discussion on the definition of the causal estimands, refer to \cite{tchetgen12}.

The causal estimands defined above are based on the potential outcomes, and thus they are not identifiable from the observed data without appropriate assumptions.
Lemma~\ref{lemma:ident} provides the identifiability of the causal estimands under the following assumptions:

\noindent
(A1) \textit{Consistency}: $Y_{ij} = \sum_{\mathbf{a}_i \in \mathcal{A}(N_i)} Y_{ij}(\mathbf{a}_{i})\mathbbm{1}(\mathbf{A}_i = \mathbf{a}_i)$

\noindent
(A2) \textit{Conditional Exchangeability}: $\mathbf{Y}_i(\mathbf{a}_i) \indep \mathbf{A}_i | \mathbf{X}_i, N_i \text{ for all } \mathbf{a}_i \in \mathcal{A}(N_i)$ 

\noindent
(A3) \textit{Positivity}: $\mathbb{P}(\mathbf{A}_i = \mathbf{a}_i|\mathbf{X}_i, N_i) \in (c,1-c) \text{ for all } \mathbf{a}_i \in \mathcal{A}(N_i) \text{ for some } c \in (0,1)$


\begin{lemma}
\label{lemma:ident}
    Under \textup{(A1) -- (A3)}, 
    the causal estimands defined above can be expressed as the form of
    $$\mathbb{E} \left\{ 
        \allowbreak
       \sum_{\mathbf{a}_i \in \mathcal{A}(N_i)} 
       \allowbreak
       w(\mathbf{a}_i, \mathbf{X}_i, N_i)^\top
       \allowbreak
       \mathbb{E} \big(\mathbf{Y}_i | \mathbf{A}_i = \mathbf{a}_i, \mathbf{X}_i, N_i \big)
       \allowbreak
       \right\}$$
    where 
    $w(\mathbf{a}_i, \mathbf{X}_i, N_i)$ is a length $N_i$ vector whose components are functions of $\mathbf{a}_i, \mathbf{X}_i, N_i$.
    Thus, the causal estimands are identifiable from the observed data.
    Specifically, 
    $w(\mathbf{a}_i, \mathbf{X}_i, N_i) 
    = 
    N_i^{-1} Q(\mathbf{a}_i| \mathbf{X}_i, N_i) \mathbf{J}_{N_i}$ 
    for $\mu(Q)$, 
    where $\mathbf{J}_{N_i}$ is a length $N_i$ column vector of ones,
    and
    $
    w(\mathbf{a}_i, \mathbf{X}_i, N_i)
    = 
    N_i^{-1}
    \big(
        \mathbbm{1}(a_{i1} = t) Q(a_{i(-1)}| \mathbf{X}_i, N_i), 
        \dots,
        \mathbbm{1}(a_{iN_i} = t) Q(a_{i(-N_i)}| \mathbf{X}_i, N_i) 
    \big)^\top $
    for $\mu_t(Q)$.
\end{lemma}

%
%
Below (A4) and (A5) are also assumed when deriving the large sample properties of the estimators proposed in the next section.

\noindent
(A4) \textit{Finite moments}: 
$\big| \mathbb{E}(Y_{ij}^p | \mathbf{A}_i, \mathbf{X}_i, N_i) \big| \le C$ 
for all $p \le 4$
and some $C < \infty$

\noindent
(A5) \textit{Finite cluster size}: $\mathbb{P}(N_i \le n_{\max}) = 1 \text{ for some } n_{\max} \in \mathbb{N}$

\section{Inference procedure}
\label{s:results}

\subsection{Nonparametric efficient influence function}

In this section, consistent, asymptotically normal, and efficient nonparametric estimators of the proposed causal estimands are constructed.
The construction of the estimators is based on nonparametric efficiency theory \citep{tsiatis06, kennedy16, hines22}, 
and in particular finding the efficient influence function (EIF) of the target causal estimand.
The proposed estimators attain the usual parametric $m^{-1/2}$ convergence rate (where $m$ is the number of clusters), even when the nuisance functions are estimated at rates slower than parametric $m^{-1/2}$ rate, as long as they are faster than $m^{-1/4}$ rate. 
This allows the utilization of a broader range of nonparametric and machine learning methods for nuisance function estimation, reducing the risk of model mis-specification.
Furthermore, each proposed estimator is nonparametric efficient since its variance equals the efficiency bound,
i.e., the lower bound of the variance of all regular and asymptotic linear estimators.
Theorem \ref{thm:eif} gives the EIFs of the target causal estimands, 
and the proof is given in the supplementary material Section A.2.

\begin{theorem}
\label{thm:eif}
Consider an estimand of the form in \textup{Lemma \ref{lemma:ident}}
\begin{align*}
    \Psi(w) 
    = 
    \mathbb{E} \left\{
        \sum_{\mathbf{a}_i \in \mathcal{A}(N_i)}
        w(\mathbf{a}_i, \mathbf{X}_i, N_i)^\top
        \mathbb{E} \big(
            \mathbf{Y}_i | 
            \mathbf{A}_i = \mathbf{a}_i, \mathbf{X}_i, N_i 
        \big) 
    \right\}.
\end{align*}
Assume the EIF of 
$w(\mathbf{a}, \mathbf{x}, n)$
is
$
\varphi_{w(\mathbf{a}, \mathbf{x}, n)}^{*}(\mathbf{O}_i) 
=
\{\mathbbm{1}(\mathbf{X}_i = \mathbf{x}, N_i = n)
/
d\mathbb{P}(\mathbf{x}, n)\}
\phi(\mathbf{A}_i, \mathbf{X}_i, N_i; \mathbf{a})$
for fixed 
$(\mathbf{a}, \mathbf{x}, n) 
\in 
\mathcal{A}(n) \times \mathcal{X}(n) \times \mathbb{N}$.
Then, the EIF of $\Psi(w)$ is
\begin{align*}
    \varphi^{*}(\mathbf{O}_i) 
    =& 
    \sum_{\mathbf{a}_i \in \mathcal{A}(N_i)}
    \big\{ w(\mathbf{a}_i, \mathbf{X}_i, N_i) + \phi(\mathbf{A}_i, \mathbf{X}_i, N_i; \mathbf{a}_i) \big\}^\top
    \mathbb{E} \big(\mathbf{Y}_i | \mathbf{A}_i = \mathbf{a}_i, \mathbf{X}_i, N_i \big) 
    \\
    &+ 
    \frac{1}{\mathbb{P}(\mathbf{A}_i|\mathbf{X}_i, N_i)}
    w(\mathbf{A}_i, \mathbf{X}_i, N_i)^\top 
    \left\{ 
        \mathbf{Y}_i - \mathbb{E} \big( \mathbf{Y}_i | \mathbf{A}_i, \mathbf{X}_i, N_i \big) 
    \right\}
    - 
    \Psi(w) 
    .
\end{align*}
\end{theorem}

In the subsequent section, 
nonparametric estimators are proposed based on the EIF by substituting the nuisance functions with their respective estimators.
Applying Theorem \ref{thm:eif} to $\mu(Q)$ yields a simplified and intuitive form of the EIF of $\mu(Q)$, which is described in Corollary \ref{cor:mu(Q)}.

\begin{corollary}
\label{cor:mu(Q)}
The EIF of the expected average potential outcome under policy $Q$, $\mu(Q)$, is
\begin{align*}
    \varphi_{\mu(Q)}^*(\mathbf{O}_i) 
    =&  
    \sum_{\mathbf{a}_i \in \mathcal{A}(N_i)}
        \big\{ 
            Q(\mathbf{a}_i | \mathbf{X}_i, N_i) 
            + 
            \phi_Q(\mathbf{A}_i, \mathbf{X}_i, N_i; \mathbf{a}_i) 
        \big\}
        \mathbb{E} \big(
            \overline{\mathbf{Y}}_{i} | \mathbf{A}_i = \mathbf{a}_i, \mathbf{X}_i, N_i 
        \big)
    \\
    &+ 
    \frac
        {Q(\mathbf{A}_{i}| \mathbf{X}_i, N_i)}
        {\mathbb{P}(\mathbf{A}_{i}|\mathbf{X}_i, N_i)}
    \left\{ 
        \overline{\mathbf{Y}}_i - \mathbb{E} \big( \overline{\mathbf{Y}}_i | \mathbf{A}_i, \mathbf{X}_i, N_i \big) 
    \right\}
    - \mu(Q)
\end{align*}
where the EIF of 
$Q(\mathbf{a} | \mathbf{x}, n)$
is
$
\varphi_{Q(\mathbf{a} | \mathbf{x}, n)}^{*}(\mathbf{O}_i)
=
\{\mathbbm{1}(\mathbf{X}_i = \mathbf{x}, N_i = n)
/
d\mathbb{P}(\mathbf{x}, n)\}
\phi_Q(\mathbf{A}_i, \mathbf{X}_i, N_i; \mathbf{a})$
for fixed 
$(\mathbf{a}, \mathbf{x}, n) 
\in 
\mathcal{A}(n) \times \mathcal{X}(n) \times \mathbb{N}$
and 
$\overline{\mathbf{Y}}_i = N_i^{-1} \sum_{j} Y_{ij}$.
\end{corollary}

Note $\varphi_{\mu(Q)}^*(\mathbf{O}_i)$ is the sum of two terms minus $\mu(Q)$.
The first term is a weighted average of the conditional expectation of the outcome 
$\mathbb{E} \big(\overline{\mathbf{Y}}_i | \mathbf{A}_i = \mathbf{a}_i, \mathbf{X}_i, N_i \big)$ 
and the second term is a bias correction term equal to a weighted residual; 
thus $\varphi_{\mu(Q)}^*(\mathbf{O}_i)$ has a form analogous to the augmented inverse propensity weighted (AIPW) estimator of the average treatment effect (ATE) under no interference setting.
Here
$\phi_Q(\mathbf{A}_i, \mathbf{X}_i, N_i; \mathbf{a}_i)$
corresponds to the estimation of the policy distribution $Q$ which may depend on the observed data distribution.
If $Q$ does not depend on the observed data distribution (e.g., type B policy),
then $\phi_Q(\mathbf{A}_i, \mathbf{X}_i, N_i; \mathbf{a}_i) = 0$,
and the resulting EIF has a similar form to \cite{park2022efficient}; see Section 4.1 for more details.
The bias correction term entails a weighted residual of $\overline{\mathbf{Y}}_i$ regressed on $\mathbf{A}_i, \mathbf{X}_i, N_i$, 
with the weight equal to the ratio of the cluster-level treatment probability in the counterfactual versus factual scenarios 
$Q(\mathbf{A}_{i}| \mathbf{X}_i, N_i)
/
\mathbb{P}(\mathbf{A}_{i}|\mathbf{X}_i, N_i)$.


\subsection{Nuisance functions}

To construct estimators based on the EIFs given in the previous section, the nuisance functions $\boldsymbol{\eta} = (G, H, w, \phi)$ which appear in the EIFs need to be estimated:
(i) cluster-level outcome regression
$G(\mathbf{a}_i, \mathbf{x}_i, n_i) 
= 
\mathbb{E} \big( 
    \mathbf{Y}_{i} | \mathbf{A}_i = \mathbf{a}_i, \mathbf{X}_i = \mathbf{x}_i, N_i = n_i 
\big)$;
(ii) cluster-level treatment probability
$H(\mathbf{a}_i, \mathbf{x}_i, n_i) 
= 
\mathbb{P} \big(
    \mathbf{A}_{i} = \mathbf{a}_{i} | \mathbf{X}_i = \mathbf{x}_i, N_i = n_i
\big)$;
(iii) weight function
$w(\mathbf{a}_i, \mathbf{x}_i, n_i)$;
and
(iv) EIF of the weight function
$\phi(\mathbf{a}'_i, \mathbf{x}_i, n_i; \mathbf{a}_i)$.
The weight function $w$ and the EIF $\phi$ are specific to the policy $Q$ of interest and the type of target estimand (whether $\mu(Q)$, $\mu_t(Q)$, or $OE(Q,Q')$, etc).
Oftentimes, $w$ and $\phi$ are functions of $H$ or $\pi$ (individual-level propensity score), 
and thus it is not always required to estimate the four nuisance functions separately.
Estimation of $G$ and $H$ may utilize various methods.
Parametric modeling of $G$ and $H$ may be employed, e.g., by using generalized linear mixed effects models, 
but at the risk of model mis-specification. 
Alternatively, flexible data-adaptive methods which allow for correlated (clustered) data can be used, 
for example, 
mixed effect machine learning \citep{ngufor19}, 
smoothed kernel regression for dependent data  \citep[Section 4.2 of][]{park2022efficient},
etc.

On the other hand, if it is reasonable to assume that $Y_{ij}$'s are conditionally independent given $\mathbf{A}_i$, $\mathbf{X}_i, N_i$,
and that
$A_{ij}$'s are conditionally independent given $\mathbf{X}_i, N_i$ (but not necessarily marginally independent),
individual-level nuisance functions may be estimated instead of cluster-level functions $G$ and $H$:
(i) individual-level outcome regression
$    g(j, \mathbf{a}_i, \mathbf{x}_i, n_i) 
    = 
    \mathbb{E} \big( Y_{ij} | \mathbf{A}_i = \mathbf{a}_i, \mathbf{X}_i = \mathbf{x}_i, N_i = n_i \big)
$;
and
(ii) individual-level propensity score
$    \pi(j, \mathbf{x}_i, n_i) 
    = 
    \mathbb{P}(A_{ij} = 1 | \mathbf{X}_i = \mathbf{x}_i, N_i = n_i)
$.
Estimates of $G$ and $H$ can then be constructed based on estimates of $g$ and $\pi$.
In practice, estimating these nuisance functions using nonparametric or machine learning methods is challenging because the input dimension varies by cluster size $n_i$, i.e., $\mathbf{a}_i$ is a $n_i$ dimensional vector and $\mathbf{x}_i$ is $n_i \times p$ dimensional.
In the simulation study and the real data analysis presented below, the nuisance functions are specified as follows:

(i)
$
    g(j, \mathbf{a}_i, \mathbf{x}_i, n_i) 
    =
    \mathbb{E} \big( Y_{ij} | A_{ij} = a_{ij}, \overline{\mathbf{A}}_{i(-j)} = \overline{\mathbf{a}}_{i(-j)}, \mathbf{X}_{ij} = \mathbf{x}_{ij} \big)
    \eqqcolon
    g^{*}(a_{ij}, \overline{\mathbf{a}}_{i(-j)}, \mathbf{x}_{ij}) 
$;

(ii)
$
    \pi(j, \mathbf{x}_i, n_i) 
    = 
    \mathbb{P}(A_{ij} = 1 | \mathbf{X}_{ij} = \mathbf{x}_{ij})
    \eqqcolon
    \pi^{*}(\mathbf{x}_{ij}) 
$

\noindent
where 
$\overline{\mathbf{A}}_{i(-j)} = (\sum_{k \ne j} A_{ik}) / (N_i-1)$
and
$\overline{\mathbf{a}}_{i(-j)} = (\sum_{k \ne j} a_{ik}) / (n_i-1)$. 
Here, $g^{*}: \{0,1\} \times (0,1) \times \mathbb{R}^p \mapsto \mathbb{R}$ 
and $\pi^{*}: \mathbb{R}^p \mapsto (0,1)$ have domains that do not vary by cluster size,
and thus standard regression methods can be used.
However, the methodological results derived below do not require this specification for the nuisance functions. 
For instance, the propensity score may be specified to depend both on the unit's covariates $\mathbf{x}_{ij}$ and on the average of other units' covariates in the same cluster $\overline{\mathbf{x}}_{i(-j)} = (\sum_{k \ne j} \mathbf{x}_{ik})/(n_i-1)$, i.e., 
$\pi(j, \mathbf{x}_i, n_i) 
=
\pi^{**}(\mathbf{x}_{ij}, \overline{\mathbf{x}}_{i(-j)})$
for some $\pi^{**}: \mathbb{R}^p \times \mathbb{R}^p \mapsto (0,1)$.

The $g$ and $\pi$ functions may be estimated using parametric models, e.g., linear or logistic regression. 
Alternatively, the nuisance functions can be estimated using flexible data-adaptive methods which are less susceptible to model mis-specification. 
For instance, in the sections below
$g$ and $\pi$ are estimated using the super learner algorithm \citep{vanderLaan07}, 
an ensemble estimator based on a library of parametric and data-adaptive (i.e., machine learning) methods.
The super learner ensemble estimator asymptotically attains the best performance of the algorithms included in the library,
while reducing the risk of model mis-specification by including a number of data-adaptive methods.

\subsection{Proposed estimators}

In this section, sample splitting estimators \citep{chernozhukov18} of the target causal estimands are proposed based on the EIFs given in the previous section.
Sample splitting estimation advantageously provides asymptotically normal and efficient estimators 
without restricting the complexity of the nuisance function estimators (for example, functions in the Donsker class).

As in Theorem \ref{thm:eif}, let 
$\Psi(w)
= 
\mathbb{E} \left\{
    \sum_{\mathbf{a}_i \in \mathcal{A}(N_i)}
    w(\mathbf{a}_i, \mathbf{X}_i, N_i)^\top
    \mathbb{E} \big(
        \mathbf{Y}_i | 
        \mathbf{A}_i = \mathbf{a}_i, \mathbf{X}_i, N_i 
    \big) 
\right\}$ 
denote a target causal estimand,
where $w$ is defined based on the target estimand.
Also, let
$\varphi^{}(\mathbf{O}_i; \boldsymbol{\eta})$ denote the uncentered EIF of $\Psi(w)$
such that
$\varphi^*(\mathbf{O}_i; \boldsymbol{\eta}) = \varphi^{}(\mathbf{O}_i; \boldsymbol{\eta}) - \Psi(w)$
is the EIF of $\Psi(w)$
and
$\mathbb{E} \{ \varphi^{}(\mathbf{O}_i; \boldsymbol{\eta}) \} \allowbreak
= \Psi(w)$.
Then, the estimator of $\Psi(w)$ is constructed as follows.
First, cluster-level data $(\mathbf{O}_1, \dots, \mathbf{O}_m)$ are randomly partitioned into $K$ disjoint groups. 
Let $S_i \in \allowbreak \{ 1, \dots, K \}, \allowbreak i=1,\dots,m$ denote group membership for cluster $i$ and
$m_k = \sum_{i=1}^{m} \mathbbm{1}(S_i=k)$ denote the size of group $k \in \{1, \dots, K\}$.
For group $k$, the nuisance functions 
$\boldsymbol{\eta} = (G, H, w, \phi)$ 
(or $\boldsymbol{\eta} = (g, \pi, w, \phi)$)
estimators are 
trained on data from groups other than group $k$;
denote this estimator by 
$\widehat{\boldsymbol{\eta}}^{(-k)} 
= 
(\widehat{G}^{(-k)},  \allowbreak
\widehat{H}^{(-k)},  \allowbreak
\widehat{w}^{(-k)}, \allowbreak
\widehat{\phi}^{(-k)})$
(or
$\widehat{\boldsymbol{\eta}}^{(-k)} 
= \allowbreak
(\widehat{g}^{(-k)}, \allowbreak
\widehat{\pi}^{(-k)}, \allowbreak
\widehat{w}^{(-k)}, \allowbreak
\widehat{\phi}^{(-k)})$).
Next, the uncentered EIFs are evaluated for clusters in group $k$ by substituting $\boldsymbol{\eta}$ with $\widehat{\boldsymbol{\eta}}^{(-k)}$
and then averaged to estimate the group-averaged uncentered EIF, 
which is denoted by $\mathbb{P}_m^k \big\{ \varphi^{}(\mathbf{O}; \widehat{\boldsymbol{\eta}}^{(-k)}) \big\} = 
m_k^{-1} \sum_{i: S_i=k} 
\varphi^{}(\mathbf{O}_i; \widehat{\boldsymbol{\eta}}^{(-k)})$.
Here, in general, 
$\mathbb{P}_m^k \big\{ f(\mathbf{O}) \big\}$ equals 
$m_k^{-1} \sum_{i: S_i=k} f(\mathbf{O}_i)$
for any function $f$ of $\mathbf{O}$.
Finally, the proposed nonparametric sample splitting 
(NSS)
estimator is obtained by averaging the estimated group-averaged uncentered EIFs over groups, i.e.,
\begin{align*}
    \widehat{\Psi}(w) 
    =
    \frac{1}{K} 
    \sum_{k=1}^{K} 
    \mathbb{P}_m^k 
    \big\{ 
        \varphi^{}(\mathbf{O}; \widehat{\boldsymbol{\eta}}^{(-k)}) 
    \big\}
    =
    \frac{1}{K} 
    \sum_{k=1}^{K} 
    \frac{1}{m_k} 
    \sum_{i: S_i=k}
        \Bigg[
        \sum_{\mathbf{a}_i \in \mathcal{A}(N_i)}
            \widehat{h}^{(-k)}(\mathbf{O}_i, \mathbf{a}_i) 
            +
            \widehat{l}^{(-k)}(\mathbf{O}_i) 
        \Bigg]
\end{align*}
where
$\widehat{h}^{(-k)}(\mathbf{O}_i, \mathbf{a}_i)
= \allowbreak
\big\{ 
    \widehat{w}^{(-k)}(\mathbf{a}_i, \mathbf{X}_i, N_i) 
    + 
    \allowbreak
    \widehat{\phi}^{(-k)}(\mathbf{A}_i, \mathbf{X}_i, N_i; 
    \allowbreak
    \mathbf{a}_i) 
\big\}^\top
\allowbreak
\widehat{G}^{(-k)} \big(
    \mathbf{a}_i, \mathbf{X}_i, N_i 
\big) $
and
$\widehat{l}^{(-k)}(\mathbf{O}_i)
\allowbreak
=
\widehat{w}^{(-k)}(\mathbf{A}_i, \mathbf{X}_i, N_i)^\top 
\big\{ 
    \mathbf{Y}_i 
    - 
    \widehat{G}^{(-k)} \big( 
        \mathbf{A}_i, \mathbf{X}_i, N_i 
    \big) 
\big\}
\big/
\widehat{H}^{(-k)}(\mathbf{A}_i,\mathbf{X}_i, N_i)$.
The variance of $\widehat{\Psi}(w)$ can be estimated by
$
    \widehat{\sigma}^2(w) 
    = 
    K^{-1}
    \sum_{k=1}^{K} 
    \mathbb{P}_m^k 
    \Big[ 
        \big\{ 
            \varphi^{}(\mathbf{O}; \widehat{\boldsymbol{\eta}}^{(-k)}) - \widehat{\Psi}(w) 
        \big\}^2
    \Big]
$.

There are two challenges when implementing the proposed estimators.
First, the summation 
$\sum_{\mathbf{a}_i \in \mathcal{A}(N_i)}
    \widehat{h}^{(-k)}(\mathbf{O}_i, \mathbf{a}_i)$ 
can be computationally intensive for moderately large $N_i$.
That said, this summation can be approximated by 
$
2^{N_i}
\sum_{q=1}^{r}
\widehat{h}^{(-k)}(\mathbf{O}_i, \mathbf{a}_i^{(q)})
/r
$,
which we refer to as the subsampling approximation,
where 
$r < 2^{N_i}$ is a user-specified number and
$\mathbf{a}_i^{(q)}$ $(q=1,\dots,r)$ is randomly sampled from $\mathcal{A}(N_i)$.
Note that this approximation is unrelated to the specification of the nuisance functions.
For example, suppose the individual-level outcome regression model depends only on $a_{ij}$ and $\overline{\mathbf{a}}_{i(-j)}$, i.e.,
$
g(j, \mathbf{a}_i, \mathbf{x}_i, n_i)
\allowbreak
=
g^{*}(a_{ij}, \overline{\mathbf{a}}_{i(-j)}, \mathbf{x}_{ij}) 
$
as in the previous section.
Then, for $\mathbf{a}_{i}, \mathbf{a}'_{i} \in \mathcal{A}(n_i)$ such that
$(a_{ij}, \overline{\mathbf{a}}_{i(-j)}) 
= 
(a'_{ij}, \overline{\mathbf{a}'}_{i(-j)})$,
it is true that
$g^{*}(a_{ij}, \overline{\mathbf{a}}_{i(-j)}, \mathbf{x}_{ij})
\allowbreak
=
g^{*}(a'_{ij}, \overline{\mathbf{a}'}_{i(-j)}, \mathbf{x}_{ij})$,
but in general it will not also be the case that
$\widehat{h}^{(-k)}(\mathbf{O}_i, \mathbf{a}_i)
=
\widehat{h}^{(-k)}(\mathbf{O}_i, \mathbf{a}'_i)$.
Therefore, 
the proposed estimator requires computation of the summation of $\widehat{h}^{(-k)}(\mathbf{O}_i, \mathbf{a}_i)$ over all binary treatment vectors $\mathbf{a}_i \in \mathcal{A}(N_i)$
despite the assumed outcome regression specification.
The large sample properties and finite sample performance of the proposed estimators under the subsampling approximation are presented in the supplementary material Sections A.10 and C.3, respectively.

Second, the sample splitting estimator depends on a specific sample split, introducing finite sample variability of the estimator. 
Therefore, one can repeat splitting the sample to construct the estimator $S$ times 
and then take the median of $S$ estimators to get a split-robust estimator
\citep{chernozhukov18}.
Larger values of $S$ are recommended to reduce variability. 
In the simulation studies presented below $S=1$ worked well, 
and in the data analysis in Section 6, 
$S=30$ yielded stable results (see supplementary material Section D.3).

\subsection{Theoretical results}

In this section, the large sample properties of the proposed estimators are derived.
Hereinafter, let $\norm{f}_{L_2(\mathbb{P})} = \{ \int f(\mathbf{o})^2 d\mathbb{P}(\mathbf{o}) \}^{1/2}$ denote the squared $L_2(\mathbb{P})$ norm, 
which is the square root of the expectation of $f(\mathbf{O})^2$,
treating the function $f$ as fixed even when it is estimated from the sample and thus random.
Let $\norm{\cdot}_2$ to be the Euclidean norm of a vector.
Let $O_\mathbb{P}$ and $o_\mathbb{P}$ denote the usual big O and little o quantities with respect to the observed data distribution $\mathbb{P}$.

We first introduce assumptions about the nuisance functions in addition to the assumptions (A1) $-$ (A5).
For all 
$\mathbf{a}, \mathbf{a}' \in \mathcal{A}(n)$,
$\mathbf{x} \in \mathcal{X}(n)$,
$n \in \mathbbm{N}$,
and $k = 1,\dots,K$,
there exist $c \in (0,1)$ and $C \in (0,\infty)$
and $r_G, r_H, r_{\phi}, r_w > 0$ satisfying
\vspace{-0.2cm}
\begin{enumerate}
    \item[(B1)] 
    \textit{Bounded} $H$ and $\widehat{H}^{(-k)}$: 
    $H(\mathbf{a}, \mathbf{x}, n) \in (c, 1-c)$
    and
    $\widehat{H}^{(-k)}(\mathbf{a}, \mathbf{x}, n) \in (c, 1-c)$
    
    \item[(B2)] 
    \textit{Bounded} $G$ and $\widehat{G}^{(-k)}$: 
    $\normt{G(\mathbf{a}, \mathbf{x}, n)} \le C$
    and
    $\normt{\widehat{G}^{(-k)}(\mathbf{a}, \mathbf{x}, n)} \le C$

    \item[(B3)] 
    \textit{Bounded} $\phi$ and $\widehat{\phi}^{(-k)}$: 
    $\normt{\phi(\mathbf{a}', \mathbf{x}, n; \mathbf{a})} \le C$
    and
    $\normt{\widehat{\phi}^{(-k)}(\mathbf{a}', \mathbf{x}, n; \mathbf{a})} \le C$

    \item[(B4)] 
    \textit{Bounded} $w$ and $\widehat{w}^{(-k)}$: 
    $\normt{w(\mathbf{a}, \mathbf{x}, n)} \le C$
    and
    $\normt{\widehat{w}^{(-k)}(\mathbf{a}, \mathbf{x}, n)} \le C$
    
    \item[(B5)] 
    \textit{Convergence rate of} $\widehat{H}^{(-k)}$: 
    $
    \normp{
        \sum_{\mathbf{a}\in \mathcal{A}(N)}
            \big|
                \big(\widehat{H}^{(-k)}-H\big)(\mathbf{a}, \mathbf{X}, N)
            \big|
    }
    =
    O_\mathbb{P}(r_H)
    $

    \item[(B6)] 
    \textit{Convergence rate of} $\widehat{G}^{(-k)}$: 
    $
    \normp{
        \sum_{\mathbf{a}\in \mathcal{A}(N)}
            \normt{
                \big(\widehat{G}^{(-k)}-G\big)(\mathbf{a}, \mathbf{X}, N)
            }
    }
    =
    O_\mathbb{P}(r_G)
    $

    \item[(B7)] 
    \textit{Convergence rate of} $\widehat{\phi}^{(-k)}$: 
    $
    \normp{
        \sum_{\mathbf{a}\in \mathcal{A}(N)}
            \normt{
                \big(
                    \widehat{\phi}^{(-k)} - \phi
                \big)
                (\mathbf{A}, \mathbf{X}, N; \mathbf{a})
            }
    }
    =
    O_\mathbb{P}(r_\phi)
    $

    \item[(B8)] 
    \textit{Second order convergence rate of} $\widehat{w}^{(-k)}$: 
    \vspace{-0.2cm}
    $$
    \norm\Bigg{
        \sum_{\mathbf{a}\in \mathcal{A}(N)}
            \norm\bigg{
                \big(
                \widehat{w}^{(-k)}
                -
                w
                \big)
                (\mathbf{a}, \mathbf{X}, N)
                +
                \sum_{\mathbf{a}' \in \mathcal{A}(N)}
                    \widehat{\phi}^{(-k)}(\mathbf{a}', \mathbf{X}, N; \mathbf{a})
                    H(\mathbf{a}', \mathbf{X}, N)
            }_2
    }_{L_2(\mathbb{P})}
    =
    O_\mathbb{P}(r_w^2)
    $$    

\end{enumerate}

\noindent
Assumptions (B1) $-$ (B4) bound the nuisance functions and their estimators, 
while (B5) $-$ (B8) specify the convergence rate of nuisance functions estimators.
Note that (B8) describes the convergence rate of $\widehat{w}^{(-k)}$ combined with $\widehat{\phi}^{(-k)}$ rather than that of $\widehat{w}^{(-k)}$ only.
The quantity in (B8) is the second order remainder term in the von Mises expansion of $w$, where $\phi$ is a pathwise derivative of $w$ \citep{fisher2021visually, kennedy22, hines22}.
Refer to the supplementary material Section A.9 for technical details.
The second order remainder is expected to have a faster convergence rate than the first order remainder $\widehat{w}^{(-k)}-w$,
and oftentimes the rate depends on the square of the difference between $\widehat{w}^{(-k)}$ and $w$,
which will be determined by the actual form of $w$ and $\phi$.
Under the assumptions above,
Theorem~\ref{thm:consistency} provides the consistency of $\widehat{\Psi}(w)$.

\begin{theorem}
\label{thm:consistency}
Assume \textup{(B1) $-$ (B8)} hold.
If
\textup{(i)} $r_w = o(1)$ and \textup{(ii)} $r_G = o(1)$ or $r_H = r_{\phi} = o(1)$ as $m \to \infty$,
then $\widehat{\Psi}(w) \overset{p}{\to} \Psi(w)$.
\end{theorem}

Unlike the AIPW estimator of ATE in no interference setting, the proposed estimator is not multiply robust in the sense that
the consistency of $\widehat{\Psi}(w)$ is only guaranteed if $w$ is consistently estimated.
Nevertheless, if $w$ is consistently estimated, the $G$ estimator need not to be consistent as long as the $H$ and $\phi$ estimators are consistent, and vice versa.
Oftentimes, 
the consistency of the $w$ and $\phi$ estimators depend upon the consistency of the $H$ estimator,
which will be clear when investigating examples in Section 4.
With stronger rate conditions on the nuisance functions estimators, $\widehat{\Psi}(w)$ is nonparametric efficient and converges in distribution to a Normal distribution, as stated in Theorem~\ref{thm:weakconv}.

\begin{theorem}
\label{thm:weakconv}
Assume \textup{(B1) $-$ (B8)} hold. 
If 
\textup{(i)} $r_G = r_H = r_{\phi} = o(1)$, 
\textup{(ii)} $r_w = o(m^{-1/4})$, 
and \textup{(iii)} $r_G(r_H + r_{\phi}) = o(m^{-1/2})$ 
as $m \to \infty$,
then
$
\sqrt{m}\{\widehat{\Psi}(w) - \Psi(w)\}
\overset{d}{\to}
N(0,\sigma^2(w))
$,
where 
$
\sigma^2(w)
=
\mathbb{E}
    \Big[ 
        \big\{ 
            \varphi^{*}(\mathbf{O}; \boldsymbol{\eta}) 
        \big\}^2
    \Big]
$
is the nonparametric efficiency bound of $\Psi(w)$.
\end{theorem}  

The rate conditions in Theorem \ref{thm:weakconv} are slower than the usual parametric $m^{-1/2}$ convergence rate since
$r_w = r_G = r_H = r_{\phi} = o(m^{-1/4})$ as $m \to \infty$ is a sufficient condition,
which can be achieved by using an ensemble of data-adaptive methods. 
Theorem~\ref{thm:varest} presents the consistency of the variance estimator
$
    \widehat{\sigma}^2(w) 
    = 
    K^{-1}
    \sum_{k=1}^{K} 
    \mathbb{P}_m^k 
    \Big[ 
        \big\{ 
            \varphi^{}(\mathbf{O}; \widehat{\boldsymbol{\eta}}^{(-k)}) - \widehat{\Psi}(w) 
        \big\}^2
    \Big]
$,
providing the basis for inference of $\Psi(w)$ which will be used in the simulation and real data analysis sections below.

\begin{theorem}
\label{thm:varest}
Assume \textup{(B1) $-$ (B8)} hold. 
If $r_w = r_G = r_H = r_{\phi} = o(1)$ as $m \to \infty$,
then $\widehat{\sigma}^2(w) \overset{p}{\to} {\sigma}^2(w)$, 
i.e.,
$\widehat{\sigma}^2(w)$ is a consistent estimator of the asymptotic variance of $\widehat{\Psi}(w)$.
In addition, if the conditions in Theorem~\ref{thm:weakconv} hold,
then
$
\sqrt{m}\{\widehat{\Psi}(w) - \Psi(w)\}/\widehat{\sigma}(w)
\overset{d}{\to}
N(0,1)
$.

\end{theorem}

In conclusion, under mild conditions, the proposed estimators are consistent, asymptotically normal, and nonparametric efficient.

\section{Examples}

In this section, the broad applicability of the proposed inferential methods is demonstrated by considering the four counterfactual policies defined in Section 2.2. 
Here, the large sample properties of $\widehat{\mu}(Q)$ are investigated for each policy $Q$.
For technical details, refer to the supplementary material Section B.

\subsection{Type B policy}

Since the type B policy distribution does not depend on the observed data distribution,
the EIF of weight function $\phi = 0$ 
and the proposed estimators are doubly robust,
i.e.,
the estimators are consistent if either $G$ or $H$,
but not necessarily both,
is consistently estimated.
In particular, as $m \to \infty$,
(i) $r_G = o(1)$ or $r_H = o(1)$ implies consistency of $\widehat{\mu}_{\scriptscriptstyle \textup{B}}(\alpha)$,
(ii) $r_G \cdot r_H = o(m^{-1/2})$ implies asymptotic normality
of $\widehat{\mu}_{\scriptscriptstyle \textup{B}}(\alpha)$,
and
(iii) $r_G = r_H = o(1)$ implies consistency of the variance estimator $\widehat{\sigma}_{\scriptscriptstyle \textup{B}}^2(\alpha)$ for each fixed $\alpha \in (0,1)$.
Furthermore, the product of the two estimators' convergence rates being $m^{-1/2}$ is sufficient for asymptotic normality,
allowing each rate to be slower than the usual parametric $m^{-1/2}$ rate,
which can be achieved using slow yet robust nonparametric estimation methods.
These results are essentially the same as those given in Section 4.2 of \cite{park2022efficient}.

If true $G$ and $H$ have a parametric form,
then \cite{liu19} proposed parametric doubly robust estimators of the type B policy estimands which are consistent and asymptotically normal if either the $G$ or $H$ model (but not necessarily both) is correctly specified.
However, the \cite{liu19} estimators will not be consistent if the true $G$ and $H$ do not have parametric forms, 
or if the analyst fails to correctly specify the $G$ and $H$ models. 
In contrast, the NSS estimator does not rely on parametric assumptions, thus mitigating the risk for model mis-specification.

\subsection{Cluster incremental propensity score policy}


Suppose the conditional independencies in Section 3.2 are assumed such that the individual-level nuisance functions $g$ and $\pi$ are estimated and then used to construct $G$, $H$, $w$, and $\phi$.
In detail, if 
$ 
\normp{
    \sum_{j=1}^{N}
        \lvert
            (\widehat{\pi}^{(-k)} - \pi) (j, \mathbf{X}, N)
        \rvert
}
=
O_\mathbb{P}(r_{\pi})
$
and
$
\normp{
    \sum_{\mathbf{a} \in \mathcal{A}(N)}
    \sum_{j=1}^{N}
        \lvert
            (\widehat{g}^{(-k)} - g) (j, \mathbf{a}, \mathbf{X}, N)
        \rvert
}
\allowbreak
=
O_\mathbb{P}(r_g)$
for some $r_{\pi}, r_g > 0$,
then
$r_H = O(r_{\pi})$,
$r_G = O(r_g)$,
$r_{\phi} = O(r_{\pi})$,
and
$r_{w} = O(r_{\pi})$.
Therefore, when $m \to \infty$,
(i) $r_{\pi} = o(1)$ implies consistency of $\widehat{\mu}_{\scriptscriptstyle \textup{CIPS}}(\delta)$,
(ii) $r_{\pi} = o(m^{-1/4})$ and $r_{\pi} \cdot r_g = o(m^{-1/2})$ imply asymptotic normality
of $\widehat{\mu}_{\scriptscriptstyle \textup{CIPS}}(\delta)$,
and
(iii) $r_{\pi} = r_g = o(1)$ implies consistency of the variance estimator 
$\widehat{\sigma}_{\scriptscriptstyle \textup{CIPS}}^2(\delta)$
for each fixed $\delta > 0$.

Unlike the type B policy estimator, 
the CIPS policy estimator is not doubly robust since 
the consistency of 
$\widehat{\mu}_{\scriptscriptstyle \textup{CIPS}}(\delta)$
is not guaranteed when $r_g = o(1)$ but not $r_{\pi} = o(1)$.
However, as long as $\pi$ is consistently estimated, $g$ needs not to be consistently estimated.
The rate condition for asymptotic normality is stronger for the propensity score model than the outcome regression model yet can be slower than the usual parametric $m^{-1/2}$ convergence rate,
i.e., $r_{\pi} = r_g = o(m^{-1/4})$ is a sufficient condition.

\subsection{Cluster multiplicative shift policy}


Similar to the CIPS policy,
the CMS policy estimands depicts the counterfactual scenario that the individual-level propensity score distribution is shifted in multiplicative scale.
It can be shown that sufficient conditions for the large sample properties (consistency, asymptotic normality) of 
$\widehat{\mu}_{\scriptscriptstyle \textup{CMS}}(\lambda)$
are the same as that of 
$\widehat{\mu}_{\scriptscriptstyle \textup{CIPS}}(\delta)$;
for example, the proposed CMS policy estimator is consistent if $\pi$ is consistently estimated, 
and $g$ need not to be consistently estimated as long as $\pi$ is consistently estimated.

Like the CIPS estimator, the CMS policy estimator lacks double robustness.
However, 
if $N_i=1$ for all $i$ (there is no interference),
then the CMS policy reduces to the multiplicative shift (MS) policy considered by \cite{wen23} and double robustness is achieved (see Example 3 in Section 6.1 of \cite{wen23}). 
On the other hand, 
if $N_i > 1$ for some $i$, 
then double robustness does not hold, 
highlighting the importance of accurate estimation of the propensity score when targeting the CMS policy estimand in the presence of clustered interference.

\subsection{Treated proportion bound policy}
Unlike the CMS and CIPS policies, the TPB policy is defined based on cluster-level treatment probability 
$H(\mathbf{a}_i, \mathbf{x}_i, n_i) 
= 
\mathbb{P} \big(
    \mathbf{A}_{i} = \mathbf{a}_{i} | \mathbf{X}_i = \mathbf{x}_i, N_i = n_i
\big)$.
It can be shown that
$r_{\phi} = O(r_H)$ and
$r_{w} = O(r_H)$,
and thus, when $m \to \infty$,
(i) $r_H = o(1)$ implies consistency
of $\widehat{\mu}_{\scriptscriptstyle \textup{TPB}}(\rho)$,
(ii) $r_{H} = o(m^{-1/4})$ and $r_{H} \cdot r_G = o(m^{-1/2})$ implies asymptotic normality
of $\widehat{\mu}_{\scriptscriptstyle \textup{TPB}}(\rho)$,
and
(iii) $r_{H} = r_G = o(1)$ implies consistency of the variance estimator 
$\widehat{\sigma}_{\scriptscriptstyle \textup{TPB}}^2(\rho)$
for each fixed $\rho \in (0,1)$.

Similar to CMS and CIPS policies, 
the TPB policy NSS estimators are not doubly robust since consistent estimation of the TPB estimands is not guaranteed if only $G$ is consistently estimated.
Nevertheless, as long as $H$ is consistently estimated, $G$ need not to be consistently estimated.
Likewise, a stronger convergence rate condition is required for $H$ than $G$ for asymptotic normality, 
reflecting the fact that the policy distribution depends on $H$. 


\subsection{Remarks}

Thus far it has been shown that the type B estimators are doubly robust, while the CIPS, CMS, and TPB estimators are not.
This difference arises from whether the policy depends on the observed data distribution or not.
Even though the large sample properties presented above are with respect to $\widehat{\mu}(Q)$ only, 
analogous results hold for the proposed NSS estimators of the other causal estimands.

Considering a collection of policies 
$Q(\mathbf{a}_i|\mathbf{X}_i, N_i;\theta)$ indexed by $\theta$,
oftentimes it is possible to obtain a stronger theoretical result than the point-wise asymptotic normality.
For example, the process 
$\{
    \widehat{\mu}(\theta):
    \theta \in \Theta
\}$ 
is nonparametric efficient at each $\theta$ and weakly converges to a Gaussian process,
as stated in the following Theorems \ref{thm:typeBweakconv}$-$\ref{thm:CMSweakconv} for the type B, CIPS, and CMS policies.
Let $\ell^{\infty} (\Theta)$ denote the function space with the supremum norm $\norm{f}_{\Theta} = \sup_{\theta \in \Theta}|f(\theta)|$.

\begin{theorem}
\label{thm:typeBweakconv}
Assume \textup{(B1) $-$ (B8)} hold with $r_{G} \cdot r_H = o(m^{-1/2})$.
Consider the collection of type B policies indexed by 
$\alpha \in \mathbb{A}
= 
[\alpha_l, \alpha_u]$, where $0< \alpha_l < \alpha_u < 1$.
Then,
$\sqrt{m}\{
    \widehat{\mu}_{\scriptscriptstyle \textup{B}}(\cdot) 
    - 
    \mu_{\scriptscriptstyle \textup{B}}(\cdot)
\}
\rightsquigarrow
\mathbb{G}(\cdot)$
in $\ell^{\infty} (\mathbb{A})$
as $m \to \infty$,
where $\mathbb{G}(\cdot)$ is a mean zero Gaussian process with covariance
$
\mathbb{E} \{ \mathbb{G}(\alpha) \mathbb{G}(\alpha') \} \allowbreak
= 
\mathbb{E} \{ 
    \varphi_{\mu_{\scriptscriptstyle \textup{B}}(\alpha)}^{*}(\mathbf{O})
    \allowbreak
    \varphi_{\mu_{\scriptscriptstyle \textup{B}}(\alpha')}^{*}(\mathbf{O}) 
\} 
$
where
$\varphi_{\mu_{\scriptscriptstyle \textup{B}}(\alpha)}^{*}(\mathbf{O})$
is the EIF of
$\mu_{\scriptscriptstyle \textup{B}}(\alpha)$.
\end{theorem}

\begin{theorem}
\label{thm:CIPSweakconv}
Assume \textup{(B1) $-$ (B8)} hold with $r_{\pi} = o(m^{-1/4})$ and $r_{\pi} \cdot r_g = o(m^{-1/2})$.
Consider the collection of CIPS policies with constant $\delta(\mathbf{X}_i, N_i) = \delta_0$ indexed by 
$\delta_0 \in \mathbb{D} 
= 
[\delta_l, \delta_u]$, where $0< \delta_l < \delta_u < \infty$.
Then,
$\sqrt{m}\{
    \widehat{\mu}_{\scriptscriptstyle \textup{CIPS}}(\cdot) 
    - 
    \mu_{\scriptscriptstyle \textup{CIPS}}(\cdot)
\}
\rightsquigarrow
\mathbb{G}(\cdot)$
in $\ell^{\infty} (\mathbb{D})$
as $m \to \infty$,
where $\mathbb{G}(\cdot)$ is a mean zero Gaussian process with covariance
$
\mathbb{E} \{ \mathbb{G}(\delta_0) \mathbb{G}(\delta_0') \} \allowbreak
= 
\mathbb{E} \{ 
    \varphi_{\mu_{\scriptscriptstyle \textup{CIPS}}(\delta_0)}^{*}(\mathbf{O})
    \allowbreak
    \varphi_{\mu_{\scriptscriptstyle \textup{CIPS}}(\delta_0')}^{*}(\mathbf{O}) 
\} 
$
where
$\varphi_{\mu_{\scriptscriptstyle \textup{CIPS}}(\delta_0)}^{*}(\mathbf{O})$
is the EIF of
$\mu_{\scriptscriptstyle \textup{CIPS}}(\delta_0)$.
\end{theorem}

\begin{theorem}
\label{thm:CMSweakconv}
Assume \textup{(B1) $-$ (B8)} hold with
$r_{\pi} = o(m^{-1/4})$ and $r_{\pi} \cdot r_g = o(m^{-1/2})$.
Consider the collection of CMS policies indexed by 
$\lambda \in \mathbb{L}
= 
[\lambda_l, \lambda_u]$, where $0< \lambda_l < \lambda_u < 1$.
Then,
$\sqrt{m}\{
    \widehat{\mu}_{\scriptscriptstyle \textup{CMS}}(\cdot) 
    - 
    \mu_{\scriptscriptstyle \textup{CMS}}(\cdot)
\}
\rightsquigarrow
\mathbb{G}(\cdot)$
in $\ell^{\infty} (\mathbb{L})$
as $m \to \infty$,
where $\mathbb{G}(\cdot)$ is a mean zero Gaussian process with covariance
$
\mathbb{E} \{ \mathbb{G}(\lambda) \mathbb{G}(\lambda') \} \allowbreak
= 
\mathbb{E} \{ 
    \varphi_{\mu_{\scriptscriptstyle \textup{CMS}}(\lambda)}^{*}(\mathbf{O})
    \allowbreak
    \varphi_{\mu_{\scriptscriptstyle \textup{CMS}}(\lambda')}^{*}(\mathbf{O}) 
\} 
$
where
$\varphi_{\mu_{\scriptscriptstyle \textup{CMS}}(\lambda)}^{*}(\mathbf{O})$
is the EIF of
$\mu_{\scriptscriptstyle \textup{CMS}}(\lambda)$.
\end{theorem}

Theorem \ref{thm:CIPSweakconv} can be viewed as an extension of Theorem 3 in \cite{kennedy19} to the clustered interference setting.
Note that there is no weak convergence result for TPB policy since the policy has discontinuity with respect to the index $\rho$.

\section{Simulation study}

To assess the finite sample performance of the proposed NSS estimators, $D = 1000$ datasets were simulated.
Each dataset consisted of $m = 500$ clusters, and the number of units $N_i$ in cluster $i = 1, \dots, m$ was randomly sampled from $\{5,6,\dots,20\}$.
For each cluster, one cluster-level covariate $C_{i}$ was generated from a standard Normal distribution $\text{N}(0,1)$.
For unit $j = 1,\dots,N_i$ in cluster $i$, two independent covariates $X_{ij1} \sim \text{N}(0,1)$ and $X_{ij2} \sim \text{Bernoulli}(0.5)$ were generated.
The treatment status was generated from 
$A_{ij} 
\sim 
\text{Bernoulli} (
    \text{expit}(0.1 + 0.2|X_{ij1}| + 0.2|X_{ij1}|X_{ij2} + 0.1\mathbbm{1}(C_i > 0))
)$,
and the outcome was generated from 
$Y_{ij} 
\sim 
\text{Bernoulli} (
    \text{expit}(3 - 2A_{ij} - \overline{\mathbf{A}}_{i(-j)} - 1.5|X_{ij1}| + 2X_{ij2} - 3|X_{ij1}|X_{ij2} - 2\mathbbm{1}(C_i > 0))
)$,
where $\text{expit}(x) = 1/(1+e^{-x})$.
The target parameters were the counterfactual means and causal effects for the CIPS policy with constant $\delta(\mathbf{X}_i, N_i) = \delta_0 \in \{0.5, 1, 2\}$ 
and 
TPB policy with $\rho \in \{0.3, 0.45, 0.6\}$.
The CIPS policy estimands with varying $\delta(\mathbf{X}_i, N_i) = \delta_0 (1+1/N_i)$ were also investigated (see supplementary material Section C.1).
For all scenarios, two kinds of estimators were constructed and compared.
First, the NSS estimators were evaluated with $K=2$, $r=100$, and $S=1$.
The individual-level propensity score model $\pi$ was fit by taking ${X}_{ij1}, {X}_{ij2}$, and $C_i$ as the covariates and $A_{ij}$ as the outcome,
while the individual-level outcome regression model $g$ was fit by taking $A_{ij}$, $\overline{\mathbf{A}}_{i(-j)}$, ${X}_{ij1}, {X}_{ij2}$, and $C_i$ as the covariates and $Y_{ij}$ as the outcome.
Both nuisance functions were fit using the ensemble of main effects only logistic regression, random forest, generalized additive model, and a single-layer neural network
\citep[SuperLearner package in R by][]{superlearner}.
For sake of comparison, parametric sample splitting (PSS) estimators were also evaluated,
which were constructed via the same procedure, yet both nuisance functions were estimated by the main effects only logistic regression model.

Denoting the estimator of an estimand $\Psi$ and its standard error (SE) estimator from the $d^{\text{th}}$ simulated dataset by $\widehat{\Psi}_{d}$ and $\widehat{\sigma}_{d}$, 
bias (Bias = $\sum_{d=1}^{D} (\widehat{\Psi}_{d} - \Psi)/D$), 
root-mean-squared error (RMSE = $ \big\{\sum_{d=1}^{D} (\widehat{\Psi}_{d} - \Psi)^{2}/D \big\}^{1/2}$), 
average SE (ASE = $\sum_{d=1}^{D} \widehat{\sigma}_{d} / D$), 
empirical SE (ESE = $\text{sd} \{ \widehat{\Psi}_{d} \}_{d=1}^D$), 
and 95$\%$ point-wise Wald confidence interval (CI) coverage (Cov \%) were computed for each scenario.
For comparison of the NSS and PSS estimators, 
the RMSE Ratio (RMSE of the nonparametric method divided by RMSE of the parametric method) was also calculated.

The simulation results are given in Tables~\ref{tab:simulCIPS} and \ref{tab:simulTPB}.
For all policies,
the NSS estimators performed well, with minimal bias, excellent agreement between the ASE and ESE, and corresponding 95\% CI coverage very close to the nominal level. 
In contrast, the PSS estimator performed poorly, with greater bias and CI coverage often well below the desired 95\% rate. 
The poor performance of the parametric estimators is not surprising, given mis-specification of the outcome regression and propensity score models.
These results demonstrate that the proposed NSS estimators
are more robust to model mis-specification by utilization of ensemble estimators of the nuisance functions.

\begin{table}[H]
\captionsetup{width=.9\textwidth}
\caption{\small Simulation results for nonparametric and parametric sample splitting estimators for CIPS policy with constant $\delta$}
\label{tab:simulCIPS}
\centering
\resizebox{0.9\textwidth}{!}{%
\begin{tabular}{cccccccccccccccc}
\hline
               &        &  & \multicolumn{5}{c}{Nonparametric}       &  & \multicolumn{5}{c}{Parametric}          &  & RMSE  \\ \cline{4-8} \cline{10-14}
Estimand       & Truth  &  & Bias   & RMSE  & ASE   & ESE   & Cov \% &  & Bias   & RMSE  & ASE   & ESE   & Cov \% &  & Ratio \\ \hline
$\mu_{\scriptscriptstyle \textup{CIPS}}(0.5)$ & 0.436 &  & 0.002 & 0.018 & 0.017 & 0.017 & 94.9\% &  & 0.012 & 0.023 & 0.019 & 0.020 & 90.3\% &  & 0.77 \\
$\mu_{\scriptscriptstyle \textup{CIPS}, \scriptstyle 1}(0.5)$ & 0.264 &  & -0.002 & 0.016 & 0.015 & 0.016 & 93.8\% &  & -0.011 & 0.020 & 0.016 & 0.017 & 87.1\% &  & 0.80 \\
$\mu_{\scriptscriptstyle \textup{CIPS}, \scriptstyle 0}(0.5)$ & 0.555 &  & 0.006 & 0.024 & 0.023 & 0.024 & 93.6\% &  & 0.035 & 0.044 & 0.025 & 0.027 & 68.6\% &  & 0.55 \\
$DE_{\scriptscriptstyle \textup{CIPS}}(0.5)$ & -0.291 &  & -0.008 & 0.025 & 0.023 & 0.024 & 93.2\% &  & -0.046 & 0.053 & 0.026 & 0.027 & 51.3\% &  & 0.48 \\
$SE_{\scriptscriptstyle \textup{CIPS}, \scriptstyle 1}(0.5, 1)$ & 0.021 &  & 0.001 & 0.012 & 0.012 & 0.012 & 94.6\% &  & 0.003 & 0.014 & 0.014 & 0.014 & 94.3\% &  & 0.86 \\
$SE_{\scriptscriptstyle \textup{CIPS}, \scriptstyle 0}(0.5, 1)$ & 0.025 &  & 0.003 & 0.020 & 0.018 & 0.020 & 94.0\% &  & 0.006 & 0.024 & 0.021 & 0.023 & 93.0\% &  & 0.85 \\
$OE_{\scriptscriptstyle \textup{CIPS}}(0.5, 1)$ & 0.072 &  & 0.003 & 0.015 & 0.013 & 0.014 & 93.7\% &  & 0.013 & 0.021 & 0.015 & 0.017 & 84.8\% &  & 0.69 \\
$TE_{\scriptscriptstyle \textup{CIPS}}(0.5, 1)$ & -0.266 &  & -0.005 & 0.017 & 0.017 & 0.017 & 94.6\% &  & -0.039 & 0.044 & 0.019 & 0.019 & 43.1\% &  & 0.40 \\ \hline
$\mu_{\scriptscriptstyle \textup{CIPS}}(1)$ & 0.364 &  & -0.001 & 0.012 & 0.012 & 0.012 & 94.1\% &  & -0.001 & 0.012 & 0.012 & 0.012 & 93.7\% &  & 1.02 \\
$\mu_{\scriptscriptstyle \textup{CIPS}, \scriptstyle 1}(1)$ & 0.242 &  & -0.003 & 0.011 & 0.011 & 0.011 & 93.4\% &  & -0.014 & 0.018 & 0.010 & 0.011 & 69.4\% &  & 0.64 \\
$\mu_{\scriptscriptstyle \textup{CIPS}, \scriptstyle 0}(1)$ & 0.530 &  & 0.003 & 0.015 & 0.015 & 0.015 & 94.3\% &  & 0.029 & 0.033 & 0.015 & 0.015 & 54.4\% &  & 0.47 \\
$DE_{\scriptscriptstyle \textup{CIPS}}(1)$ & -0.287 &  & -0.006 & 0.014 & 0.013 & 0.012 & 92.7\% &  & -0.043 & 0.045 & 0.014 & 0.014 & 13.5\% &  & 0.31 \\ \hline
$\mu_{\scriptscriptstyle \textup{CIPS}}(2)$ & 0.300 &  & -0.003 & 0.020 & 0.021 & 0.020 & 94.6\% &  & -0.010 & 0.022 & 0.020 & 0.020 & 89.7\% &  & 0.91 \\
$\mu_{\scriptscriptstyle \textup{CIPS}, \scriptstyle 1}(2)$ & 0.224 &  & -0.004 & 0.021 & 0.021 & 0.021 & 93.4\% &  & -0.017 & 0.026 & 0.020 & 0.020 & 82.5\% &  & 0.81 \\
$\mu_{\scriptscriptstyle \textup{CIPS}, \scriptstyle 0}(2)$ & 0.507 &  & 0.002 & 0.023 & 0.023 & 0.023 & 96.0\% &  & 0.023 & 0.033 & 0.025 & 0.024 & 84.8\% &  & 0.68 \\
$DE_{\scriptscriptstyle \textup{CIPS}}(2)$ & -0.283 &  & -0.006 & 0.023 & 0.022 & 0.023 & 93.3\% &  & -0.040 & 0.046 & 0.023 & 0.023 & 57.9\% &  & 0.50 \\
$SE_{\scriptscriptstyle \textup{CIPS}, \scriptstyle 1}(2, 1)$ & -0.018 &  & -0.001 & 0.017 & 0.017 & 0.017 & 94.1\% &  & -0.003 & 0.017 & 0.016 & 0.016 & 93.1\% &  & 1.04 \\
$SE_{\scriptscriptstyle \textup{CIPS}, \scriptstyle 0}(2, 1)$ & -0.022 &  & -0.001 & 0.018 & 0.018 & 0.018 & 95.7\% &  & -0.005 & 0.020 & 0.019 & 0.019 & 94.6\% &  & 0.89 \\
$OE_{\scriptscriptstyle \textup{CIPS}}(2, 1)$ & -0.063 &  & -0.002 & 0.017 & 0.017 & 0.016 & 94.7\% &  & -0.009 & 0.019 & 0.016 & 0.016 & 88.9\% &  & 0.89 \\
$TE_{\scriptscriptstyle \textup{CIPS}}(2, 1)$ & -0.306 &  & -0.007 & 0.024 & 0.022 & 0.023 & 93.3\% &  & -0.046 & 0.051 & 0.022 & 0.023 & 44.7\% &  & 0.47 \\ \hline
\end{tabular}%
}
\captionsetup{width=.9\textwidth}
\caption*{\footnotesize RMSE: root mean squared error, ASE: average standard error estimates, ESE: standard deviation of estimates, Cov \%: 95\% CI coverage, RMSE Ratio: RMSE ratio of nonparametric and parametric estimators}
\end{table}
\vspace{-0.5cm}

Several additional simulation studies were conducted. 
First, the NSS estimators were compared with the parametric inverse probability weighted (IPW) estimators proposed by \cite{Papadogeorgou19} and \cite{barkley20}. 
The results (see supplementary material Section C.2) show the NSS estimators had smaller bias, smaller ESE, and 95\% CI coverage closer to the nominal level compared to the IPW estimator.
Next, the finite sample performance of the NSS estimators for different values of $r$ (degree of the subsampling approximation) was evaluated.
The results in supplementary material Section C.3 show that the finite sample bias of the NSS estimator was insensitive to $r$, 
but the empirical SE decreased in $r$, 
while the 95\% CI coverage achieved the nominal level regardless of $r$.
Finally, 
simulations were conducted for different distributions of $N_i$ (see the supplementary material Section C.4); as expected, the proposed methods performed well since no assumptions are made about the distribution of $N_i$.

\begin{table}[H]
\captionsetup{width=.9\textwidth}
\caption{\small Simulation results for nonparametric and parametric sample splitting estimators for TPB policy}
\label{tab:simulTPB}
\centering
\resizebox{.9\textwidth}{!}{%
\begin{tabular}{cccccccccccccccc}
\hline
               &        &  & \multicolumn{5}{c}{Nonparametric}       &  & \multicolumn{5}{c}{Parametric}          &  & RMSE  \\ \cline{4-8} \cline{10-14}
Estimand       & Truth  &  & Bias   & RMSE  & ASE   & ESE   & Cov \% &  & Bias   & RMSE  & ASE   & ESE   & Cov \% &  & Ratio \\ \hline
$\mu_{\scriptscriptstyle \textup{TPB}}(0.3)$ & 0.361 &  & -0.002 & 0.010 & 0.009 & 0.010 & 93.1\% &  & -0.003 & 0.010 & 0.009 & 0.010 & 93.1\% &  & 0.99 \\
$\mu_{\scriptscriptstyle \textup{TPB}, \scriptstyle 1}(0.3)$ & 0.243 &  & -0.004 & 0.010 & 0.009 & 0.009 & 91.4\% &  & -0.015 & 0.018 & 0.009 & 0.009 & 59.8\% &  & 0.58 \\
$\mu_{\scriptscriptstyle \textup{TPB}, \scriptstyle 0}(0.3)$ & 0.531 &  & 0.002 & 0.012 & 0.012 & 0.012 & 95.6\% &  & 0.026 & 0.029 & 0.013 & 0.013 & 44.0\% &  & 0.41 \\
$DE_{\scriptscriptstyle \textup{TPB}}(0.3)$ & -0.288 &  & -0.006 & 0.012 & 0.011 & 0.011 & 92.5\% &  & -0.042 & 0.043 & 0.013 & 0.013 & 10.0\% &  & 0.29 \\
$SE_{\scriptscriptstyle \textup{TPB}, \scriptstyle 1}(0.3, 0.45)$ & 0.005 &  & 0.000 & 0.003 & 0.003 & 0.003 & 93.5\% &  & 0.001 & 0.003 & 0.003 & 0.003 & 93.6\% &  & 0.88 \\
$SE_{\scriptscriptstyle \textup{TPB}, \scriptstyle 0}(0.3, 0.45)$ & 0.006 &  & 0.000 & 0.003 & 0.003 & 0.003 & 93.7\% &  & 0.000 & 0.004 & 0.004 & 0.004 & 94.6\% &  & 0.84 \\
$OE_{\scriptscriptstyle \textup{TPB}}(0.3, 0.45)$ & 0.017 &  & 0.000 & 0.003 & 0.003 & 0.003 & 92.9\% &  & 0.002 & 0.005 & 0.004 & 0.004 & 92.6\% &  & 0.74 \\
$TE_{\scriptscriptstyle \textup{TPB}}(0.3, 0.45)$ & -0.282 &  & -0.007 & 0.013 & 0.012 & 0.011 & 92.2\% &  & -0.041 & 0.043 & 0.014 & 0.014 & 13.3\% &  & 0.30 \\ \hline
$\mu_{\scriptscriptstyle \textup{TPB}}(0.45)$ & 0.344 &  & -0.002 & 0.010 & 0.009 & 0.009 & 93.3\% &  & -0.005 & 0.011 & 0.010 & 0.010 & 91.7\% &  & 0.90 \\
$\mu_{\scriptscriptstyle \textup{TPB}, \scriptstyle 1}(0.45)$ & 0.238 &  & -0.005 & 0.011 & 0.010 & 0.010 & 91.3\% &  & -0.016 & 0.019 & 0.010 & 0.010 & 60.8\% &  & 0.58 \\
$\mu_{\scriptscriptstyle \textup{TPB}, \scriptstyle 0}(0.45)$ & 0.525 &  & 0.002 & 0.012 & 0.013 & 0.012 & 95.2\% &  & 0.026 & 0.029 & 0.013 & 0.013 & 49.0\% &  & 0.43 \\
$DE_{\scriptscriptstyle \textup{TPB}}(0.45)$ & -0.287 &  & -0.007 & 0.014 & 0.012 & 0.012 & 92.6\% &  & -0.042 & 0.044 & 0.014 & 0.014 & 14.1\% &  & 0.31 \\ \hline
$\mu_{\scriptscriptstyle \textup{TPB}}(0.6)$ & 0.316 &  & -0.004 & 0.011 & 0.010 & 0.011 & 92.4\% &  & -0.009 & 0.015 & 0.011 & 0.011 & 86.0\% &  & 0.77 \\
$\mu_{\scriptscriptstyle \textup{TPB}, \scriptstyle 1}(0.6)$ & 0.229 &  & -0.005 & 0.012 & 0.011 & 0.011 & 90.8\% &  & -0.017 & 0.021 & 0.011 & 0.012 & 65.7\% &  & 0.60 \\
$\mu_{\scriptscriptstyle \textup{TPB}, \scriptstyle 0}(0.6)$ & 0.514 &  & 0.002 & 0.014 & 0.014 & 0.014 & 94.4\% &  & 0.024 & 0.029 & 0.016 & 0.016 & 65.6\% &  & 0.49 \\
$DE_{\scriptscriptstyle \textup{TPB}}(0.6)$ & -0.285 &  & -0.007 & 0.017 & 0.015 & 0.015 & 91.4\% &  & -0.041 & 0.045 & 0.017 & 0.017 & 29.5\% &  & 0.37 \\
$SE_{\scriptscriptstyle \textup{TPB}, \scriptstyle 1}(0.6, 0.45)$ & -0.009 &  & -0.001 & 0.005 & 0.005 & 0.005 & 95.0\% &  & -0.001 & 0.006 & 0.006 & 0.006 & 94.3\% &  & 0.86 \\
$SE_{\scriptscriptstyle \textup{TPB}, \scriptstyle 0}(0.6, 0.45)$ & -0.011 &  & 0.000 & 0.006 & 0.006 & 0.006 & 94.5\% &  & -0.002 & 0.008 & 0.007 & 0.007 & 94.5\% &  & 0.82 \\
$OE_{\scriptscriptstyle \textup{TPB}}(0.6, 0.45)$ & -0.028 &  & -0.001 & 0.006 & 0.005 & 0.006 & 93.7\% &  & -0.004 & 0.008 & 0.007 & 0.007 & 89.9\% &  & 0.71 \\
$TE_{\scriptscriptstyle \textup{TPB}}(0.6, 0.45)$ & -0.296 &  & -0.007 & 0.015 & 0.013 & 0.013 & 92.2\% &  & -0.043 & 0.046 & 0.015 & 0.016 & 20.7\% &  & 0.33 \\ \hline
\end{tabular}%
}
\captionsetup{width=.9\textwidth}
\caption*{\footnotesize RMSE: root mean squared error, ASE: average standard error estimates, ESE: standard deviation of estimates, Cov \%: 95\% CI coverage, RMSE Ratio: RMSE ratio of nonparametric and parametric estimators}
\end{table}
\vspace{-0.5cm}

\section{Application to Senegal DHS data}

The goal of the analysis presented in this section was to evaluate the effect of WASH facilities on diarrheal diseases among children. 
The proposed method was applied to 2015 – 2019 Senegal DHS data, which provides sociodemographic, environmental, and health-related information on household members.
Census blocks were considered as clusters and households as treatment units,
allowing clustered interference within census blocks.
Definitions of the outcome, treatment, and pre-treatment covariates were adapted from \citet{park21}.
For household $j$ in census block $i$,
the outcome $Y_{ij}$ was set to 1 if all children did not have diarrhea within two weeks from the survey data; 
otherwise $Y_{ij}=0$. 
The treatment $A_{ij}$ equalled 1 if the household had a private WASH facility, 
and $A_{ij}=0$ otherwise.
The pre-treatment covariates were unit-level characteristics consisting of household size, number of children, whether parents had a job, whether parents ever attended school, mother's age, average age of children,
as well as cluster-level characteristics including cluster size and whether the cluster was in an urban area.
Restricting the sample to households with complete data resulted in 1,074 clusters with 4,565 households.

The proposed method was employed to evaluate
CIPS policies with constant $\delta(\mathbf{X}_i, N_i) = \delta_0 \in [0.5, 2]$ 
and TPB policies with $\rho \in [0, 0.5]$.
Note the factual (observed) scenario corresponds to the CIPS policy when $\delta_0 = 1$  and the TPB policy when $\rho = 0$.
For each policy, 
$K = 5, S = 30$,
and no subsampling approximation was used because the cluster sizes were small (maximum 12).
Assuming the conditional independencies described in Section 3.2, individual-level nuisance functions $g$ and $\pi$ were estimated instead of $G$ and $H$.
Nuisance functions were fit using the ensemble of logistic regression, logistic lasso/elastic net, spline regression, generalized additive model, gradient boosting machine, random forest, and neural net estimators using the super learner algorithm in R \citep{superlearner}.

\begin{figure}[h]
 \centerline{\includegraphics[width = \textwidth]{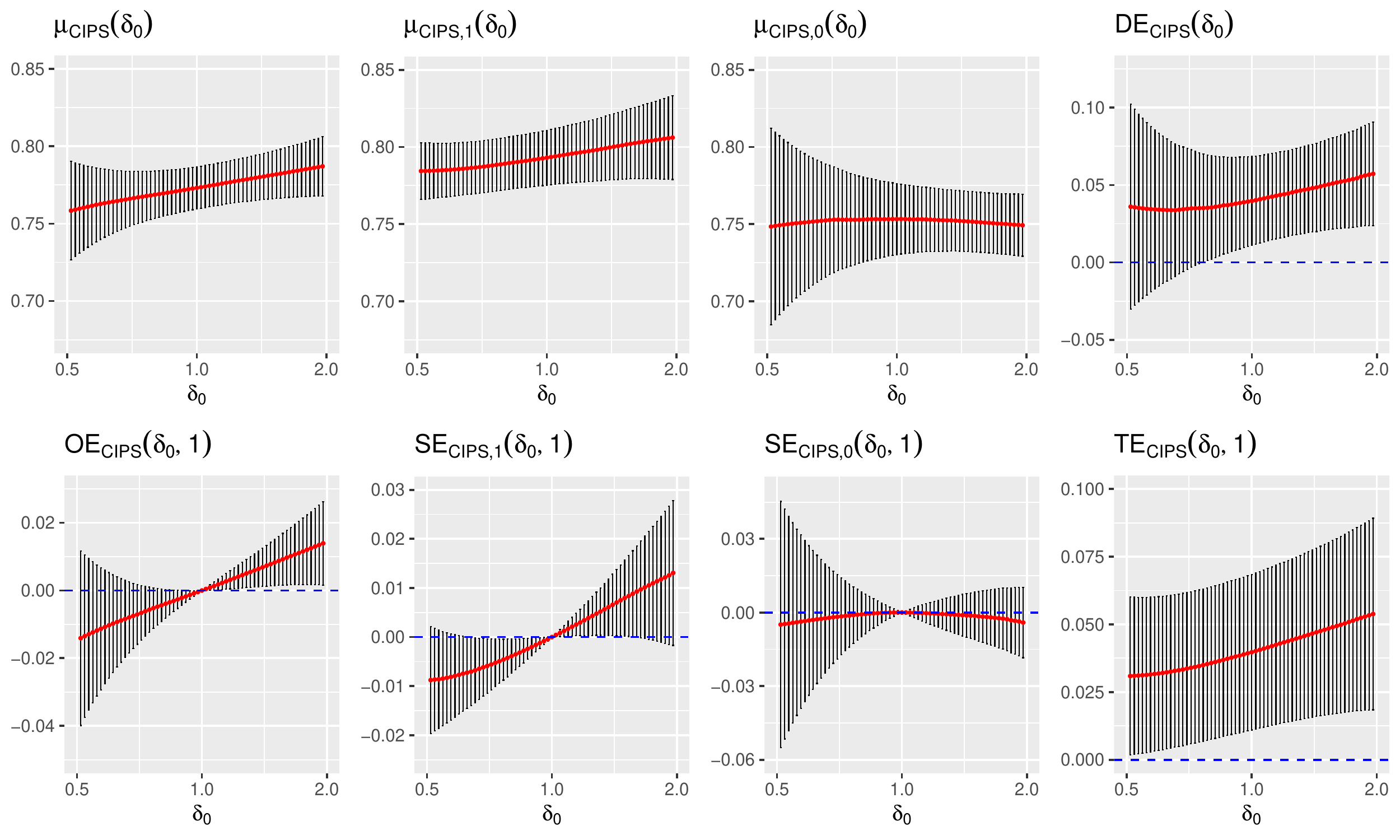}}
\caption{\footnotesize 
Nonparametric sample splitting (NSS) estimators of CIPS estimands
from the analysis of the Senegal DHS. 
Red dots and black lines indicate the point estimates and 95\% CIs, 
and blue dashed horizontal lines indicate the null value of the effects. $x$-axis ($\delta$) is on the log scale.}
\label{fig:95ci_CIPS}
\end{figure}

NSS estimates and 95\% confidence intervals for the CIPS policy estimands are shown in Figure~\ref{fig:95ci_CIPS}.
The estimates of $\mu_{\scriptscriptstyle \textup{CIPS}}(\delta_0)$ increased as $\delta_0$ increased,
indicating lower risk of diarrhea at the population level under policies where the odds of having WASH facilities are greater.
The overall effect estimates $\widehat{OE}_{\scriptscriptstyle \textup{CIPS}}(\delta_0,1)$ were negative for $\delta_0 < 1$ and positive for $\delta_0 > 1$,
and corresponding 95\% CIs did not include zero when $\delta_0 > 1$.
For example, $\widehat{OE}_{\scriptscriptstyle \textup{CIPS}}(2,1)=0.014$ (95\% CI: [0.001, 0.028]) indicated that if the odds of having a WASH facility were doubled, 14 more diarrhea-free households per 1000 households would be expected.
Estimates of $\mu_{\scriptscriptstyle \textup{CIPS}, \scriptstyle 0}(\delta_0)$ exhibited minimal variability over the range of $\delta_0$, 
and spillover effect estimates when untreated, 
$\widehat{SE}_{\scriptscriptstyle \textup{CIPS}, \scriptstyle 0}(\delta_0,1)$, were close to zero with wide corresponding CIs, 
indicating no or minimal effect of WASH facilities in neighboring households on non-WASH households.
On the other hand, estimates of $\mu_{\scriptscriptstyle \textup{CIPS}, \scriptstyle 1}(\delta_0)$ increased with $\delta_0$, 
suggesting a protective spillover effect when a household has a WASH facility.
The spillover estimates when treated, 
$\widehat{SE}_{\scriptscriptstyle \textup{CIPS}, \scriptstyle 1}(\delta_0,1)$, were negative for $\delta_0 < 1$ and positive for $\delta_0 > 1$,
and corresponding 95\% CIs did not include zero when $\delta_0 \in [0.6, 1.6]$.
Direct effect estimates, $\widehat{DE}_{\scriptscriptstyle \textup{CIPS}}(\delta_0)$, and corresponding 95\% CIs indicated that WASH facilities had a protective effect on preventing diarrhea, 
and direct effect was greater when $\delta_0$ was greater.
Estimates of total effect exhibited a similar trend as the direct effect estimates, 
because there was no or modest spillover effect when untreated.

\begin{figure}[h]
 \centerline{\includegraphics[width = \textwidth]{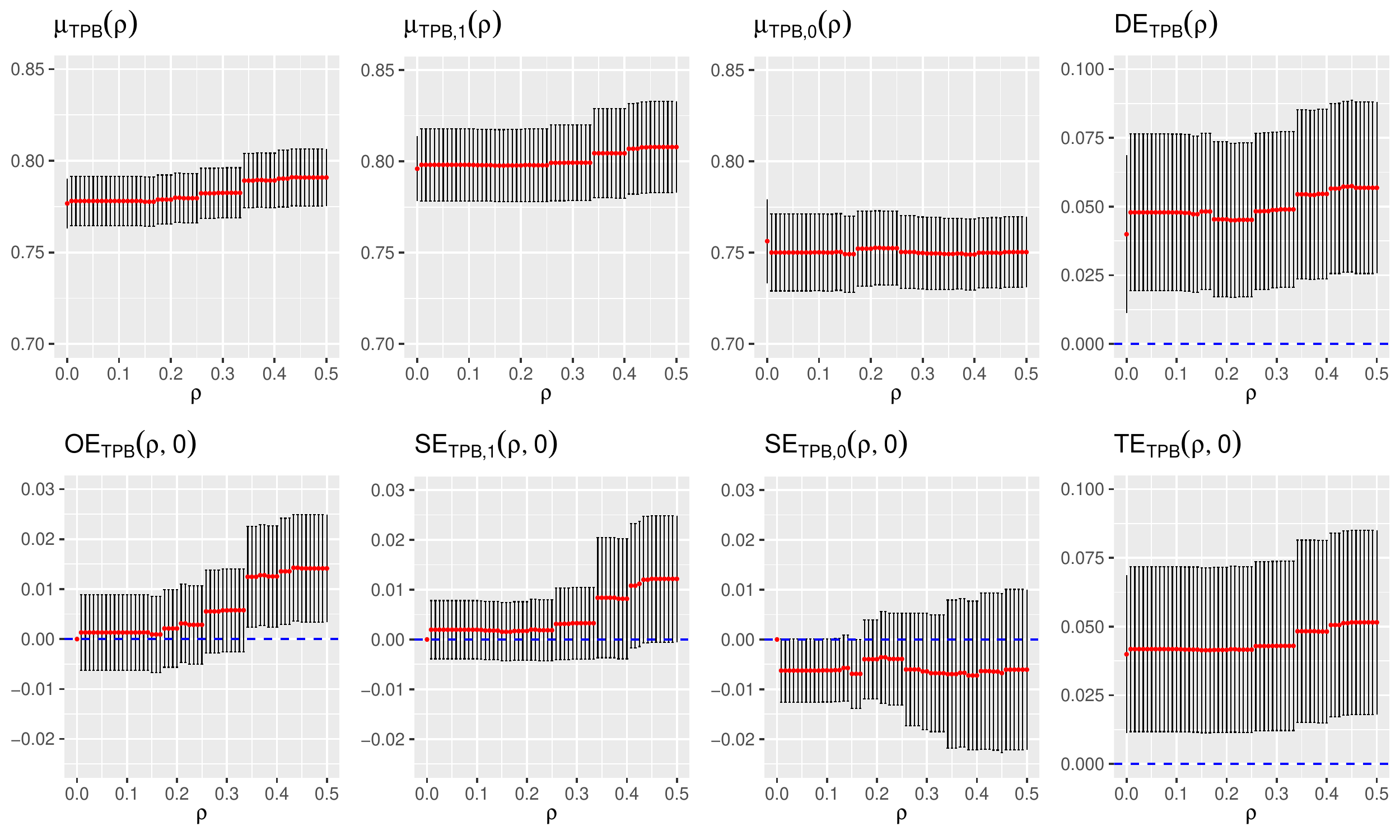}}
\caption{\footnotesize 
Nonparametric sample splitting (NSS) estimators of TPB estimands
from the analysis of the Senegal DHS. 
Red dots and black lines indicate the point estimates and 95\% CIs, 
and blue dashed horizontal lines indicate the null value of the effects.}
\label{fig:95ci_TPB}
\end{figure}

Results for the TPB policy are shown in Figure \ref{fig:95ci_TPB}. 
Note the treated proportion $\overline{\mathbf{a}}_i$ can take on a finite number of possible values, 
in particular 
$\overline{\mathbf{a}}_i \in \mathcal{S} = \{s/n: 0 \le s \le n, 2 \le n \le 12\}$.
Thus, the TPB policy $Q_{\scriptscriptstyle \textup{TPB}}(\mathbf{a}_i | \mathbf{X}_i, N_i; \rho)$ is constant for different values of $\rho$ within intervals defined by the unique ordered values of $\mathcal{S}$. 
Consequently, the point estimates and 95\% CIs in Figure \ref{fig:95ci_TPB} are piecewise constant. 
The
$\mu_{\scriptscriptstyle \textup{TPB}}(\rho)$ 
and 
$\mu_{\scriptscriptstyle \textup{TPB}, \scriptstyle 1}(\rho)$
estimates were increasing in $\rho$, 
implying a larger protective effect when the treated proportion is larger,
while
$\mu_{\scriptscriptstyle \textup{TPB}, \scriptstyle 0}(\rho)$
estimates changed minimally with $\rho$.
Accordingly, 
$OE_{\scriptscriptstyle \textup{TPB}}(\rho, 0)$ 
and 
$SE_{\scriptscriptstyle \textup{TPB}, \scriptstyle 1}(\rho, 0)$
estimates were positive and increasing in $\rho$, 
while 
$SE_{\scriptscriptstyle \textup{TPB}, \scriptstyle 0}(\rho, 0)$
estimates were approximately constant with wide corresponding CIs including zero, 
suggesting no or minimal protective spillover effects when a household does not have a WASH facility.
For $\rho \in [0.35, 0.5]$, 95\% CIs of 
$OE_{\scriptscriptstyle \textup{TPB}}(\rho, 0)$ 
did not include the null value.
Direct and total effect estimates were positive and increasing
with corresponding 95\% CIs excluding zero,
suggesting that WASH facilities had a protective effect on preventing diarrhea.
For example, 
$\widehat{DE}_{\scriptscriptstyle \textup{TPB}}(0.2) = 0.045$ (95\% CI: [0.017, 0.074]), indicating that 
we would expect approximately 45 more diarrhea-free households per 1000 WASH households compared to non-WASH households
if the proportion of WASH facilities were at least 0.2 for all clusters,
while
$\widehat{DE}_{\scriptscriptstyle \textup{TPB}}(0.5) = 0.056$ (95\% CI: [0.026, 0.088]), suggesting that 
if the proportion were at least 0.5, 
then we would expect approximately 56 more diarrhea-free households per 1000 WASH households compared to non-WASH households.
Overall, the results for TPB estimands were similar to that for CIPS estimands.

\citet{park21} also analyzed the Senegal DHS data, 
with the goal of estimating the optimal allocation of WASH facilities needed to have at least $\mathcal{T}$ proportion of household in a census block diarrhea free (for different values of $\mathcal{T}$). 
Because the goal of the analysis presented here differed from \citet{park21}, direct comparison of the results is challenging. 
Nonetheless, Section D.4 of the supplementary material provides some discussion regarding the results in this manuscript 
compared to \cite{park21}. 


In summary, 
household WASH facilities appear to protect children from diarrhea, 
and the protective effects increase when neighboring households also have WASH facilities.
Unfortunately, however, children from households without WASH facilities do not appear to benefit from such spillover effects.
Therefore, it is important to install as many WASH facilities as possible to prevent diarrhea among children.

\section{Discussion}
\label{s:discuss}

In this paper, nonparametric methods are developed which can be used to draw inference about treatment effects in the presence of clustered interference.
The methods are general in the sense that they can be applied to
any treatment allocation policy. 
Four example policies are used to illustrate the methods: 
(i) type B policy, 
(ii) cluster incremental propensity score
policy, 
(iii) cluster multiplicative shift policy, 
and (iv) treated proportion bound policy.
The class of target causal estimands considered may be relevant in many settings because they allow for the units' probability of receiving treatment to vary by their covariates and are not based on parametric models.
%
%
Furthermore, the proposed nonparametric efficient sample splitting estimators exploit a variety of data-adaptive methods, 
and therefore are robust to model mis-specification compared to parametric estimators.
The simulations demonstrated that the proposed nonparametric estimators perform well in finite samples.
The application to the Senegal DHS data suggested that having a private water source or flushable toilet decreases the risk of diarrhea among children,
and that children from WASH households may receive an additional protective spillover effect from neighboring WASH households.
While applied to the Senegal DHS data,
the proposed methods can be applied in other settings where clustered interference may be present.

There are several possible avenues of future research related to this paper.
First, the methods developed here rely on asymptotic regimens as the number of clusters $m$ grows large. Thus, the methods may not perform well for data sets with a small number of clusters. Future research could develop methods 
suitable for settings where there are only a few (but possibly large) clusters. 
Second, extensions of the method to multivalued or continuous treatment settings could be considered. 
In these settings, shifts in the propensity score distribution may be defined in a variety of ways because the treatment is no longer binary.
Next, extensions to censored survival times or longitudinal outcomes could also be considered.
For instance, methods could be developed to assess the effect of WASH facilities on the time until diarrhea among individuals within a household.
Finally, allowing for more general network interference would be interesting.
In this case, stronger assumptions may be needed to develop valid inferential methods for the causal estimands of interest.

\bibliographystyle{agsm}
\bibliography{bibliography.bib}

@article{hudgens08,
    author = {Michael G Hudgens and M. Elizabeth Halloran},
    title = {Toward Causal Inference With Interference},
    journal = {Journal of the American Statistical Association},
    volume = {103},
    number = {482},
    pages = {832-842},
    year  = {2008},
    publisher = {Taylor & Francis}
}

@article{sobel06,
    author = {Michael E Sobel},
    title = {What Do Randomized Studies of Housing Mobility Demonstrate? {C}ausal inference in the face of interference},
    journal = {Journal of the American Statistical Association},
    volume = {101},
    number = {476},
    pages = {1398-1407},
    year  = {2006},
    publisher = {Taylor & Francis}
}

@article{barkley20,
  title={Causal inference from observational studies with clustered interference, with application to a cholera vaccine study},
  author={Barkley, Brian G and Hudgens, Michael G and Clemens, John D and Ali, Mohammad and Emch, Michael E},
  journal={The Annals of Applied Statistics},
  volume={14},
  number={3},
  pages={1432--1448},
  year={2020},
  publisher={Institute of Mathematical Statistics}
}

@article{tchetgen12,
    author = {Tchetgen Tchetgen, Eric J and Tyler J VanderWeele},
    title ={On causal inference in the presence of interference},
    journal = {Statistical Methods in Medical Research},
    volume = {21},
    number = {1},
    pages = {55-75},
    year = {2012}
}

@article{Papadogeorgou19,
    author = {Papadogeorgou, Georgia and Mealli, Fabrizia and Zigler, Corwin M.},
    title = {Causal inference with interfering units for cluster and population level treatment allocation programs},
    journal = {Biometrics},
    volume = {75},
    number = {3},
    pages = {778-787},
    year = {2019}
}

@article{kennedy19,
    author = {Edward H. Kennedy},
    title = {Nonparametric Causal Effects Based on Incremental Propensity Score Interventions},
    journal = {Journal of the American Statistical Association},
    volume = {114},
    number = {526},
    pages = {645-656},
    year  = {2019},
    publisher = {Taylor & Francis}
}

@article{perez14,
    author = {Perez-Heydrich, Carolina and Hudgens, Michael G. and Halloran, M. Elizabeth and Clemens, John D. and Ali, Mohammad and Emch, Michael E.},
    title = {Assessing effects of cholera vaccination in the presence of interference},
    journal = {Biometrics},
    volume = {70},
    number = {3},
    pages = {731-741},
    year = {2014}
}

@article{park2022efficient,
  title={Efficient semiparametric estimation of network treatment effects under partial interference},
  author={Park, Chan and Kang, Hyunseung},
  journal={Biometrika},
  volume={109},
  number={4},
  pages={1015--1031},
  year={2022},
  publisher={Oxford University Press}
}

@article{liu19,
  title={Doubly robust estimation in observational studies with partial interference},
  author={Liu, Lan and Hudgens, Michael G and Saul, Bradley and Clemens, John D and Ali, Mohammad and Emch, Michael E},
  journal={Stat},
  volume={8},
  number={1},
  pages={e214},
  year={2019},
  publisher={Wiley Online Library}
}

@article{chernozhukov18,
    author = {Chernozhukov, Victor and Chetverikov, Denis and Demirer, Mert and Duflo, Esther and Hansen, Christian and Newey, Whitney and Robins, James},
    title = "{Double/debiased machine learning for treatment and structural parameters}",
    journal = {The Econometrics Journal},
    volume = {21},
    number = {1},
    pages = {C1-C68},
    year = {2018},
    month = {01},
    issn = {1368-4221}
}

@article{vanderLaan07,
title = {Super Learner},
author = {Mark J. van der Laan and Eric C Polley and Alan E. Hubbard},
year = {2007},
volume = {6},
pages = {1--21},
journal = {Statistical Applications in Genetics and Molecular Biology},
lastchecked = {2022-09-16}
}

@book{tsiatis06,
  title={Semiparametric Theory and Missing Data},
  author={Tsiatis, Anastasios A},
  year={2006},
  publisher={New York, NY: Springer}
}

@incollection{kennedy16,  
    title     = {Semiparametric Theory and Empirical Processes in Causal Inference},
    author    = {Kennedy, Edward H.},                       
    pages     = {141--167},                                 
    crossref  = {kennedybook16}                      
}

@book{kennedybook16, 
    year      = {2016},                                     
    editor    = {He, Hua and Wu, Pan and {Chen, DG}},
    title     = {Statistical Causal Inferences and Their Applications in Public Health Research}, 
    booktitle = {Statistical Causal Inferences and Their Applications in Public Health Research}, 
    author    = {Kennedy, Edward H.},            
    publisher = {Springer International Publishing},              
    address   = {Cham}                   
}

@article{hines22,
    author = {Oliver Hines and Oliver Dukes and Karla Diaz-Ordaz and Stijn Vansteelandt},
    title = {Demystifying Statistical Learning Based on Efficient Influence Functions},
    journal = {The American Statistician},
    volume = {76},
    number = {3},
    pages = {292-304},
    year  = {2022},
    publisher = {Taylor & Francis}
}

@article{park21,
  author={Park, Chan and Chen, Guanhua and Yu, Menggang and Kang, Hyunseung},
  title={Optimal Allocation of Water and Sanitation Facilities To Prevent Communicable Diarrheal Diseases in {S}enegal Under Partial Interference},
  year={2021},
  journal={arXiv preprint arXiv:2004.08950}
}

@Manual{superlearner,
    title = {SuperLearner: Super Learner Prediction},
    author = {Eric Polley and Erin LeDell and Chris Kennedy and Mark {van der Laan}},
    year = {2021},
    note = {R package version 2.0-28}
  }

@online{un21, 
    title={Summary progress update 2021: SDG 6 - water and sanitation for all}, 
    journal={UN}, 
    author={UN-Water}, 
    year={2021},
    note = {https://www.unwater.org/publications/summary-progress-update-2021-sdg-6-water-and-sanitation-all} 
}

@article{thiam17,
  title={Prevalence of diarrhoea and risk factors among children under five years old in {M}bour, {S}enegal: a cross-sectional study},
  author={Thiam, Sokhna and Di{\`e}ne, Aminata N and Fuhrimann, Samuel and Winkler, Mirko S and Sy, Ibrahima and Ndione, Jacques A and others},
  journal={Infectious Diseases of Poverty},
  volume={6},
  number={04},
  pages={43--54},
  year={2017},
  publisher={Editorial Office of Infectious Diseases of Poverty, National Institute of~…}
}

@misc{dhs20,
    author={{Agence Nationale de la Statistique et de la Démographie (ANSD)} and ICF},
    shortauthor={ANSD and ICF},
    title={Senegal: Enquête {D}émographique et de {S}anté {C}ontinue ({EDS}-{C}ontinue) 2019},
    year={2020},
    publisher={ANSD/ICF},
    note={https://www.dhsprogram.com/pubs/pdf/FR368/FR368.pdf}
}

@article{munoz12,
  title={Population intervention causal effects based on stochastic interventions},
  author={Mu{\~n}oz, Iv{\'a}n D{\'\i}az and Van Der Laan, Mark},
  journal={Biometrics},
  volume={68},
  number={2},
  pages={541--549},
  year={2012},
  publisher={Wiley Online Library}
}

@article{salo22,
  title={The indirect effect of {mRNA-based COVID-19} vaccination on healthcare workers’ unvaccinated household members},
  author={Salo, Jussipekka and H{\"a}gg, Milla and Kortelainen, Mika and Leino, Tuija and Saxell, Tanja and Siikanen, Markku and S{\"a}{\"a}ksvuori, Lauri},
  journal={Nature Communications},
  volume={13},
  number={1},
  pages={1--7},
  year={2022},
  publisher={Nature Publishing Group}
}

@article{prunas22,
  title={Vaccination with {BNT162b2} reduces transmission of {SARS-CoV-2} to household contacts in {Israel}},
  author={Prunas, Ottavia and Warren, Joshua L and Crawford, Forrest W and Gazit, Sivan and Patalon, Tal and Weinberger, Daniel M and Pitzer, Virginia E},
  journal={Science},
  volume={375},
  number={6585},
  pages={1151--1154},
  year={2022},
  publisher={American Association for the Advancement of Science}
}

@article{barrera11,
  title={Improving the design of conditional transfer programs: Evidence from a randomized education experiment in {C}olombia},
  author={Barrera-Osorio, Felipe and Bertrand, Marianne and Linden, Leigh L and Perez-Calle, Francisco},
  journal={American Economic Journal: Applied Economics},
  volume={3},
  number={2},
  pages={167--95},
  year={2011}
}

@article{kilpatrick21,
  title={G-formula for observational studies with partial interference, with application to bed net use on malaria},
  author={Kilpatrick, Kayla W and Hudgens, Michael G},
  journal={arXiv preprint arXiv:2102.01155},
  year={2021}
}

@article{kennedy22,
  title={Semiparametric doubly robust targeted double machine learning: a review},
  author={Kennedy, Edward H},
  journal={arXiv preprint arXiv:2203.06469},
  year={2022}
}

@book{hernan20,
  title={Causal Inference: What If},
  author={Hernan, M. A. and Robins, J. M.},
  year={2020},
  publisher={Boca Raton: Chapman \& Hall/CRC}
}

@article{wen23,
  title={Intervention treatment distributions that depend on the observed treatment process and model double robustness in causal survival analysis},
  author={Wen, Lan and Marcus, Julia L and Young, Jessica G},
  journal={Statistical Methods in Medical Research},
  volume={32},
  number={3},
  pages={509--523},
  year={2023},
  publisher={SAGE Publications Sage UK: London, England}
}

@article{ngufor19,
  title={Mixed effect machine learning: A framework for predicting longitudinal change in hemoglobin A1c},
  author={Ngufor, Che and Van Houten, Holly and Caffo, Brian S and Shah, Nilay D and McCoy, Rozalina G},
  journal={Journal of Biomedical Informatics},
  volume={89},
  pages={56--67},
  year={2019},
  publisher={Elsevier}
}

@article{fisher2021visually,
  title={Visually communicating and teaching intuition for influence functions},
  author={Fisher, Aaron and Kennedy, Edward H},
  journal={The American Statistician},
  volume={75},
  number={2},
  pages={162--172},
  year={2021},
  publisher={Taylor \& Francis}
}

@article{majerek2005conditional,
  title={Conditional strong law of large number},
  author={Majerek, Dariusz and Nowak, Wioletta and Zieba, Wieslaw},
  journal={Int. J. Pure Appl. Math},
  volume={20},
  number={2},
  pages={143--156},
  year={2005}
}

@article{benjamin2018randomized,
  title={A randomized controlled trial to measure spillover effects of a combined water, sanitation, and handwashing intervention in rural Bangladesh},
  author={Benjamin-Chung, Jade and Amin, Nuhu and Ercumen, Ayse and Arnold, Benjamin F and Hubbard, Alan E and Unicomb, Leanne and Rahman, Mahbubur and Luby, Stephen P and Colford Jr, John M},
  journal={American Journal of Epidemiology},
  volume={187},
  number={8},
  pages={1733--1744},
  year={2018},
  publisher={Oxford University Press}
}


\pagebreak
\begin{center}
    \textbf{\LARGE Supplemental Materials}
\end{center}
\setcounter{section}{0}
\setcounter{equation}{0}
\setcounter{figure}{0}
\setcounter{table}{0}
\setcounter{page}{1}
\makeatletter
\renewcommand{\theequation}{S\arabic{equation}}
\renewcommand{\thefigure}{S\arabic{figure}}
\renewcommand{\thetable}{S\arabic{table}}
\renewcommand{\bibnumfmt}[1]{[S#1]}
\renewcommand{\citenumfont}[1]{S#1}
\renewcommand{\thesection}{\Alph{section}}

\section{Theoretical properties and proofs}

In this section, the proofs of Lemma 1 and Theorems 1 -- 7, a note on Assumption (B8), and the large sample properties under the subsampling approximation are presented.

\subsection{Proof of Lemma 1}

First, note that
\begin{align*}
\mu(Q) 
&= 
\mathbb{E} \left\{ 
    \frac{1}{N_i} 
    \sum_{j = 1}^{N_i} 
        \sum_{\mathbf{a}_i \in \mathcal{A} (N_i)} 
            Y_{ij}(\mathbf{a}_i) 
            Q(\mathbf{a}_i | \mathbf{X}_i, N_i) 
\right\}   
\\
&= 
\mathbb{E} \left\{ 
    \sum_{\mathbf{a}_i \in \mathcal{A} (N_i)}
        w(\mathbf{a}_i, \mathbf{X}_i, N_i)^\top
        \mathbf{Y}_i(\mathbf{a}_i) 
\right\}   
\\
&= 
\mathbb{E} \left[
    \sum_{\mathbf{a}_i \in \mathcal{A} (N_i)}
    w(\mathbf{a}_i, \mathbf{X}_i, N_i)^\top
    \mathbb{E} \left\{ 
        \mathbf{Y}_i(\mathbf{a}_i)
        \middle| \mathbf{X}_i, N_i
    \right\}   
\right] 
&& \because \text{Iterated expectation}
\\
&= 
\mathbb{E} \left\{
    \sum_{\mathbf{a}_i \in \mathcal{A} (N_i)}
        w(\mathbf{a}_i, \mathbf{X}_i, N_i)^\top
        \mathbb{E} \left( 
            \mathbf{Y}_i
            \middle| 
            \mathbf{A}_i = \mathbf{a}_i, \mathbf{X}_i, N_i
        \right)   
\right\}
&&
\because
\begin{tabular}[t]{@{}l@{}}
    Conditional exchangeability\\
    and causal consistency
\end{tabular}
\end{align*}
where 
$w(\mathbf{a}_i, \mathbf{X}_i, N_i) 
= 
N_i^{-1} Q(\mathbf{a}_i| \mathbf{X}_i, N_i) \mathbf{J}_{N_i}$ 
and $\mathbf{J}_{N_i}$ is a length $N_i$ vector of ones.

Similarly, for $t \in \{0,1\}$,
\begin{align*}
\mu_t(Q) 
&= 
\mathbb{E} \left\{ 
    \frac{1}{N_i} 
    \sum_{j = 1}^{N_i} 
    \sum_{a_{i(-j)} \in \mathcal{A} (N_i-1)} 
        Y_{ij}(t, \mathbf{a}_{i(-j)})) 
        Q(\mathbf{a}_{i(-j)} | \mathbf{X}_i, N_i)
\right\} 
\\
&= 
\mathbb{E} \left\{ 
    \frac{1}{N_i} 
    \sum_{j = 1}^{N_i} 
    \sum_{a_{i} \in \mathcal{A} (N_i)} 
        Y_{ij}(\mathbf{a}_i) 
        \mathbbm{1}(a_{ij} = t)
        Q(\mathbf{a}_{i(-j)} | \mathbf{X}_i, N_i)
\right\} 
\\
&= 
\mathbb{E} \left\{
       \sum_{\mathbf{a}_i \in \mathcal{A} (N_i)}
       w_t(\mathbf{a}_i, \mathbf{X}_i, N_i)^\top
       \mathbb{E} \left( 
           \mathbf{Y}_i
           \middle| \mathbf{A}_i = \mathbf{a}_i, \mathbf{X}_i, N_i
       \right)   
   \right\}
\end{align*}
where
$
w_t(\mathbf{a}_i, \mathbf{X}_i, N_i)
= 
N_i^{-1}
\big(
    \mathbbm{1}(a_{i1} = t) Q(\mathbf{a}_{i(-1)}| \mathbf{X}_i, N_i), 
    \dots, 
    \mathbbm{1}(a_{iN_i} = t) Q(\mathbf{a}_{i(-N_i)}| \mathbf{X}_i, N_i) 
\big)^\top $.

Let $w(\mathbf{a}_i, \mathbf{X}_i, N_i; Q)$ and
$w_t(\mathbf{a}_i, \mathbf{X}_i, N_i; Q)$ denote
the weight functions for $\mu(Q)$ and $\mu_t(Q)$, respectively.
Direct, spillover, overall, and total effect are defined by a difference between 
$\mu(Q)$, $\mu_t(Q)$, $\mu(Q')$, $\mu_t(Q')$,
and thus they can be expressed in the form 
$
\mathbb{E} \{
    \sum_{\mathbf{a}_i \in \mathcal{A} (N_i)}
    \widetilde{w}(\mathbf{a}_i, \mathbf{X}_i, N_i)^\top
    \allowbreak
    \mathbb{E} \left( 
        \mathbf{Y}_i
        \middle| 
        \allowbreak
        \mathbf{A}_i = \mathbf{a}_i, \mathbf{X}_i, N_i
    \right)   
\}
$
where the corresponding
$\widetilde{w}(\mathbf{a}_i, \mathbf{X}_i, N_i)$'s are given by a difference between 
$w(\mathbf{a}_i, \mathbf{X}_i, N_i; Q)$, 
$w_t(\mathbf{a}_i, \mathbf{X}_i, N_i; Q)$, 
$w(\mathbf{a}_i, \mathbf{X}_i, N_i; Q')$, 
and $w_t(\mathbf{a}_i, \mathbf{X}_i, N_i; Q')$
as given in Table \ref{tab:lemma1}.

\begin{table}[H]
\centering
\caption{Causal estimands and corresponding $\widetilde{w}$}
\label{tab:lemma1}
\begin{tabular}{@{}ccc@{}}
\toprule
Estimand    & Definition            & Weight function $\widetilde{w}$                                              \\ \midrule
$\mu(Q)$    & $\mu(Q)$              & $w(\mathbf{a}_i, \mathbf{X}_i, N_i; Q)$                                             \\
$\mu_t(Q)$  & $\mu_t(Q)$            & $w_t(\mathbf{a}_i, \mathbf{X}_i, N_i; Q)$                                           \\
$DE(Q)$     & $\mu_1(Q) - \mu_0(Q)$ & $w_1(\mathbf{a}_i, \mathbf{X}_i, N_i; Q) - w_0(\mathbf{a}_i, \mathbf{X}_i, N_i; Q)$ \\
$SE_t(Q, Q')$ & $\mu_t(Q) - \mu_t(Q')$ & $w_t(\mathbf{a}_i, \mathbf{X}_i, N_i; Q) - w_t(\mathbf{a}_i, \mathbf{X}_i, N_i; Q')$ \\
$OE(Q, Q')$ & $\mu(Q) - \mu(Q')$    & $w(\mathbf{a}_i, \mathbf{X}_i, N_i; Q) - w(\mathbf{a}_i, \mathbf{X}_i, N_i; Q')$    \\
$TE(Q, Q')$   & $\mu_1(Q) - \mu_0(Q')$ & $w_1(\mathbf{a}_i, \mathbf{X}_i, N_i; Q) - w_0(\mathbf{a}_i, \mathbf{X}_i, N_i; Q')$ \\ \bottomrule
\end{tabular}
\end{table}

\subsection{Proof of Theorem 1}
\label{proof:thm1}

The proof follows an approach similar to \citet{kennedy16} and \citet{park2022efficient}. 
For simplicity, the subscript expressing cluster index is omitted in the following.
The density of observed data 
$\mathbf{O} = (\mathbf{Y},\mathbf{A},\mathbf{X},N)$ at
$\mathbf{o} = (\mathbf{y},\mathbf{a},\mathbf{x},n) \in \mathcal{Y}(n) \times \mathcal{A}(n) \times \mathcal{X}(n) \times \mathbb{N}$ is 
$
    d\mathbb{P}(\mathbf{o}) 
    = 
    \allowbreak
    d\mathbb{P}(\mathbf{y}|\mathbf{a},\mathbf{x},n)
    \allowbreak
    d\mathbb{P}(\mathbf{a}|\mathbf{x},n)
    \allowbreak
    d\mathbb{P}(\mathbf{x}|n)
    \allowbreak
    d\mathbb{P}(n)
$.
Define a smooth regular parametric submodel parametrized by $\varepsilon \in \mathbb{R}$ by
$
    d\mathbb{P}(\mathbf{o};\varepsilon) 
    = 
    \allowbreak
    d\mathbb{P}(\mathbf{y}|\mathbf{a},\mathbf{x},n;\varepsilon)
    \allowbreak
    d\mathbb{P}(\mathbf{a}|\mathbf{x},n;\varepsilon)
    \allowbreak
    d\mathbb{P}(\mathbf{x}|n;\varepsilon)
    \allowbreak
    d\mathbb{P}(n;\varepsilon) ,
$
where the density of the parametric submodel is assumed to equal the observed data density at $\varepsilon = 0$.
Then, the parametric submodel score evaluated at $\varepsilon = 0$ is given by
$
    l_{\varepsilon}'(\mathbf{o};0)
    = l_{\varepsilon}'(\mathbf{y}|\mathbf{a},\mathbf{x},n;0)
    + l_{\varepsilon}'(\mathbf{a}|\mathbf{x},n;0)
    + l_{\varepsilon}'(\mathbf{x}|n;0)
    + l_{\varepsilon}'(n;0)
$
where in general $l(s|t;\varepsilon) = \text{log } d\mathbb{P}(s|t;\varepsilon)$ for any $s,t$
and
$l_{\varepsilon}'(s|t;0) 
= 
\left. 
    \left\{ 
        \partial l(s|t;\varepsilon) /\partial \varepsilon
    \right\}
\right\rvert_{ \varepsilon = 0}$.
From the parametric submodel, the tangent space is constructed as the mean closure of linear combinations of scores, given by
\begin{align*}
    \mathcal{T} 
    = 
    \Big\{
        L(\mathbf{y}, \mathbf{a}, \mathbf{x}, n) 
        \in \mathbb{R}
        \Big| 
        & L(\mathbf{y}, \mathbf{a}, \mathbf{x}, n)
        = L_{\mathbf{Y}}(\mathbf{y}, \mathbf{a}, \mathbf{x}, n) 
        + L_{\mathbf{A}}(\mathbf{a}, \mathbf{x}, n)
        + L_{\mathbf{X}}(\mathbf{x}, n)
        + L_{N}(n)
        \in \mathbb{R}, 
        \\
        & 
        \mathbb{E} \left\{ 
            L_{\mathbf{Y}}(\mathbf{Y}, \mathbf{A}, \mathbf{X}, N) 
            \middle| \mathbf{A} = \mathbf{a}, \mathbf{X} = \mathbf{x}, N = n 
        \right\} = 0, 
        \\
        &
        \mathbb{E} \left\{ 
            L_{\mathbf{A}}(\mathbf{A}, \mathbf{X}, N) 
            \middle| \mathbf{X} = \mathbf{x}, N = n 
        \right\} = 0,
        \\
        & 
        \mathbb{E} \left\{ 
            L_{\mathbf{X}}(\mathbf{X}, N) 
            \middle| N = n 
        \right\} = 0, 
        \
        \mathbb{E} \left\{ 
            L_{\mathbf{N}}(N) 
        \right\} = 0,
        \\
        &
        \text{ for all } (\mathbf{y},\mathbf{a},\mathbf{x},n) \in \mathcal{Y}(n) \times \mathcal{A}(n) \times \mathcal{X}(n) \times \mathbb{N}
    \Big\}.
\end{align*}
We will use the fact that
$\left. 
    \left\{ 
        \partial d\mathbb{P}(s|t;\varepsilon) /\partial \varepsilon
    \right\}
\right\rvert_{ \varepsilon = 0} 
= l_{\varepsilon}'(s|t;0) d\mathbb{P}(s|t)$
from the chain rule,
and
$\mathbb{E}\{l_{\varepsilon}'(S|T;0)|T\} = 0$
from the usual property of score functions in the following proof.
For simplicity, we slightly abuse notation by suppressing the dependency on random variables if there is no ambiguity,
for example, 
$\mathbb{E} \left( 
        \mathbf{Y}
        \middle| \mathbf{a}, \mathbf{X}, N
    \right)
=
\mathbb{E} \left( 
        \mathbf{Y}
        \middle| \mathbf{A} = \mathbf{a}, \mathbf{X}, N
    \right)$.

Now the estimand $\Psi(w) = \Psi(w; \mathbb{P})$ is a functional of the distribution $\mathbb{P}$, given by
\begin{align*}
\Psi(w; \mathbb{P}) 
&= 
\mathbb{E} \left\{
    \sum_{\mathbf{a} \in \mathcal{A} (N)}
    w(\mathbf{a}, \mathbf{X}, N)^\top
    \mathbb{E} \left( 
        \mathbf{Y}
        \middle| 
        \mathbf{a}, \mathbf{X}, N
    \right)
\right\} 
\\
&= 
\sum_{n \in \mathbb{N}}
\int_{\mathcal{X}(n)} 
\sum_{\mathbf{a} \in \mathcal{A} (n)}
\int_{\mathcal{Y}(n)} 
    w(\mathbf{a}, \mathbf{x}, n)^\top
    \mathbf{y}
    d\mathbb{P}(\mathbf{y}|\mathbf{a},\mathbf{x},n)
    d\mathbb{P}(\mathbf{x}|n)
    d\mathbb{P}(n)
.
\end{align*}
Then, the estimand $\Psi(w)$ at parameter $\varepsilon$ in the regular parametric submodel, 
denoted by $\Psi(w; \mathbb{P}_{\varepsilon})$, is
\begin{align*}
\Psi(w; \mathbb{P}_{\varepsilon}) 
= 
\sum_{n \in \mathbb{N}}
\int_{\mathcal{X}(n)} 
\sum_{\mathbf{a} \in \mathcal{A} (n)}
\int_{\mathcal{Y}(n)} 
    w(\mathbf{a}, \mathbf{x}, n; \mathbb{P}_\varepsilon)^\top
    \mathbf{y}
    d\mathbb{P}(\mathbf{y}|\mathbf{a},\mathbf{x},n; \varepsilon)
    d\mathbb{P}(\mathbf{x}|n; \varepsilon)
    d\mathbb{P}(n; \varepsilon),
\end{align*}
where $\varepsilon$ in 
$w(\mathbf{a}, \mathbf{x}, n; \mathbb{P}_\varepsilon)$ expresses the possibility that 
$w(\mathbf{a}, \mathbf{x}, n)$ would depend on the observed data distribution.

To prove that
\begin{align*}
    \varphi^{*}(\mathbf{O}) 
    =& 
    \sum_{\mathbf{a} \in \mathcal{A}(N)}
    \big\{ w(\mathbf{a}, \mathbf{X}, N) + \phi(\mathbf{A}, \mathbf{X}, N; \mathbf{a}) \big\}^\top
    \mathbb{E} \big(\mathbf{Y} | \mathbf{A} = \mathbf{a}, \mathbf{X}, N \big) 
    \\
    &+ 
    \frac{1}{\mathbb{P}(\mathbf{A}|\mathbf{X}, N)}
    w(\mathbf{A}, \mathbf{X}, N)^\top 
    \left\{ 
        \mathbf{Y} - \mathbb{E} \big( \mathbf{Y} | \mathbf{A}, \mathbf{X}, N \big) 
    \right\}
    - 
    \Psi(w) 
\end{align*}
is the efficient influence function (EIF) of $\Psi(w)$, 
we need to show
(A)
$
\left. 
\left\{ 
        \partial \Psi(w;\mathbb{P}_\varepsilon) /\partial \varepsilon
    \right\}
\right\rvert_{ \varepsilon = 0}
=
\mathbb{E} \left[ 
    \varphi^{*}(\mathbf{O})
    \times
    l_{\varepsilon}'(\mathbf{Y}, \mathbf{A}, \mathbf{X}, N;0)  
\right]
$,
(B)
$
\mathbb{E} \left\{ 
    \varphi^{*}(\mathbf{O})
\right\}
\allowbreak
=
0
$,
(C)
$
\text{Var} \left\{ 
    \varphi^{*}(\mathbf{O})
\right\}
<
\infty
$,
and
(D)
$\varphi^{*}(\mathbf{O})$ is included in the tangent space $\mathcal{T}$,
where 
$
\left. 
    \left\{ 
        \partial \Psi(w;\mathbb{P}_\varepsilon) /\partial \varepsilon
    \right\}
\right\rvert_{ \varepsilon = 0} $
is a path-wise derivative of $\Psi(w;\mathbb{P}_\varepsilon)$ evaluated at $\varepsilon = 0$.
It is easy to show (B) -- (D), so we only prove (A) here.

First, the path-wise derivative of $\Psi(w;\mathbb{P}_\varepsilon)$ at $\varepsilon = 0$ is
\begin{align*}
    &
    \left. 
        \frac
            {\partial \Psi(w;\mathbb{P}_\varepsilon)}
            {\partial \varepsilon}
    \right\rvert_{ \varepsilon = 0}
    \\
    = &
    \sum_{n \in \mathbb{N}}
    \int_{\mathcal{X}(n)} 
    \sum_{\mathbf{a} \in \mathcal{A} (n)}
    \int_{\mathcal{Y}(n)} 
        \left. 
            \frac
                {\partial w(\mathbf{a}, \mathbf{x}, n; \mathbb{P}_\varepsilon)}
                {\partial \varepsilon}
        \right\rvert_{ \varepsilon = 0}^\top
        \mathbf{y}
        d\mathbb{P}(\mathbf{y}|\mathbf{a},\mathbf{x},n)
        d\mathbb{P}(\mathbf{x}|n)
        d\mathbb{P}(n)
    \\
    & +
    \sum_{n \in \mathbb{N}}
    \int_{\mathcal{X}(n)} 
    \sum_{\mathbf{a} \in \mathcal{A} (n)}
    \int_{\mathcal{Y}(n)} 
        w(\mathbf{a}, \mathbf{x}, n)^\top
        \mathbf{y}
        \{
            l_{\varepsilon}'(\mathbf{y}|\mathbf{a},\mathbf{x},n;0)
            + 
            l_{\varepsilon}'(\mathbf{x},n;0)
        \}
        d\mathbb{P}(\mathbf{y}|\mathbf{a},\mathbf{x},n)
        d\mathbb{P}(\mathbf{x}|n)
        d\mathbb{P}(n)
    .
\end{align*}
Here, the path-wise derivative of $w(\mathbf{a}, \mathbf{x}, n;\mathbb{P}_\varepsilon)$ at $\varepsilon = 0$ is
\begin{align*}
    \left. 
        \frac
            {\partial w(\mathbf{a}, \mathbf{x}, n; \mathbb{P}_\varepsilon)}
            {\partial \varepsilon}
    \right\rvert_{ \varepsilon = 0}
    &= 
    \mathbb{E} \left\{ 
        \frac
            {\mathbbm{1}(\mathbf{X} = \mathbf{x}, N = n)}
            {d\mathbb{P}(\mathbf{x}, n)}
        \phi(\mathbf{A}, \mathbf{X}, N; \mathbf{a})
        \times
        l_{\varepsilon}'(\mathbf{Y}, \mathbf{A}, \mathbf{X}, N;0)  
    \right\}
    \\
    &=
    \mathbb{E} \left\{ 
        \phi(\mathbf{A}, \mathbf{X}, N; \mathbf{a})
        \times
        l_{\varepsilon}'(\mathbf{Y}, \mathbf{A}, \mathbf{X}, N;0)
        \middle| 
        \mathbf{X} = \mathbf{x}, N = n
    \right\} 
\end{align*}
from the definition of $\phi$ and taking iterated expectation.

Therefore, the first term in 
$\left. 
\left\{ 
        \partial \Psi(w;\mathbb{P}_\varepsilon) /\partial \varepsilon
    \right\}
\right\rvert_{ \varepsilon = 0}$
equals
\begin{align*}
    &
    \sum_{n \in \mathbb{N}}
    \int_{\mathcal{X}(n)} 
    \sum_{\mathbf{a} \in \mathcal{A} (n)}
        \left. 
            \frac
                {\partial w(\mathbf{a}, \mathbf{x}, n; \mathbb{P}_\varepsilon)}
                {\partial \varepsilon}
        \right\rvert_{ \varepsilon = 0}^\top
    \int_{\mathcal{Y}(n)} 
        \mathbf{y}
        d\mathbb{P}(\mathbf{y}|\mathbf{a},\mathbf{x},n)
        d\mathbb{P}(\mathbf{x}|n)
        d\mathbb{P}(n)
    \\
    =&
    \sum_{n \in \mathbb{N}}
    \int_{\mathcal{X}(n)} 
    \sum_{\mathbf{a} \in \mathcal{A} (n)}
        \mathbb{E} \left\{ 
            \phi(\mathbf{A}, \mathbf{X}, N; \mathbf{a})
            \times
            l_{\varepsilon}'(\mathbf{Y}, \mathbf{A}, \mathbf{X}, N;0)
            \middle| 
            \mathbf{x},n
        \right\} ^\top
    \mathbb{E} \left( 
        \mathbf{Y}
        \middle| 
        \mathbf{a}, \mathbf{x}, n
    \right)
        d\mathbb{P}(\mathbf{x}|n)
        d\mathbb{P}(n)
    \\
    =&
    \sum_{n \in \mathbb{N}}
    \int_{\mathcal{X}(n)} 
    \mathbb{E} \left\{ 
        \sum_{\mathbf{a} \in \mathcal{A} (N)}
                \phi(\mathbf{A}, \mathbf{X}, N; \mathbf{a})
        ^\top
        \mathbb{E} \left( 
            \mathbf{Y}
            \middle| 
            \mathbf{a}, \mathbf{X}, N
        \right)
        \times
        l_{\varepsilon}'(\mathbf{Y}, \mathbf{A}, \mathbf{X}, N;0)
        \middle| 
        \mathbf{x},n
    \right\}
        d\mathbb{P}(\mathbf{x}|n)
        d\mathbb{P}(n)
    \\
    =&
    \mathbb{E} \left\{ 
        \sum_{\mathbf{a} \in \mathcal{A} (N)}
                \phi(\mathbf{A}, \mathbf{X}, N; \mathbf{a})
        ^\top
        \mathbb{E} \left( 
            \mathbf{Y}
            \middle| 
            \mathbf{a}, \mathbf{X}, N
        \right)
        \times
        l_{\varepsilon}'(\mathbf{Y}, \mathbf{A}, \mathbf{X}, N;0)
    \right\}
    ,
\end{align*}
and the second term equals to
\begin{align*}
    &
    \sum_{n \in \mathbb{N}}
    \int_{\mathcal{X}(n)} 
    \sum_{\mathbf{a} \in \mathcal{A} (n)}
    \int_{\mathcal{Y}(n)} 
        w(\mathbf{a}, \mathbf{x}, n)^\top
        \mathbf{y}
        \{
            l_{\varepsilon}'(\mathbf{y}|\mathbf{a},\mathbf{x},n;0)
            + 
            l_{\varepsilon}'(\mathbf{x},n;0)
        \}
        d\mathbb{P}(\mathbf{y}|\mathbf{a},\mathbf{x},n)
        d\mathbb{P}(\mathbf{x}|n)
        d\mathbb{P}(n)
    \\
    = &
    \sum_{n \in \mathbb{N}}
    \int_{\mathcal{X}(n)} 
    \sum_{\mathbf{a} \in \mathcal{A} (n)}
    \int_{\mathcal{Y}(n)} 
        \frac
            {w(\mathbf{a}, \mathbf{x}, n)^\top
            \mathbf{y}}
            {\mathbb{P}(\mathbf{a}|\mathbf{x},n)}
        \{
            l_{\varepsilon}'(\mathbf{y}|\mathbf{a},\mathbf{x},n;0)
            + 
            l_{\varepsilon}'(\mathbf{x},n;0)
        \}
        d\mathbb{P}(\mathbf{y},\mathbf{a},\mathbf{x},n)
    \\
    = &
    \mathbb{E} \left[
        \frac
            {w(\mathbf{A}, \mathbf{X}, N)^\top 
            \mathbf{Y}}
            {\mathbb{P}(\mathbf{A}|\mathbf{X}, N)}
        \times
        \left\{ 
            l_{\varepsilon}'(\mathbf{Y}|\mathbf{A},\mathbf{X},N;0)
            + 
            l_{\varepsilon}'(\mathbf{X},N;0)
        \right\}
    \right]
    \\
    = &
    \mathbb{E} \left\{
        \frac
            {w(\mathbf{A}, \mathbf{X}, N)^\top 
            \mathbf{Y}}
            {\mathbb{P}(\mathbf{A}|\mathbf{X}, N)}
        \times
        l_{\varepsilon}'(\mathbf{Y},\mathbf{A},\mathbf{X},N;0)
    \right\}
    -
    \mathbb{E} \left\{
        \frac
            {w(\mathbf{A}, \mathbf{X}, N)^\top 
            \mathbb{E} \left( 
                \mathbf{Y}
                \middle| 
                \mathbf{A}, \mathbf{X}, N
            \right)}
            {\mathbb{P}(\mathbf{A}|\mathbf{X}, N)}
        \times
        l_{\varepsilon}'(\mathbf{A},\mathbf{X},N;0)
    \right\}
    \\
    & +
    \mathbb{E} \left[
        \mathbb{E} \left\{
            \frac
                {w(\mathbf{A}, \mathbf{X}, N)^\top 
                \mathbb{E} \left( 
                    \mathbf{Y}
                    \middle| 
                    \mathbf{A}, \mathbf{X}, N
                \right)}
                {\mathbb{P}(\mathbf{A}|\mathbf{X}, N)}
            \middle| 
            \mathbf{X},N
        \right\}
        \times
        l_{\varepsilon}'(\mathbf{X},N;0)
    \right]
    \\
    = &
    \mathbb{E} \left\{
        \frac
            {w(\mathbf{A}, \mathbf{X}, N)^\top 
            \mathbf{Y}}
            {\mathbb{P}(\mathbf{A}|\mathbf{X}, N)}
        \times
        l_{\varepsilon}'(\mathbf{Y},\mathbf{A},\mathbf{X},N;0)
    \right\}
    -
    \mathbb{E} \left\{
        \frac
            {w(\mathbf{A}, \mathbf{X}, N)^\top 
            \mathbb{E} \left( 
                \mathbf{Y}
                \middle| 
                \mathbf{A}, \mathbf{X}, N
            \right)}
            {\mathbb{P}(\mathbf{A}|\mathbf{X}, N)}
        \times
        l_{\varepsilon}'(\mathbf{Y},\mathbf{A},\mathbf{X},N;0)
    \right\}
    \\
    & +
    \mathbb{E} \left\{
        \sum_{\mathbf{a} \in \mathcal{A} (N)}
            w(\mathbf{a}, \mathbf{X}, N)^\top 
            \mathbb{E} \left( 
                \mathbf{Y}
                \middle| 
                \mathbf{a}, \mathbf{X}, N
            \right)
        \times
        l_{\varepsilon}'(\mathbf{Y},\mathbf{A},\mathbf{X},N;0)
    \right\}
\end{align*}
from the decomposition of the score 
$l_{\varepsilon}'(\mathbf{Y},\mathbf{A},\mathbf{X},N;0)$,
taking iterated expectation,
and using the property of the score such as 
$\mathbb{E} \left\{ 
    l_{\varepsilon}'(\mathbf{Y} | \mathbf{A}, \mathbf{X}, N;0)
    \middle| 
    \mathbf{A}, \mathbf{X}, N
\right\}
=0$.

Thus, we have
\begin{align*}
    \left. 
        \frac
            {\partial \Psi(w;\mathbb{P}_\varepsilon)}
            {\partial \varepsilon}
    \right\rvert_{ \varepsilon = 0}
    =& 
    \mathbb{E} \Bigg(
        \begin{aligned}[t]
        \Bigg[
            &
            \sum_{\mathbf{a} \in \mathcal{A}(N)}
                \big\{ 
                    w(\mathbf{a}, \mathbf{X}, N) + \phi(\mathbf{A}, \mathbf{X}, N; \mathbf{a}) 
                \big\}
                ^\top
                \mathbb{E} \big(
                    \mathbf{Y} | \mathbf{a}, \mathbf{X}, N 
                \big) 
            \\
            &+ 
            \frac
                {w(\mathbf{A}, \mathbf{X}, N)
                ^\top 
                \left\{ 
                    \mathbf{Y} 
                    -
                    \mathbb{E} \big( 
                        \mathbf{Y} | \mathbf{A}, \mathbf{X}, N 
                    \big) 
                \right\}}
                {\mathbb{P}(\mathbf{A}|\mathbf{X}, N)}
            \Bigg]
            \times
            l_{\varepsilon}'(\mathbf{Y},\mathbf{A},\mathbf{X},N;0)
            \Bigg)
        \end{aligned}
    \\
    =&
    \mathbb{E} \left[ 
        \varphi^{*}(\mathbf{O})
        \times
        l_{\varepsilon}'(\mathbf{Y}, \mathbf{A}, \mathbf{X}, N;0)  
    \right]
\end{align*}
which follows from
$\mathbb{E} \left[ 
    \Psi(w)
    \times
    l_{\varepsilon}'(\mathbf{Y}, \mathbf{A}, \mathbf{X}, N;0)  
\right] = 0$,
which proves (A).

\subsection{Proof of Theorem 2}
\label{proof:thm2}

Consider the estimand
$\Psi(w) 
= 
\mathbb{E} \big\{
    \sum_{\mathbf{a} \in \mathcal{A} (N)}
    w(\mathbf{a}, \mathbf{X}, N)^\top
    \allowbreak
    \mathbb{E} \left( 
        \mathbf{Y}
        \middle| 
        \mathbf{a}, \mathbf{X}, N
    \right)
\big\}$
and its uncentered EIF
\begin{align*}
    \varphi(\mathbf{O}; \boldsymbol{\eta}) 
    =& 
    \sum_{\mathbf{a} \in \mathcal{A}(N)} 
        \big\{ 
            w(\mathbf{a}, \mathbf{X}, N) + \phi(\mathbf{A}, \mathbf{X}, N;\mathbf{a}) 
        \big\}^\top 
        G(\mathbf{a}, \mathbf{X}, N)
    + 
    \frac
        {w(\mathbf{A}, \mathbf{X}, N)^\top \left\{ \mathbf{Y} - G(\mathbf{A}, \mathbf{X}, N) \right\}}
        {H(\mathbf{A}, \mathbf{X}, N)}
    .
\end{align*}
Hereinafter, 
let $\mathbb{P}f(\mathbf{O}) 
= 
\int f(\mathbf{o}) d\mathbb{P}(\mathbf{o})$ denote the expectation of $f(\mathbf{O})$,
treating the function $f$ as fixed even when it is estimated from the sample and thus random.
For example, if $\widehat{f}$ is a prediction model (e.g., random forest prediction model) trained on the data $D$ which is independent of $\mathbf{O}$,
then
$\mathbb{P}\widehat{f}(\mathbf{O}) 
= 
\int \widehat{f}(\mathbf{o}) d\mathbb{P}(\mathbf{o})
=
\mathbb{E}\left\{
    \widehat{f}(\mathbf{O}) 
    \middle| 
    D 
\right\}
$.
Then for
$    
\Psi(w) = \allowbreak
    \mathbb{P}
    \big\{ 
        \varphi(\mathbf{O}; \boldsymbol{\eta}) 
    \big\}
$
and its estimator
$    
\widehat{\Psi}(w) = \allowbreak
    K^{-1} \allowbreak 
    \sum_{k=1}^{K} \allowbreak
    \mathbb{P}_m^k 
    \big\{ 
        \varphi(\mathbf{O}; \widehat{\boldsymbol{\eta}}^{(-k)}) 
    \big\}
$,
we have the following decomposition:
\begin{align*}
    &\widehat\Psi(w) - \Psi(w)
    \\
    &=
    \frac{1}{K}
    \sum_{k=1}^{K}
    \left[
        (\mathbb{P}_m^k - \mathbb{P})
            \varphi(\mathbf{O}; \boldsymbol{\eta}) 
        +
        (\mathbb{P}_m^k - \mathbb{P}) \big\{ \varphidiffk \big\}
        +
        \mathbb{P} \big\{ \varphidiffk \big\}
    \right]
\end{align*}
The first term $(\mathbb{P}_m^k - \mathbb{P}) \varphi(\mathbf{O}; \boldsymbol{\eta})$ is $O_\mathbb{P}(m^{-1/2})$ from the central limit theorem, 
and the second term is
\begin{align*}
(\mathbb{P}_m^k - \mathbb{P})
\big\{ \varphidiffk \big\}
=
O_\mathbb{P} \left( 
    \frac
    {|| \varphidiffk ||}
    {m_k^{1/2} }
\right)
\end{align*}
from Lemma 1 in \citet{kennedy22}.
It remains to analyze 
$||\varphidiffk||$
and
$\mathbb{P} \big\{ \varphidiffk \big\}$.
For notational convenience, omit superscript $(-k)$ in $\widehat{\boldsymbol{\eta}}^{(-k)}$ and let $D$ denote the data that $\widehat{\boldsymbol{\eta}}^{(-k)}$ was trained on and $\mathbf{O} = (\mathbf{Y}, \mathbf{A}, \mathbf{X}, N)$ denote a new cluster-level observation which is independent of $D$.
Also, let
$\widehat{G}(\mathbf{a}) = \widehat{G}(\mathbf{a}, \mathbf{X}, N)$,
$\widehat{G}(\mathbf{A}) = \widehat{G}(\mathbf{A}, \mathbf{X}, N)$,
$\widehat{H}(\mathbf{a}) = \widehat{H}(\mathbf{a}, \mathbf{X}, N)$,
$\widehat{H}(\mathbf{A}) = \widehat{H}(\mathbf{A}, \mathbf{X}, N)$,
$\widehat{w}(\mathbf{a}) = \widehat{w}(\mathbf{a}, \mathbf{X}, N)$,
$\widehat{w}(\mathbf{A}) = \widehat{w}(\mathbf{A}, \mathbf{X}, N)$,
$\widehat{\phi}(\mathbf{a}) = \widehat{\phi}(\mathbf{A}, \mathbf{X}, N; \mathbf{a})$,
and define 
$G(\mathbf{a})$,
$G(\mathbf{A})$,
$H(\mathbf{a})$,
$H(\mathbf{A})$,
$w(\mathbf{a})$,
$w(\mathbf{A})$,
$\phi(\mathbf{a})$
similarly.

First, we have the following decomposition of
$
\mathbb{P} \big\{ \varphidiff \big\}
$
from iterated expectation:
\begin{align}
    &
    \mathbb{P} \big\{ \varphidiff \big\}
    \nonumber
    \\
    = &
    \ED{[}{]}{\varphidiff}
    \nonumber
    \\
    = &
    \ED{[}{]}{
        \suma \big\{ \wh{a} + \phih \big\}^\top \Gh{a}
    }
    \nonumber
    +
    \ED{[}{]}{
        \frac
            {\wh{A}^\top \{ \Gt{A} - \Gh{A} \}}
            {\Hh{A}}
    }
    -
    \ED{[}{]}{\suma \wt{a} ^\top \Gt{a}}
    \nonumber
    \\
    = &
    \ED{(}{)}{
        \suma \left[ \wh{a} - \wt{a} 
            + \E{\{}{\}}{\phih \middle| D, \mathbf{X}, N} 
        \right]^\top \Gt{a}
    }
    \nonumber
    \\
    & +
    \ED{[}{]}{
        \suma 
            \big\{ \wh{a} + \phih \big\}^\top 
            \big\{\Gh{a} - \Gt{a}\big\}
    }
    \nonumber
    \\
    & +
    \ED{[}{]}{
        \suma
        \frac
            {\wh{a}^\top \{ \Gt{a} - \Gh{a} \}}
            {\Hh{a}}
        \Ht{a}
    }
    \nonumber
    \\
    = &
    \ED{[}{]}{
        \suma \left\{ 
            \wh{a} - \wt{a} 
            + 
            \sum_{\mathbf{a}' \in \mathcal{A}(N)}
                \widehat{\phi}(\mathbf{a}', \mathbf{X}, N; \mathbf{a})
                H(\mathbf{a}, \mathbf{X}, N)
        \right\}^\top \Gt{a}
    }
    \label{thm:2:1}
    \\
    & +
    \ED{[}{]}{
        \suma 
            \wh{a}^\top 
            \big\{\Gh{a} - \Gt{a}\big\}
            \left\{ 
                1
                -
                \frac{\Ht{a}}{\Hh{a}}
            \right\}
    }
    \label{thm:2:2}
    \\
    & +
    \ED{[}{]}{
        \suma 
            \big\{\phih - \phit \big\}^\top 
            \big\{\Gh{a} - \Gt{a}\big\}
    }
    \label{thm:2:3}
\end{align}
where
the last equality follows from
$
\mathbb{E} \left\{ 
    \phi(\mathbf{A}, \mathbf{X}, N; \mathbf{a})
    \middle| 
    \mathbf{X}, N
\right\}  = 0
$.
Each term in the above decomposition is bounded by
\begin{align*}
    (\ref{thm:2:1}) 
    \lesssim
    \ED{[}{]}{
        \suma \bigg|\bigg|
            \wh{a} - \wt{a} 
            +
            \sum_{\mathbf{a}' \in \mathcal{A}(N)}
                \widehat{\phi}(\mathbf{a}', \mathbf{X}, N; \mathbf{a})
                H(\mathbf{a}, \mathbf{X}, N)
        \bigg|\bigg|_2
    }
    =
    \OP{r_w^2}
\end{align*}
\begin{align*}
    (\ref{thm:2:2}) 
    \lesssim
    \ED{[}{]}{
        \suma \normt{\Gh{a} - \Gt{a}}^2
    }^{1/2}
    \ED{[}{]}{
        \suma | \Hh{a} - \Ht{a} |^2
    }^{1/2}
    =
    \OP{r_G r_H}
\end{align*}
\begin{align*}
    (\ref{thm:2:3}) 
    \lesssim
    \ED{[}{]}{
        \suma \normt{\phih - \phit}^2
    }^{1/2}
    \ED{[}{]}{
        \suma \normt{\Gh{a} - \Gt{a}}^2
    }^{1/2}
    =
    \OP{r_\phi r_G}
\end{align*}
where $\alpha \lesssim \beta$ if and only if there exist a constant $C$ such that $\alpha \le C \beta$
from Cauchy-Schwarz inequality and boundedness of nuisance functions.
Therefore, we have
\begin{align}
    \mathbb{P} \big\{ \varphidiff \big\}
    =
    \OP{r_w^2 + r_G r_H + r_G r_\phi}
    .
    \label{thm:2:4}
\end{align}

Next, consider $||\varphidiff||$.
We have the following decomposition:
\begin{align}
    ||\varphidiff||^2
    &=
    \int 
    \left\{
        \varphi(\mathbf{o}; \widehat{\boldsymbol{\eta}}) 
        -
        \varphi(\mathbf{o}; \boldsymbol{\eta})
    \right\}^2
    d\mathbb{P}(\mathbf{o})
    \nonumber
    \\
    &=
    \ED{[}{]}{ \{\varphidiff\}^2 }
    \nonumber
    \\
    &=
    \ED{[}{]}{ 
        \VDA{\{}{\}}{ 
            \varphidiff 
        } 
    }
    \label{thm:2:5}
    \\
    & \ \ \ +
    \VD{[}{]}{
        \EDA{\{}{\}}{
            \varphidiff
        }
    }
    \label{thm:2:6}
    \\
    & \ \ \ +
    \left[
        \ED{\{}{\}}{\varphidiff}
    \right]^2
    \label{thm:2:7}
    .
\end{align}

Note that 
\begin{align*}
    \varphidiff
    =
    \left\{
        \frac
            {\wh{A}}
            {\Hh{A}}
        -
        \frac
            {\wt{A}}
            {\Ht{A}}
    \right\}^\top
    \mathbf{Y}
    +
    T(D, \mathbf{A}, \mathbf{X}, N)
\end{align*}
where 
\begin{align*}
    T(D, \mathbf{A}, \mathbf{X}, N)
    =
    \sum_{\mathbf{a} \in \mathcal{A}(N)}
    \left[
        \big\{ \wh{a} + \phih \big\}^\top \Gh{a}
        -
        \big\{ \wt{a} + \phit \big\}^\top \Gt{a}
    \right]
    -
    \frac
        {\wh{A}^\top \Gh{A}}
        {\Hh{A}}
    +
    \frac
        {\wt{A}^\top \Gt{A}}
        {\Ht{A}}
\end{align*}
is a function of $D, \mathbf{A}, \mathbf{X}, N$, but not $\mathbf{Y}$.
Thus, the first term in the above decomposition is bounded as follows:
\begin{align*}
    (\ref{thm:2:5}) 
    = &
    \ED{(}{)}{
        \VDA{[}{]}{
            \left\{
                \frac
                    {\wh{A}}
                    {\Hh{A}}
                -
                \frac
                    {\wt{A}}
                    {\Ht{A}}
            \right\}^\top
            \mathbf{Y}
            +
            T(D, \mathbf{A}, \mathbf{X}, N)
        }
    }
    \\
    = &
    \ED{[}{]}{
        \left\{
            \frac
                {\wh{A}}
                {\Hh{A}}
            -
            \frac
                {\wt{A}}
                {\Ht{A}}
        \right\}^\top
        \V{(}{)}{
            \mathbf{Y}
            \middle|
            \mathbf{A}, \mathbf{X}, N
        }
        \left\{
            \frac
                {\wh{A}}
                {\Hh{A}}
            -
            \frac
                {\wt{A}}
                {\Ht{A}}
        \right\}
    }
    \\
    \lesssim &
    \ED{(}{)}{
        \suma
            \normt{\wh{a} - \wt{a}}^2 + |\Hh{a} - \Ht{a}|^2
    }
    \\
    = &
    \OP{r_w^4 + r_\phi^2 + r_H^2}
\end{align*}
where
the third line follows from bounded 
$\V{(}{)}{
    \mathbf{Y}
    \middle|
    \mathbf{A}, \mathbf{X}, N
}$ and nuisance functions,
\begin{align*}
    \norm{
        \frac
            {\wh{A}}
            {\Hh{A}}
        -
        \frac
            {\wt{A}}
            {\Ht{A}}
    }_2^2
    = &
    \norm{
        \frac
            {\wh{A}-\wt{A}}
            {\Hh{A}}
        +
        \frac
            {\Ht{A} - \Hh{A}}
            {\Hh{A}\Ht{A}}
        \wt{A}
    }_2^2
    \\
    & \lesssim
    \normt{\wh{A} - \wt{A}}^2 + |\Hh{A} - \Ht{A}|^2
    ,
\end{align*}
and the following relationship
\begin{align*}
    &
    \ED{(}{)}{
        \suma \normt{\wh{a} - \wt{a}}^2
    }
    \\
    \lesssim &
    \ED{[}{]}{
        \suma \normt{\wh{a} - \wt{a} + 
        \mathbb{E}\{ \phih | D, \mathbf{X}, N \}}^2
    }
    +
    \ED{[}{]}{
        \suma \normt{\mathbb{E}\{ \phih | D, \mathbf{X}, N \}}^2
    }
    \\
    = &
    \ED{[}{]}{
        \suma \normt{\wh{a} - \wt{a} + 
        \mathbb{E}\{ \phih | D, \mathbf{X}, N \}}^2
    }
    +
    \ED{[}{]}{
        \suma \normt{\mathbb{E}\{ \phih - \phit | D, \mathbf{X}, N \}}^2
    }
    \\
    \lesssim &
    \ED{[}{]}{
        \suma \normt{\wh{a} - \wt{a} + 
        \mathbb{E}\{ \phih | D, \mathbf{X}, N \}}^2
    }
    +
    \ED{[}{]}{
        \suma \normt{ 
            \phih - \phit
        }^2
    }
    \\
    = &
    \OP{r_w^4 + r_\phi^2}
\end{align*}
from
$\E{\{}{\}}{
    \normt{
        \E{(}{)}{
            V | W
        }
    }^2
}
\le
\E{(}{)}{
    \normt{V}^2
}
$
and
$\mathbb{E}\{\phit | D, \mathbf{X}, N \} 
= 
\mathbb{E} \left\{ 
    \phi(\mathbf{A}, \mathbf{X}, N; \mathbf{a})
    \middle| 
    \mathbf{X}, N
\right\}
=
0$.\\

On the other hand, from
\begin{align*}
    \EDA{\{}{\}}{
        \varphidiff
    }
    = &
    \suma \{\phih - \phit\}^\top \Gh{a}
    + 
    \suma \{\wh{a} - \wt{a}\}^\top \Gh{a}
    \\
    & +
    \suma \{\wt{a} + \phit\}^\top \{\Gh{a} - \Gt{a}\}
    +
    \frac
        {\wh{A}^\top \{\Gt{A} - \Gh{A}\}}
        {\Hh{A}}
    ,
\end{align*}
the second term is bounded by
\begin{align*}
    (\ref{thm:2:6}) 
    = &
    \VD{[}{]}{
        \EDA{\{}{\}}{
            \varphidiff
        }
    }
    \\    
    \lesssim &
    \ED{[}{]}{
        \suma \normt{\phih - \phit}^2
    }
     +
    \ED{[}{]}{
        \suma \normt{\wh{a} - \wt{a}}^2
    }
     +
    \ED{[}{]}{
        \suma \normt{\Gh{a} - \Gt{a}}^2
    }
    \\
    = &
    \OP{r_\phi^2 + r_w^4 + r_G^2}
\end{align*}
from $\textup{Var}(Z_1 + \dots + Z_n) \lesssim \mathbb{E}Z_1^2 + \dots + \mathbb{E}Z_n^2$ 
and boundedness of nuisance functions.

Finally, from (\ref{thm:2:4}),
$
    (\ref{thm:2:7})
    =
    \left[
        \ED{\{}{\}}{\varphidiff}
    \right]^2
    =
    \OP{r_w^4 + r_G^2r_H^2 + r_G^2r_\phi^2}
$,
and thus
\begin{align}
    ||\varphidiff||
    =
    \OP{r_w^2 + r_G + r_H + r_\phi}
    =
    \OP{1}
    .
    \label{thm:2:8}
\end{align}

In conclusion, from (\ref{thm:2:4}) and (\ref{thm:2:8}),
\begin{align*}
    \widehat\Psi(w) - \Psi(w)
    =
    O_\mathbb{P} (m^{-1/2} + r_w^2 + r_G r_H + r_G r_\phi)
,
\end{align*}
which proves the consistency of $\widehat{\Psi}(w)$.

\subsection{Proof of Theorem 3}

Assume sample splitting is done approximately uniformly, such that $\frac{m_k}{m} = \frac{1}{K} + O(\frac{1}{m})$.
From the proof of Theorem 2, we have
\begin{align*}
    \widehat\Psi(w) - \Psi(w)
    = &
    \frac{1}{K}
    \sum_{k=1}^{K}
    \left[
        (\mathbb{P}_m^k - \mathbb{P})
            \varphi(\mathbf{O}; \boldsymbol{\eta}) 
    \right]
    +
    m^{-1/2} O_\mathbb{P} (r_w^2 + r_G + r_H + r_\phi)
    +
    O_\mathbb{P} (r_w^2 + r_G r_H + r_G r_\phi)
    \\
    = &
    (\mathbb{P}_m - \mathbb{P})
    \varphi(\mathbf{O}; \boldsymbol{\eta})
    +
    \OP{m^{-1}}
    +
    m^{-1/2} O_\mathbb{P} (r_w^2 + r_G + r_H + r_\phi)
    +
    O_\mathbb{P} (r_w^2 + r_G r_H + r_G r_\phi)
\end{align*}
from
\begin{align*}
    \frac{1}{K}
    \sum_{k=1}^{K}
    \left[
        (\mathbb{P}_m^k - \mathbb{P})
            \varphi(\mathbf{O}; \boldsymbol{\eta}) 
    \right]
    =
    (\mathbb{P}_m - \mathbb{P})
    \varphi(\mathbf{O}; \boldsymbol{\eta})
    +
    \frac{1}{K}
    \sum_{k=1}^K
        \mathbb{P}_m^k
        \left\{
            \varphi(\mathbf{O}; \boldsymbol{\eta})
        \right\}
        \left(
        1
        -
        \frac{K m_k}{m}
        \right)
\end{align*}
and
\begin{align*}
    \mathbb{P}_m^k
        \left\{
            \varphi(\mathbf{O}; \boldsymbol{\eta})
        \right\}
        \left(
        1
        -
        \frac{K m_k}{m}
        \right)
    =
    \{\Psi(w) + \oP{1}\}O(m^{-1})
    =
    \OP{m^{-1}}
    .
\end{align*}
Therefore, if
\textup{(i)} $r_w = r_G = r_H = r_{\phi} = o(1)$,
\textup{(ii)} $r_w = o(m^{-1/4})$,
and \textup{(iii)} $r_G(r_H + r_{\phi}) = o(m^{-1/2})$
as $m \to \infty$,
then
\begin{align*}
    \sqrt{m}\{\widehat{\Psi}(w) - \Psi(w)\}
    \overset{d}{\to}
    N(0,\sigma^2(w))
\end{align*}
where 
$
\sigma^2(w)
=
\textup{Var}\Big\{
    \varphi(\mathbf{O}; \boldsymbol{\eta}) 
\Big\}
=
\mathbb{E}
    \Big[ 
        \big\{ 
            \varphi^{*}(\mathbf{O}; \boldsymbol{\eta}) 
        \big\}^2
    \Big]
$
is the nonparametric efficiency bound of $\Psi(w)$.

\subsection{Proof of Theorem 4}
\label{proof:thm4}

First, note that
$$
    \widehat{\sigma}^2(w) 
    = 
    \frac{1}{K} \sum_{k=1}^{K} 
    \mathbb{P}_m^k \allowbreak
    \Big[ 
        \big\{ \allowbreak
            \varphi(\mathbf{O}; \widehat{\boldsymbol{\eta}}^{(-k)}) \allowbreak
        \big\}^2
    \Big]
    - 
    \widehat{\Psi}(w)^2
$$
and
$$
    {\sigma}^2(w) 
    = 
    \mathbb{P} \allowbreak
    \Big[ 
        \big\{ \allowbreak
            \varphi(\mathbf{O}; {\boldsymbol{\eta}}) \allowbreak
        \big\}^2
    \Big]
    - 
    \Psi(w)^2
    .
$$
Since 
$\widehat{\Psi}(w) \overset{p}{\to} \Psi(w)$ from Theorem 2,
it suffices to show
$$
\mathbb{P}_m^k \allowbreak
    \Big[ 
        \big\{ \allowbreak
            \varphi(\mathbf{O}; \widehat{\boldsymbol{\eta}}^{(-k)}) \allowbreak
        \big\}^2
    \Big]
\overset{p}{\to}
\mathbb{P} \allowbreak
    \Big[ 
        \big\{ \allowbreak
            \varphi(\mathbf{O}; {\boldsymbol{\eta}}) \allowbreak
        \big\}^2
    \Big]
.
$$
From the conditional law of large numbers 
(Theorem 4.2. in \cite{majerek2005conditional}),
$$
\mathbb{P}_m^k \allowbreak
    \Big[ 
        \big\{ \allowbreak
            \varphi(\mathbf{O}; \widehat{\boldsymbol{\eta}}^{(-k)}) \allowbreak
        \big\}^2
    \Big]
-
\mathbb{P} \allowbreak
    \Big[ 
        \big\{ \allowbreak
            \varphi(\mathbf{O}; \widehat{\boldsymbol{\eta}}^{(-k)}) \allowbreak
        \big\}^2
    \Big]
=\oP{1}
.
$$
Thus, it suffices to show
$$
\mathbb{P} \allowbreak
    \Big[ 
        \big\{ \allowbreak
            \varphi(\mathbf{O}; \widehat{\boldsymbol{\eta}}^{(-k)}) \allowbreak
        \big\}^2
        -
        \big\{ \allowbreak
            \varphi(\mathbf{O}; \boldsymbol{\eta}) \allowbreak
        \big\}^2
    \Big]
=\oP{1}
.
$$
Using some algebra and from Cauchy-Schwarz inequality,
\begin{align*}
    &
    \mathbb{P} \Big[ 
        \big\{
            \varphi(\mathbf{O}; \widehat{\boldsymbol{\eta}}^{(-k)}) 
        \big\}^2
        -
        \big\{
            \varphi(\mathbf{O}; \boldsymbol{\eta})
        \big\}^2
    \Big]
    \\
    = &
    \mathbb{P} \Big[ 
        \big\{
            \varphidiffk
        \big\}^2
    \Big]
    +
    2
    \mathbb{P} \Big[ 
        \big\{
            \varphidiffk
        \big\}
        \varphi(\mathbf{O}; \boldsymbol{\eta})
    \Big]
    \\
    \le &
    \mathbb{P} \Big[ 
        \big\{
            \varphidiffk
        \big\}^2
    \Big]
    +
    2
    \left(
        \mathbb{P} \Big[ 
            \big\{
                \varphidiffk
            \big\}^2
        \Big]
    \right)^{\frac{1}{2}}
    \left(
        \mathbb{P} \Big[ 
            \big\{
                \varphi(\mathbf{O}; \boldsymbol{\eta})
            \big\}^2
        \Big]
    \right)^{\frac{1}{2}}
    .
\end{align*}
Here,
\begin{align*}
\left(
    \mathbb{P} \Big[ 
        \big\{
            \varphidiffk
        \big\}^2
    \Big]
\right)^{\frac{1}{2}}
    =
    \normp{ \varphidiffk }
    =
    \OP{r_w^2 + r_G + r_H + r_\phi}
\end{align*}
and
\begin{align*}
    \left(
        \mathbb{P} \Big[ 
            \big\{
                \varphi(\mathbf{O}; \boldsymbol{\eta})
            \big\}^2
        \Big]
    \right)^{\frac{1}{2}}
    =
    \sigma(w)
    =\OP{1}
\end{align*}
which finishes the proof.

\subsection{Proof of Theorem 5}

To prove the weak convergence to the Gaussian process of the proposed estimator $\widehat{\mu}_{\scriptscriptstyle \textup{B}}(\alpha)$,
we follow an approach similar to \citet{kennedy19}.
For notational convenience, omit subscripts in $\mu_{\scriptscriptstyle \textup{B}}(\alpha)$, 
i.e., let $\mu(\alpha)$ denote $\mu_{\scriptscriptstyle \textup{B}}(\alpha)$.

First, define the processes
\begin{align*}
    \widetilde{\Phi}_{m}(\alpha) 
    =&
    \sqrt{m}\{\widehat{\mu}(\alpha) - \mu(\alpha)\}
    \\
    \Phi_{m}(\alpha) 
    =& 
    \sqrt{m}(\mathbb{P}_m - \mathbb{P}) \{\varphi_{\mu(\alpha)}(\mathbf{O}; \boldsymbol{\eta})\}
    =
    \mathbb{G}_m\{\varphi_{\mu(\alpha)}(\mathbf{O}; \boldsymbol{\eta})\}
\end{align*}
where
$\mathbb{G}_m = \sqrt{m}(\mathbb{P}_m - \mathbb{P})$ 
is the empirical process on the full sample.
Also, let 
$||f||_{\mathbb{A}} = \sup_{\alpha \in \mathbb{A}} |f(\alpha)|$
denote the supremum norm over $\mathbb{A} = [\alpha_l, \alpha_u]$,
and
$\mathbb{G}(\cdot)$ denote the mean zero Gaussian process with covariance
$
\mathbb{E} \{ \mathbb{G}(\alpha_1) \mathbb{G}(\alpha_2) \} \allowbreak
= \allowbreak
\mathbb{E} \{ 
    \varphi^*_{\mu(\alpha_1)}(\mathbf{O}; \boldsymbol{\eta}) \allowbreak
    \varphi^*_{\mu(\alpha_2)}(\mathbf{O}; \boldsymbol{\eta})
\}
$,
where
$
\varphi^*_{\mu(\alpha)}(\mathbf{O}; \boldsymbol{\eta}) 
=
\varphi_{\mu(\alpha)}(\mathbf{O}; \boldsymbol{\eta})
-
\mu(\alpha)
$ is the EIF of $\mu(\alpha)$.
We will prove that 
(i) the process $\Phi_m(\cdot)$ weakly converges to the Gaussian process $\mathbb{G}(\cdot)$, i.e., 
$
\Phi_m(\cdot) 
\rightsquigarrow
\mathbb{G}(\cdot)
\text{ in }
\ell^{\infty} (\mathbb{A})
$
and
(ii)
$
||\widetilde{\Phi}_m - \Phi_m||_{\mathbb{A}} = o_{\mathbb{P}}(1)
$,
which gives the desired result,
$\widetilde{\Phi}_m 
\rightsquigarrow
\mathbb{G}(\cdot)
\text{ in }
\ell^{\infty} (\mathbb{A})$.

First, (i) holds from the fact that the function class 
$\mathcal{F}_{{\boldsymbol{\eta}}} = 
\{ \varphi_{\mu(\alpha)}^{}(\cdot; {\boldsymbol{\eta}}) : \alpha \in \mathbb{A} \}$
is Lipschitz for any fixed ${\boldsymbol{\eta}}$,
and thus it is Donsker.
To show 
$\mathcal{F}_{{\boldsymbol{\eta}}}$
is Lipschitz, we show that 
$\varphi_{\mu(\alpha)}(\mathbf{O}; \boldsymbol{\eta})$ 
is a sum of products of Lipschitz functions with respect to $\alpha$.
From Section \ref{example:TypeB},
\begin{align*}
    \varphi_{\mu(\alpha)}(\mathbf{O}; \boldsymbol{\eta})
    =
    \frac{1}{N}
    \sum_{j=1}^{N}
    \begin{aligned}[t]
        \Bigg[  
        &
        \sum_{\mathbf{a} \in \mathcal{A}(N)}
            G_j \big(
                \mathbf{a}, \mathbf{X}, N 
            \big)
            \prod_{l=1}^{N} \alpha^{a_{l}} (1-\alpha)^{1-a_{l}}
        \\
        & + 
        \frac
            {\prod_{l=1}^{N} \alpha^{A_{l}} (1-\alpha)^{1-A_{l}}}
            {H(\mathbf{A},\mathbf{X}, N)}
        \left\{ 
            Y_{j} 
            - 
            G_j \big( 
                \mathbf{A}, \mathbf{X}, N 
            \big) 
        \right\}
        \Bigg]
    \end{aligned}
\end{align*}
where $G_j$ is $j$-th component of $G$.
The following derivative with respect to $\alpha$ is bounded:
\begin{align*}
    \left|
        \frac{\partial}{\partial \alpha}
        \left\{
            \prod_{l=1}^{N} \alpha^{a_{l}} (1-\alpha)^{1-a_{l}}
        \right\}
    \right|
    =
    \alpha^{T-1}
    (1-\alpha)^{N-T-1}
    \left|
        T - N\alpha
    \right|
    \le 
    N
\end{align*}
where $T = \sum_{l}a_l \in \{0,1,\dots,N\}$,
which implies that 
$\varphi_{\mu(\alpha)}(\mathbf{O}; \boldsymbol{\eta})$
is a Lipschitz function.
Therefore, $\mathcal{F}_{{\boldsymbol{\eta}}}$ is Donsker and thus (i) holds.

Next, to show (ii), we assume that sample splitting is done approximately uniformly, such that $\frac{m_k}{m} = \frac{1}{K} + O(\frac{1}{m})$. Then,
\begin{align*}
    \widetilde{\Phi}_m(\alpha) - \Phi_m(\alpha)
    =
    B_{m,1}(\alpha) + B_{m,2}(\alpha) + B_{m,3}(\alpha)
\end{align*}
where
\begin{align*}
    B_{m,1}(\alpha) 
    =& 
    \frac{\sqrt{m}}{K}
        \sum_{k=1}^K
            (\mathbb{P}_m^k - \mathbb{P})
            \left\{
                \varphi_{\mu(\alpha)}^{}(\mathbf{O}; \widehat{\boldsymbol{\eta}}^{(-k)})
                -
                \varphi_{\mu(\alpha)}^{}(\mathbf{O}; \boldsymbol{\eta})
            \right\}
    ,
    \\
    B_{m,2}(\alpha) 
    =&
    \frac{\sqrt{m}}{K}
    \sum_{k=1}^K
        \mathbb{P}
        \left\{
            \varphi_{\mu(\alpha)}^{}(\mathbf{O}; \widehat{\boldsymbol{\eta}}^{(-k)})
            -
            \varphi_{\mu(\alpha)}^{}(\mathbf{O}; \boldsymbol{\eta})
        \right\}
    ,
    \\
    B_{m,3}(\alpha)
    =&
    \frac{\sqrt{m}}{K}
    \sum_{k=1}^K
        \mathbb{P}_m^k
        \left\{
            \varphi_{\mu(\alpha)}^{}(\mathbf{O}; \boldsymbol{\eta})
        \right\}
        \left(
        1
        -
        \frac{K m_k}{m}
        \right)
    .
\end{align*}
First, from the proof of Theorem 2, we have
\begin{align*}
    \sup_{\alpha \in \mathbb{A}}
    ||\varphi_{\mu(\alpha)}(\mathbf{O}; \widehat{\boldsymbol{\eta}}^{(-k)}) 
    -
    \varphi_{\mu(\alpha)}(\mathbf{O}; \boldsymbol{\eta})||
    = 
    O_{\mathbb{P}}(r_w^2 + r_G + r_H + r_\phi)
    =
    o_{\mathbb{P}}(1)
    .
\end{align*}
Thus,
\begin{align*}
    \sup_{\alpha \in \mathbb{A}}
    | (\mathbb{P}_m^k - \mathbb{P})
    \big\{ 
        \varphi_{\mu(\alpha)}(\mathbf{O}; \widehat{\boldsymbol{\eta}}^{(-k)}) 
        -
        \varphi_{\mu(\alpha)}(\mathbf{O}; \boldsymbol{\eta}) 
    \big\} |
    &=
    O_\mathbb{P} \left( 
        \frac
        {\sup_{\alpha \in \mathbb{A}}
        ||
            \varphi_{\mu(\alpha)}(\mathbf{O}; \widehat{\boldsymbol{\eta}}^{(-k)}) 
            -
            \varphi_{\mu(\alpha)}(\mathbf{O}; \boldsymbol{\eta}) 
        ||}
        {m_k^{1/2} }
    \right)
    \\
    &= 
    o_\mathbb{P} (m^{-1/2})
    .
\end{align*}
Also, from the proof of Theorem 2,
\begin{align*}
    \sup_{\alpha \in \mathbb{A}}
    \mathbb{P} \left\{
        \varphi_{\mu(\alpha)}^{}(\mathbf{O}; \widehat{\boldsymbol{\eta}}^{(-k)})
        -
        \varphi_{\mu(\alpha)}^{}(\mathbf{O}; \boldsymbol{\eta})
    \right\}
    =
    O_{\mathbb{P}}(r_w^2 + r_G r_H + r_G r_\phi)
    =
    o_\mathbb{P}(m^{-1/2})
    .
\end{align*}
Finally, 
\begin{align*}
    \sup_{\alpha \in \mathbb{A}}
    \mathbb{P}_m^k
        \left\{
            \varphi_{\mu(\alpha)}^{}(\mathbf{O}; \boldsymbol{\eta})
        \right\}
    \times
    \sqrt{m}
        \left(
        1
        -
        \frac{K m_k}{m}
        \right)
    =
    \sup_{\alpha \in \mathbb{A}}
    \left\{
        \mu(\alpha) 
        +
        o_{\mathbb{P}}(1)
    \right\}
    o(1)
    =
    o_{\mathbb{P}}(1)
    .
\end{align*}
Therefore, 
$
||B_{m,1}||_{\mathbb{A}} 
= 
||B_{m,2}||_{\mathbb{A}} 
= 
||B_{m,3}||_{\mathbb{A}} 
= 
o_{\mathbb{P}}(1)$,
which yields the desired result.


\subsection{Proof of Theorem 6}

Proof of Theorem 6 is similar to that of Theorem 5. 
For notational convenience, omit subscripts in $\mu_{\scriptscriptstyle \textup{CIPS}}(\delta_0)$, 
i.e., let $\mu(\delta)$ denote $\mu_{\scriptscriptstyle \textup{CIPS}}(\delta_0)$.
Similar to the proof of Theorem 5,
define the processes
\begin{align*}
    \widetilde{\Phi}_{m}(\delta) 
    =&
    \sqrt{m}\{\widehat{\mu}(\delta) - \mu(\delta)\}
    \\
    \Phi_{m}(\delta) 
    =& 
    \sqrt{m}(\mathbb{P}_m - \mathbb{P}) \{\varphi_{\mu(\delta)}(\mathbf{O}; \boldsymbol{\eta})\}
    =
    \mathbb{G}_m\{\varphi_{\mu(\delta)}(\mathbf{O}; \boldsymbol{\eta})\}
\end{align*}
and prove that
(i) the process $\Phi_m(\cdot)$ weakly converges to the Gaussian process $\mathbb{G}(\cdot)$, i.e., 
$
\Phi_m(\cdot) 
\rightsquigarrow
\mathbb{G}(\cdot)
\text{ in }
\ell^{\infty} (\mathbb{D})
$
and
(ii)
$
||\widetilde{\Phi}_m - \Phi_m||_{\mathbb{D}} = o_{\mathbb{P}}(1)
$.

First, (i) holds from the fact that the function class 
$\mathcal{F}_{{\boldsymbol{\eta}}} = 
\{ \varphi_{\mu(\delta)}^{}(\cdot; {\boldsymbol{\eta}}) : \delta \in \mathbb{D} \}$
is Lipschitz for any fixed ${\boldsymbol{\eta}}$,
and thus it is Donsker.
To show 
$\mathcal{F}_{{\boldsymbol{\eta}}}$
is Lipschitz, we show that 
$\varphi_{\mu(\delta)}(\mathbf{O}; \boldsymbol{\eta})$
is a sum of products of Lipschitz functions.
From Section \ref{example:CIPS},
\begin{align*}
    \varphi_{\mu(\delta)}(\mathbf{O}; \boldsymbol{\eta})
    =
    \frac{1}{N}
    \sum_{j=1}^{N}
    \begin{aligned}[t]
        \Bigg[  
        &
        \sum_{\mathbf{a} \in \mathcal{A}(N)}
            G_j \big(
                \mathbf{a}, \mathbf{X}, N 
            \big)
            \times
            Q_{\scriptscriptstyle \textup{CIPS}}(\mathbf{a} | \mathbf{X}, N; \delta)
            \\
            & \qquad \quad
            \times
            \left\{
                1 
                +
                \sum_{l=1}^{N}
                \frac
                    {(2a_{l}-1) \delta (A_{l} - \pi_{l})}
                    {(\pi_{l,\delta})^{a_{l}} (1 - \pi_{l,\delta})^{1-a_{l}}
                    (\delta\pi_{l} + 1 - \pi_{l})^2}
            \right\}
        \\
        & + 
        \frac
            {Q_{\scriptscriptstyle \textup{CIPS}}(\mathbf{A} | \mathbf{X}, N; \delta)}
            {H(\mathbf{A},\mathbf{X}, N)}
        \left\{ 
            Y_{j} 
            - 
            G_j \big( 
                \mathbf{A}, \mathbf{X}, N 
            \big) 
        \right\}
        \Bigg]
    \end{aligned}
\end{align*}
where $G_j$ is $j$ th component of $G$
and
$Q_{\scriptscriptstyle \textup{CIPS}}(\mathbf{a} | \mathbf{X}, N; \delta) 
=
\prod_{j = 1}^{N} 
    (\pi_{j,\delta})^{a_{j}}
    (1 - \pi_{j,\delta})^{1-a_{j}} 
,
$
where
$\pi_{j, \delta} \allowbreak
= \delta
\pi_{j} / \allowbreak
\{ \delta \pi_{j} + 1 - \allowbreak  \pi_{j} \}$
denotes the shifted propensity score.
From the assumption that $\pi_{l} \in (c, 1-c)$
and the fact that $\delta \pi_{l} + 1 - \pi_{l} \in [\delta_l, \delta_u]$,
the following derivatives with respect to $\delta$ are all bounded:
\begin{gather*}
    \left|
        \frac{\partial}{\partial \delta}
        \left\{
            (\pi_{j,\delta})^{a_{j}}
            (1 - \pi_{j,\delta})^{1-a_{j}} 
        \right\}
    \right|
    =
    \frac
        {\pi_{j}(1-\pi_{j})}
        {( \delta \pi_{j} + 1 - \pi_{j} )^2}
    \le 
    \frac{1}{\delta_l^2}
    ,
    \\
    \left|
        \frac{\partial}{\partial \delta}
        \left[
            \frac
                {(2a_{l}-1) 
                \delta 
                \left( A_{l}-\pi_{l} \right)}
                {(\pi_{l,\delta})^{a_{l}}
                (1 - \pi_{l,\delta})^{1-a_{l}} 
                ( \delta \pi_{l} + 1 - \pi_{l} )^2}
        \right]
    \right|
    =
    \frac
        {|A_{l}-\pi_{l}|}
        {( \delta \pi_{l} + 1 - \pi_{l} )^2}
    \le 
    \frac{1}{\delta_l^2}
    ,
\end{gather*}
which implies that 
$\varphi_{\mu(\delta)}(\mathbf{O}; \boldsymbol{\eta})$
is a Lipschitz function.
Therefore, $\mathcal{F}_{{\boldsymbol{\eta}}}$ is Donsker and thus (i) holds.
The proof of (ii) is the same as the proof of Theorem 5, which is omitted here.


\subsection{Proof of Theorem 7}

Proof of Theorem 7 is similar to that of Theorem 5. 
For notational convenience, omit subscripts in $\mu_{\scriptscriptstyle \textup{CMS}}(\lambda)$, 
i.e., let $\mu(\lambda)$ denote $\mu_{\scriptscriptstyle \textup{CMS}}(\lambda)$.
Similar to the proof of Theorem 5,
define the processes
\begin{align*}
    \widetilde{\Phi}_{m}(\lambda) 
    =&
    \sqrt{m}\{\widehat{\mu}(\lambda) - \mu(\lambda)\}
    \\
    \Phi_{m}(\lambda) 
    =& 
    \sqrt{m}(\mathbb{P}_m - \mathbb{P}) \{\varphi_{\mu(\lambda)}(\mathbf{O}; \boldsymbol{\eta})\}
    =
    \mathbb{G}_m\{\varphi_{\mu(\lambda)}(\mathbf{O}; \boldsymbol{\eta})\}
\end{align*}
and prove that
(i) the process $\Phi_m(\cdot)$ weakly converges to the Gaussian process $\mathbb{G}(\cdot)$, i.e., 
$
\Phi_m(\cdot) 
\rightsquigarrow
\mathbb{G}(\cdot)
\text{ in }
\ell^{\infty} (\mathbb{L})
$
and
(ii)
$
||\widetilde{\Phi}_m - \Phi_m||_{\mathbb{L}} = o_{\mathbb{P}}(1)
$.

First, (i) holds from the fact that the function class 
$\mathcal{F}_{{\boldsymbol{\eta}}} = 
\{ \varphi_{\mu(\lambda)}^{}(\cdot; {\boldsymbol{\eta}}) : \lambda \in \mathbb{L} \}$
is Lipschitz for any fixed ${\boldsymbol{\eta}}$,
and thus it is Donsker.
To show 
$\mathcal{F}_{{\boldsymbol{\eta}}}$
is Lipschitz, we show that 
$\varphi_{\mu(\lambda)}(\mathbf{O}; \boldsymbol{\eta})$
is a sum of products of Lipschitz functions.
From Section \ref{example:CMS},
\begin{align*}
    \varphi_{\mu(\lambda)}(\mathbf{O}; \boldsymbol{\eta})
    = &
    \frac{1}{N}
    \sum_{j=1}^{N}
    \begin{aligned}[t]
        \Bigg[  
        &
        \sum_{\mathbf{a} \in \mathcal{A}(N)}
            G_j \big(
                \mathbf{a}, \mathbf{X}, N 
            \big)
            \times
            Q_{\scriptscriptstyle \textup{CMS}}(\mathbf{a} | \mathbf{X}, N; \lambda)
            \\
            & \qquad \quad
            \times
            \left\{
                1 
                +
                \sum_{l=1}^{N}
                \frac
                    {(2a_{l}-1) (A_{l} - \pi_{l}) (X_l^* \lambda + 1 - X_l^*)}
                    {(\pi_{l,\lambda})^{a_{l}} (1 - \pi_{l,\lambda})^{1-a_{l}}}
            \right\}
        \\
        & + 
        \frac
            {Q_{\scriptscriptstyle \textup{CMS}}(\mathbf{A} | \mathbf{X}, N; \lambda)}
            {H(\mathbf{A},\mathbf{X}, N)}
        \left\{ 
            Y_{j} 
            - 
            G_j \big( 
                \mathbf{A}, \mathbf{X}, N 
            \big) 
        \right\}
        \Bigg]
    \end{aligned}
\end{align*}
where $G_j$ is $j$ th component of $G$
and
$Q_{\scriptscriptstyle \textup{CMS}}(\mathbf{a} | \mathbf{X}, N; \lambda) 
=
\prod_{j = 1}^{N} 
    (\pi_{j,\lambda})^{a_{j}}
    (1 - \pi_{j,\lambda})^{1-a_{j}} 
,
$
where
$\pi_{j, \lambda} \allowbreak
= \mathbb{P}_{\lambda}(A_{j} = 1 | \mathbf{X}, N) \allowbreak
= (1-\lambda) X_j^* + \pi_{j} (X_j^* \lambda + 1 - X_j^*)$
is the shifted propensity score.

From the assumption that $\pi_{l} \in (c, 1-c)$,
the following derivatives with respect to $\lambda$ are all bounded:
\begin{gather*}
    \left|
        \frac{\partial}{\partial \lambda}
        \left\{
            (\pi_{j,\lambda})^{a_{j}}
            (1 - \pi_{j,\lambda})^{1-a_{j}} 
        \right\}
    \right|
    =
    (1-\pi_j)X_j^*
    \le 
    1
    ,
    \\
    \left|
        \frac{\partial}{\partial \lambda}
        \left\{
            \frac
                {
                (2a_{l}-1) 
                \left(
                    A_{l}-\pi_{l}
                \right)
                (X_l^*\lambda + 1 - X_l^*)
                }
                {
                (\pi_{l,\lambda})^{a_{l}} (1 - \pi_{l,\lambda})^{1-a_{l}}
                }
        \right\}
    \right|
    =
    \frac
        {a_l X_l^*\left| A_{l}-\pi_{l} \right|}
        {(1-\lambda+\pi_l \lambda)^2}
    \le 
    \frac{1}{(1-\lambda_u)^2}
    ,
\end{gather*}
which implies that 
$\varphi_{\mu(\lambda)}(\mathbf{O}; \boldsymbol{\eta})$
is a Lipschitz function.
Therefore, $\mathcal{F}_{{\boldsymbol{\eta}}}$ is Donsker and thus (i) holds.
The proof of (ii) is the same as the proof of Theorem 5, which is omitted here.


\subsection{Note on Assumption (B8)}

In this section, we show how the quantity in (B8) equals the second order remainder term in the von Mises expansion. 
First, note that for fixed
$(\mathbf{a}, \mathbf{x}, n) 
\in 
\mathcal{A}(n) \times \mathcal{X}(n) \times \mathbb{N}$,
$w(\mathbf{a}, \mathbf{x}, n) 
= 
w(\mathbf{a}, \mathbf{x}, n; \mathbb{P})$ 
is a functional of a distribution $\mathbb{P}$ on 
$\mathbf{O} = (\mathbf{Y},\mathbf{A},\mathbf{X},N)$.
For another fixed 
$(\mathbf{a'}, \mathbf{x'}, n') 
\in 
\mathcal{A}(n) \times \mathcal{X}(n) \times \mathbb{N}$,
$\phi(\mathbf{a'}, \mathbf{x'}, n'; \mathbf{a}) 
= 
\phi(\mathbf{a'}, \mathbf{x'}, n'; \mathbf{a}; \mathbb{P})$
is a functional of a distribution $\mathbb{P}$,
where the EIF of 
$w(\mathbf{a}, \mathbf{x}, n; \mathbb{P})$
is
$
\{\mathbbm{1}(\mathbf{X} = \mathbf{x}, N = n)
/
d\mathbb{P}(\mathbf{x}, n)\}
\phi(\mathbf{A}, \mathbf{X}, N; \mathbf{a}; \mathbb{P})$
and
$\phi(\mathbf{a'}, \mathbf{x'}, n'; \mathbf{a}; \mathbb{P})$
is $\phi(\mathbf{A}, \mathbf{X}, N; \mathbf{a}; \mathbb{P})$
evaluated at 
$(\mathbf{A}, \mathbf{X}, N) = (\mathbf{a'}, \mathbf{x'}, n')$.
Similarly, $G$ and $H$ are also functionals of $\mathbb{P}$.

Now, let $\widehat{\mathbb{P}}$ be the distribution that generates the estimators of nuisance functions $\widehat{\boldsymbol{\eta}}$. 
That is, $\widehat{\mathbb{P}}$ is a probability distribution on 
$\mathbf{O} = (\mathbf{Y},\mathbf{A},\mathbf{X},N)$ which satisfies
$\widehat{G}(\mathbf{a}, \mathbf{x}, n) 
= 
G(\mathbf{a}, \mathbf{x}, n; \widehat{\mathbb{P}})
=
\mathbb{E}_{\widehat{\mathbb{P}}}(\mathbf{Y} 
| 
\mathbf{A} = \mathbf{a}, \mathbf{X} = \mathbf{x}, N = n)$,
$\widehat{H}(\mathbf{a}, \mathbf{x}, n) 
= 
H(\mathbf{a}, \mathbf{x}, n; \widehat{\mathbb{P}})
=
\widehat{\mathbb{P}}(\mathbf{A} = \mathbf{a}
|
\mathbf{X} = \mathbf{x}, N = n)$,
$\widehat{w}(\mathbf{a}, \mathbf{x}, n) 
= 
w(\mathbf{a}, \mathbf{x}, n; \widehat{\mathbb{P}})$,
and
$\widehat{\phi}(\mathbf{a'}, \mathbf{x'}, n'; \mathbf{a})
=
\phi(\mathbf{a'}, \mathbf{x'}, n'; \mathbf{a}; \widehat{\mathbb{P}})$.
Under this setting, the von Mises expansion \citep{fisher2021visually, kennedy22, hines22} of $w$  is given as follows:
\begin{align*}
    w(\mathbf{a}, \mathbf{x}, n; \widehat{\mathbb{P}})
    -
    w(\mathbf{a}, \mathbf{x}, n; \mathbb{P})
    = &
    \int
        \frac
            {\mathbbm{1}(\mathbf{x'} = \mathbf{x}, n' = n)}
            {d\widehat{\mathbb{P}}(\mathbf{x}, n)}
        \phi(\mathbf{a'}, \mathbf{x'}, n'; \mathbf{a}; \widehat{\mathbb{P}})
    d(\widehat{\mathbb{P}} - \mathbb{P})(\mathbf{a'}, \mathbf{x'}, n')
    +
    {R}_2(\widehat{\mathbb{P}}, \mathbb{P})(\mathbf{a}, \mathbf{x}, n)
    \\
    = &
    -
    \int
        \phi(\mathbf{a'}, \mathbf{x}, n; \mathbf{a}; \widehat{\mathbb{P}})
    d\mathbb{P}(\mathbf{a'} | \mathbf{x}, n)
    +
    {R}_2(\widehat{\mathbb{P}}, \mathbb{P})(\mathbf{a}, \mathbf{x}, n)
    \\
    = &
    -
    \sum_{\mathbf{a}' \in \mathcal{A}(n)}
        \phi(\mathbf{a}', \mathbf{x}, n; \mathbf{a}; \widehat{\mathbb{P}})
        H(\mathbf{a'}, \mathbf{x}, n)
    +
    {R}_2(\widehat{\mathbb{P}}, \mathbb{P})(\mathbf{a}, \mathbf{x}, n)
\end{align*}
where 
${R}_2$ is a second order remainder term.
This implies
\begin{align*}
    \widehat{w}(\mathbf{a}, \mathbf{X}, N)
    -
    w(\mathbf{a}, \mathbf{X}, N)
    +
    \sum_{\mathbf{a}' \in \mathcal{A}(N)}
        \widehat{\phi}(\mathbf{a}', \mathbf{X}, N; \mathbf{a})
        H(\mathbf{a'}, \mathbf{X}, N)
    =
    {R}_2(\widehat{\mathbb{P}}, \mathbb{P})(\mathbf{a}, \mathbf{X}, N)
    .
\end{align*}



\subsection{Large sample property under subsampling approximation}

Here, the large sample properties of the proposed estimators are presented when subsampling approximation is applied.

We first define some notations. 
First, note that the uncentered EIF of an estimand $\Psi(w)$ can be expressed by
\begin{align*}
    \varphi(\mathbf{O}; \boldsymbol{\eta})  
    =
    \suma h(\mathbf{O}, \mathbf{a}) 
    +
    l(\mathbf{O}) 
    ,
\end{align*}
where
$h(\mathbf{O}, \mathbf{a}) 
= 
\big\{ 
    w(\mathbf{a}, \mathbf{X}, N) + \phi(\mathbf{A}, \mathbf{X}, N;\mathbf{a}) 
\big\}^\top 
G(\mathbf{a}, \mathbf{X}, N)$ 
and
\begin{align*}
    l(\mathbf{O}) 
    =
    \frac
        {w(\mathbf{A}, \mathbf{X}, N)^\top \left\{ \mathbf{Y} - G(\mathbf{A}, \mathbf{X}, N) \right\}}
        {H(\mathbf{A}, \mathbf{X}, N)}
    .
\end{align*}
Let
$\widehat{h}(\mathbf{O}, \mathbf{a})$ and $\widehat{l}(\mathbf{O})$ 
denote
$h(\mathbf{O}, \mathbf{a})$ and $l(\mathbf{O})$ 
when nuisance functions $\boldsymbol{\eta} = (G, H, w, \phi)$  are substituted by their estimators
$\widehat{\boldsymbol{\eta}} 
= 
(\widehat{G}, \widehat{H}, \widehat{w},\widehat{\phi})$.
Then, the proposed estimator without subsampling approximation is
\begin{align*}
    \widehat{\Psi}(w) 
    =
    \frac{1}{K} 
    \sum_{k=1}^{K} 
    \mathbb{P}_m^k 
    \big\{ 
        \varphi^{}(\mathbf{O}; \widehat{\boldsymbol{\eta}}^{(-k)}) 
    \big\}
    =
    \frac{1}{K} 
    \sum_{k=1}^{K} 
    \mathbb{P}_m^k 
    \left\{ 
        \suma \widehat{h}^{(-k)}(\mathbf{O}, \mathbf{a}) 
        +
        \widehat{l}^{(-k)}(\mathbf{O}) 
    \right\}
    .
\end{align*}
Subsampling approximation is used since the summation $\suma \widehat{h}^{(-k)}(\mathbf{O}, \mathbf{a})$ in the estimator can be computationally intensive.

Now consider a probability distribution $f$ on $\mathcal{A}(N)$
such that
$\suma f(\mathbf{a}) = 1$ 
and 
$f(\mathbf{a}) > 0$ for all $\mathbf{a} \in \mathcal{A}(N)$.
Let
$\mathbf{a}^{*} 
= 
\big(\mathbf{a}^{(1)}, \dots,  \mathbf{a}^{(r)}\big)$
denote a random sample from $f$ such that
$\mathbf{a}^{(q)} \overset{\mathrm{iid}}{\sim} f(\cdot), q=1,\dots,r$,
which is independent of an observed data $\mathbf{O}$.
Then, 
\begin{align*}
    \suma \widehat{h}(\mathbf{O}, \mathbf{a})
    =
    \suma 
        \frac{\widehat{h}(\mathbf{O}, \mathbf{a})}{f(\mathbf{a})}
        f(\mathbf{a})
    =
    \mathbb{E}_{\mathbf{a}^{(\cdot)}} \left\{
        \frac
            {\widehat{h}(\mathbf{O}, \mathbf{a}^{(\cdot)})}
            {f(\mathbf{a}^{(\cdot)})}
        \middle| D, \mathbf{O}
    \right\}
\end{align*}
which can be approximated by 
\begin{align*}
    \frac{1}{r}
    \sum_{q=1}^{r}
        \frac
            {\widehat{h}(\mathbf{O}, \mathbf{a}^{(q)})}
            {f(\mathbf{a}^{(q)})}
    ,
\end{align*}
where $D$ denotes the data independent of $\mathbf{O}$ that $\widehat{\boldsymbol{\eta}}$ was trained on as in the proof of Theorem 2 (Supplementary material Section \ref{proof:thm2}).
In the main text,
$f$ was chosen to be the uniform distribution on $\mathcal{A}(N)$ such that $f(\mathbf{a}) = 2^{-N}$ for all $\mathbf{a} \in \mathcal{A}(N)$.

Let 
$\mathbf{Z} 
= (\mathbf{O}, \mathbf{a}^{*}) 
= (\mathbf{Y}, \mathbf{A}, \mathbf{X}, N, \mathbf{a}^{*})$
and assume the random sample $\mathbf{a}^{*}$ is observed besides of $\mathbf{O}$.
Also, define the approximated EIF by
\begin{align*}
    \varphi(\mathbf{Z}; \boldsymbol{\eta}) 
    =
    \varphi(\mathbf{O}, \mathbf{a}^{*}; \boldsymbol{\eta})
    =
    \sumaq \frac{h(\mathbf{O}, \aq)}{f(\aq)}
    +
    l(\mathbf{O}) 
\end{align*}
and the estimated approximated EIF by
\begin{align*}
    \varphi(\mathbf{Z}; \widehat{\boldsymbol{\eta}}) 
    =
    \varphi(\mathbf{O}, \mathbf{a}^{*}; \widehat{\boldsymbol{\eta}})
    =
    \sumaq \frac{\widehat{h}(\mathbf{O}, \aq)}{f(\aq)}
    +
    \widehat{l}(\mathbf{O}) 
    .
\end{align*}
Note that
$\mathbb{E}\{
    \varphi(\mathbf{Z}; \boldsymbol{\eta}) 
    | \mathbf{O}
\}
=
\varphi(\mathbf{O}; \boldsymbol{\eta})$
and 
$\mathbb{E}\{
    \varphi(\mathbf{Z}; \widehat{\boldsymbol{\eta}}) 
    | D, \mathbf{O}
\}
=
\varphi(\mathbf{O}; \widehat{\boldsymbol{\eta}})$.
Then, the proposed estimator under subsampling approximation is
\begin{align*}
    \widehat{\Psi}^{\textup{app}}(w) 
    =
    \frac{1}{K} 
    \sum_{k=1}^{K} 
    \mathbb{P}_m^k 
    \big\{ 
        \varphi(\mathbf{Z}; \widehat{\boldsymbol{\eta}}^{(-k)}) 
    \big\}
    =
    \frac{1}{K} 
    \sum_{k=1}^{K} 
    \mathbb{P}_m^k 
    \left\{ 
        \sumaq \frac{\widehat{h}^{(-k)}(\mathbf{O}, \mathbf{\aq}) }{f(\aq)}
        +
        \widehat{l}^{(-k)}(\mathbf{O}) 
    \right\}
    ,
\end{align*}
and the estimand is 
$\Psi(w)
=
\mathbb{P}\{ \varphi(\mathbf{Z}; \boldsymbol{\eta}) \}$,
which follows from the iterated expectation.
Also, the variance estimator of $\widehat{\Psi}^{\textup{app}}(w)$ is given by
$
    \widehat{\sigma}_{\textup{app}}^2(w) 
    = 
    K^{-1}
    \sum_{k=1}^{K} 
    \mathbb{P}_m^k 
    \Big[ 
        \big\{ 
            \varphi(\mathbf{Z}; \widehat{\boldsymbol{\eta}}^{(-k)}) - \widehat{\Psi}^\textup{app}(w) 
        \big\}^2
    \Big]
$.

The following theorems give the large sample properties of the proposed estimator under subsampling approximation.

\begin{theorem}
\label{thm:app:consistency}
Assume \textup{(B1) -- (B8)} in the main text hold.
Under the same conditions for the consistency of $\widehat{\Psi}(w)$ as stated in Theorem 2 in the main text, 
$\widehat{\Psi}^{\textup{app}}(w) \overset{p}{\to} \Psi(w)$.
\end{theorem}

\begin{theorem}
\label{thm:app:weakconv}
Assume \textup{(B1) -- (B8)} in the main text hold. 
Under the same conditions for the asymptotic normality of $\widehat{\Psi}(w)$ as stated in Theorem 3 in the main text,
$
\sqrt{m}\{\widehat{\Psi}^{\textup{app}}(w) - \Psi(w)\}
\overset{d}{\to}
N(0,\sigma_\textup{app}^{2}(w))
$,
where 
\begin{align*}
    \sigma_\textup{app}^{2}(w)
    =
    \textup{Var}
        \big\{ 
            \varphi(\mathbf{Z}; \boldsymbol{\eta}) 
        \big\}
    =
    \sigma^2(w)
    +
    \frac{1}{r}
    \mathbb{E}\left[
        \textup{Var}\left\{
            \frac
                {h(\mathbf{O}, \mathbf{a}^{(\cdot)})}
                {f(\mathbf{a}^{(\cdot)})}
            \middle| \mathbf{O}
        \right\}
    \right]
    .
\end{align*}
\end{theorem}  

\begin{theorem}
\label{thm:app:varest}
Assume \textup{(B1) -- (B8)} in the main text hold. 
Under the same conditions for the consistency of $\widehat{\sigma}^2(w)$ as stated in Theorem 4 in the main text,
$\widehat{\sigma}_{\textup{app}}^2(w)$ is a consistent estimator of the asymptotic variance of $\widehat{\Psi}^{\textup{app}}(w)$.
Therefore,
$
\sqrt{m}\{
    \widehat{\Psi}^{\textup{app}}(w) 
    - 
    \Psi(w)
\}
/
\widehat{\sigma}_{\textup{app}}(w)
\overset{d}{\to}
N(0,1)
$.
\end{theorem}

In conclusion, under mild conditions, 
the proposed estimator under subsampling approximation is consistent and asymptotically normal, 
with asymptotic variance approaching the nonparametric efficiency bound as $r \to \infty$.
The decreasing finite sample variance of $\widehat{\Psi}^{\textup{app}}(w)$ is numerically illustrated in Section \ref{simul:r}.
The proof of the above Theorems are as follows.

\subsubsection{Proof of Theorem \ref{thm:app:consistency}}

We can derive the decomposition of 
$\widehat{\Psi}^{\textup{app}}(w) - \Psi(w)$ similar to the proof of Theorem 2 (Supplementary material Section \ref{proof:thm2}), given by
\begin{align*}
    &\widehat{\Psi}^{\textup{app}}(w) - \Psi(w)
    \\
    &=
    \frac{1}{K}
    \sum_{k=1}^{K}
    \left[
        (\mathbb{P}_m^k - \mathbb{P})
            \varphi(\mathbf{Z}; \boldsymbol{\eta}) 
        +
        (\mathbb{P}_m^k - \mathbb{P}) \big\{ 
            \varphidiffkZ
        \big\}
        +
        \mathbb{P} \big\{ \varphidiffkZ \big\}
    \right]
    .
\end{align*}
The first term $(\mathbb{P}_m^k - \mathbb{P}) \varphi(\mathbf{Z}; \boldsymbol{\eta})$ is $O_\mathbb{P}(m^{-1/2})$ from the central limit theorem, 
and the second term is
\begin{align*}
(\mathbb{P}_m^k - \mathbb{P})
\big\{ \varphidiffkZ \big\}
=
O_\mathbb{P} \left( 
    \frac
    {|| \varphidiffkZ ||}
    {m_k^{1/2} }
\right)
\end{align*}
from Lemma 1 in \citet{kennedy22}.
It remains to analyze 
$||\varphidiffkZ||$
and
$\mathbb{P} \big\{ \varphidiffkZ \big\}$.
As previous, omit superscript $(-k)$ in $\widehat{\boldsymbol{\eta}}^{(-k)}$ for notational convenience.
First, we have
\begin{align*}
    \mathbb{P} \big\{ \varphidiffZ \big\}
    = &
    \ED{\{}{\}}{ \varphidiffZ }
    \\
    = &
    \ED{[}{]}{ 
        \E{\{}{\}}{
            \varphidiffZ |D, \mathbf{O}
        }
    }
    \\
    = &
    \ED{\{}{\}}{ \varphidiff }
    \\
    = & 
    \mathbb{P} \big\{ \varphidiff \big\}
    \\
    = &
    \OP{r_w^2 + r_gr_H + r_Gr_\phi}
\end{align*}
from Theorem 2.
Next, we have the following decomposition
\begin{align}
    ||\varphidiffZ||^2
    &=
    \ED{[}{]}{ \{\varphidiffZ\}^2 }
    \nonumber
    \\
    &=
    \ED{[}{]}{ 
        \V{\{}{\}}{ 
            \varphidiffZ
            | D, \mathbf{A}, \mathbf{X}, N, \mathbf{a}^* 
        } 
    }
    \label{thm:app:1}
    \\
    & \ \ \ +
    \VD{[}{]}{
        \E{\{}{\}}{
            \varphidiffZ
            | D, \mathbf{A}, \mathbf{X}, N, \mathbf{a}^* 
        }
    }
    \label{thm:app:2}
    \\
    & \ \ \ +
    \left[
        \ED{\{}{\}}{\varphidiffZ}
    \right]^2
    \label{thm:app:3}
    .
\end{align}
The analysis of each term in the above decomposition is similar to that in the proof of Theorem 2.
Using the fact that
\begin{align*}
    \varphidiff
    =
    \left\{
        \frac
            {\wh{A}}
            {\Hh{A}}
        -
        \frac
            {\wt{A}}
            {\Ht{A}}
    \right\}^\top
    \mathbf{Y}
    +
    T(D, \mathbf{A}, \mathbf{X}, N, \as)
\end{align*}
where 
\begin{align*}
    T(D, \mathbf{A}, \mathbf{X}, N, \as)
    = &
    \sumaq
    \left[
        \frac
            {\big\{ \wh{\aq} + \widehat{\phi}(\aq) \big\}^\top \Gh{\aq}
            -
            \big\{ \wt{\aq} + \phi(\aq) \big\}^\top \Gt{\aq}}
            {f(\aq)}
    \right]
    \\
    & -
    \frac
        {\wh{A}^\top \Gh{A}}
        {\Hh{A}}
    +
    \frac
        {\wt{A}^\top \Gt{A}}
        {\Ht{A}}
\end{align*}
is a function of $D, \mathbf{A}, \mathbf{X}, N, \as$, but not $\mathbf{Y}$,
we have $(\ref{thm:app:1}) = \OP{r_w^4 + r_\phi^2 + r_H^2}$.
On the other hand, from
\begin{align*}
    \E{\{}{\}}{
        \varphidiffZ
        | D, \mathbf{A}, \mathbf{X}, N, \mathbf{a}^* 
    }
    = &
    \sumaq
        \frac
            {\big\{ \widehat{\phi}(\aq) - \phi(\aq) \big\}^\top \Gh{\aq}}
            {f(\aq)}
    \\
    & +
    \sumaq
        \frac
            {\big\{ \wh{\aq} - \wt{\aq} \big\}^\top \Gh{\aq}}
            {f(\aq)}
    \\
    & +
    \sumaq
        \frac
            {\big\{ \wt{\aq} + \phi(\aq) \big\}^\top
            \big\{ \Gh{\aq} - \Gt{\aq} \big\}}
            {f(\aq)}
    \\
    & +
    \frac
        {\wh{A}^\top \{\Gt{A} - \Gh{A}\}}
        {\Hh{A}}
    ,
\end{align*}
we have
\begin{align*}
    (\ref{thm:app:2}) 
    \lesssim &
    \ED{[}{]}{
        \suma \normt{\phih - \phit}^2
    }
     +
    \ED{[}{]}{
        \suma \normt{\wh{a} - \wt{a}}^2
    }
     +
    \ED{[}{]}{
        \suma \normt{\Gh{a} - \Gt{a}}^2
    }
    \\
    = &
    \OP{r_\phi^2 + r_w^4 + r_G^2}
    .
\end{align*}
Thus,
\begin{align*}
    ||\varphidiffZ||
    =
    \OP{r_w^2 + r_G + r_H + r_\phi}
    \label{thm:2:8}
\end{align*}
which is the same rate as $||\varphidiff||$.
In conclusion,
\begin{align*}
    \widehat{\Psi}^{\textup{app}}(w) - \Psi(w)
    =
    O_\mathbb{P} (m^{-1/2} + r_w^2 + r_G r_H + r_G r_\phi)
,
\end{align*}
which is the same rate as $\widehat{\Psi}(w) - \Psi(w)$,
proving the consistency of $\widehat{\Psi}^{\textup{app}}(w)$ under the same condition required for the consistency of $\widehat{\Psi}(w)$.

\subsubsection{Proof of Theorem \ref{thm:app:weakconv}}

Assume sample splitting is done approximately uniformly, such that $\frac{m_k}{m} = \frac{1}{K} + O(\frac{1}{m})$.
From the proof of Theorem \ref{thm:app:consistency}, we have
\begin{align*}
    \widehat{\Psi}^{\textup{app}}(w) - \Psi(w)
    = &
    \frac{1}{K}
    \sum_{k=1}^{K}
    \left[
        (\mathbb{P}_m^k - \mathbb{P})
            \varphi(\mathbf{Z}; \boldsymbol{\eta}) 
    \right]
    +
    m^{-1/2} O_\mathbb{P} (r_w^2 + r_G + r_H + r_\phi)
    +
    O_\mathbb{P} (r_w^2 + r_G r_H + r_G r_\phi)
    \\
    = &
    (\mathbb{P}_m - \mathbb{P})
    \varphi(\mathbf{Z}; \boldsymbol{\eta})
    +
    \OP{m^{-1}}
    +
    m^{-1/2} O_\mathbb{P} (r_w^2 + r_G + r_H + r_\phi)
    +
    O_\mathbb{P} (r_w^2 + r_G r_H + r_G r_\phi)
\end{align*}
from
\begin{align*}
    \frac{1}{K}
    \sum_{k=1}^{K}
    \left[
        (\mathbb{P}_m^k - \mathbb{P})
            \varphi(\mathbf{Z}; \boldsymbol{\eta}) 
    \right]
    =
    (\mathbb{P}_m - \mathbb{P})
    \varphi(\mathbf{Z}; \boldsymbol{\eta})
    +
    \frac{1}{K}
    \sum_{k=1}^K
        \mathbb{P}_m^k
        \left\{
            \varphi(\mathbf{Z}; \boldsymbol{\eta})
        \right\}
        \left(
        1
        -
        \frac{K m_k}{m}
        \right)
\end{align*}
and
\begin{align*}
    \mathbb{P}_m^k
        \left\{
            \varphi(\mathbf{Z}; \boldsymbol{\eta})
        \right\}
        \left(
        1
        -
        \frac{K m_k}{m}
        \right)
    =
    \{\Psi(w) + \oP{1}\}O(m^{-1})
    =
    \OP{m^{-1}}
    .
\end{align*}
Therefore, under the same condition in Theorem 3,
we have
\begin{align*}
    \sqrt{m}\{\widehat{\Psi}^{\textup{app}}(w) - \Psi(w)\}
    \overset{d}{\to}
    N(0,\sigma_{\textup{app}}^2(w))
\end{align*}
where 
\begin{align*}
    \sigma_\textup{app}^{2}(w)
    = &
    \textup{Var}
        \big\{ 
            \varphi(\mathbf{Z}; \boldsymbol{\eta}) 
        \big\}
    \\
    =&
    \V{[}{]}{
        \E{\{}{\}}{
            \varphi(\mathbf{Z}; \boldsymbol{\eta}) 
            | \mathbf{O}
        }
    }
    +
    \E{[}{]}{ 
        \V{\{}{\}}{ 
            \varphi(\mathbf{Z}; \boldsymbol{\eta}) 
            | \mathbf{O}
        } 
    }
    \\
    =&
    \V{\{}{\}}{
        \varphi(\mathbf{O}; \boldsymbol{\eta}) 
    }
    +
    \E{[}{]}{ 
        \V{\{}{\}}{ 
            \sumaq \frac{h(\mathbf{O}, \aq)}{f(\aq)}
            +
            l(\mathbf{O})
            \middle| \mathbf{O}
        } 
    }
    \\
    = &
    \sigma^2(w)
    +
    \frac{1}{r}
    \mathbb{E}\left[
        \textup{Var}\left\{
            \frac
                {h(\mathbf{O}, \mathbf{a}^{(\cdot)})}
                {f(\mathbf{a}^{(\cdot)})}
            \middle| \mathbf{O}
        \right\}
    \right]
    .
\end{align*}
and
$\sigma^2(w)
=
\textup{Var}\Big\{
    \varphi(\mathbf{O}; \boldsymbol{\eta}) 
\Big\}$
is the nonparametric efficiency bound of $\Psi(w)$.

\subsubsection{Proof of Theorem \ref{thm:app:varest}}

First, note that the variance estimator of $\widehat{\Psi}^{\textup{app}}(w)$ is
\begin{align*}
    \widehat{\sigma}_{\textup{app}}^2(w) 
    = 
    \frac{1}{K} \sum_{k=1}^{K} 
    \mathbb{P}_m^k \allowbreak
    \Big[ 
        \big\{ \allowbreak
            \varphi(\mathbf{Z}; \widehat{\boldsymbol{\eta}}^{(-k)}) \allowbreak
        \big\}^2
    \Big]
    - 
    \widehat{\Psi}^{\textup{app}}(w)^2
\end{align*}
and the asymptotic variance of $\widehat{\Psi}^{\textup{app}}(w)$ is
\begin{align*}
    \sigma_\textup{app}^{2}(w)
    =
    \textup{Var}
        \big\{ 
            \varphi(\mathbf{Z}; \boldsymbol{\eta}) 
        \big\}
    = 
    \mathbb{P} \allowbreak
    \Big[ 
        \big\{ \allowbreak
            \varphi(\mathbf{Z}; {\boldsymbol{\eta}}) \allowbreak
        \big\}^2
    \Big]
    - 
    \Psi(w)^2
    .
\end{align*}
Since 
$\widehat{\Psi}^{\textup{app}}(w) \overset{p}{\to} \Psi(w)$ from Theorem 2,
it suffices to show
$$
\mathbb{P}_m^k \allowbreak
    \Big[ 
        \big\{ \allowbreak
            \varphi(\mathbf{Z}; \widehat{\boldsymbol{\eta}}^{(-k)}) \allowbreak
        \big\}^2
    \Big]
\overset{p}{\to}
\mathbb{P} \allowbreak
    \Big[ 
        \big\{ \allowbreak
            \varphi(\mathbf{Z}; {\boldsymbol{\eta}}) \allowbreak
        \big\}^2
    \Big]
.
$$
This can be shown exactly the same as explained in the proof of Theorem 4 (Section \ref{proof:thm4}), using conditional law of large numbers and Cauchy-Schwarz inequality.

\newpage

\section{Details on example policies}

This section presents details on the large sample properties of the proposed estimators under the example policies.

We first define some additional notation.
For fixed 
$(\mathbf{a}, \mathbf{x}, n) 
\in 
\mathcal{A}(n) \times \mathcal{X}(n) \times \mathbb{N}$,
assume that the EIF of 
$Q(\mathbf{a} | \mathbf{x}, n)$
is given by
$
\{\mathbbm{1}(\mathbf{X} = \mathbf{x}, N = n)
/
d\mathbb{P}(\mathbf{x}, n)\}
\phi_Q(\mathbf{A}, \mathbf{X}, N; \mathbf{a})$,
where $\phi_Q(\mathbf{A}, \mathbf{X}, N; \mathbf{a})$ is a function of 
$(\mathbf{A}, \mathbf{X}, N, \mathbf{a}) 
\in 
\mathcal{A}(N) \times \mathcal{X}(N) \times \mathbb{N} \times \mathcal{A}(n)$.
For notational convenience, let 
$w(\mathbf{a}, \mathbf{x}, n)$ and
$w_t(\mathbf{a}, \mathbf{x}, n)$ for $t \in \{0,1\}$
denote 
the weight function $w$ for $\mu(Q)$ and $\mu_t(Q)$, respectively,
which are given by 
\begin{align*}
    w(\mathbf{a}, \mathbf{X}, N) 
    = &
    \frac{1}{N} Q(\mathbf{a}| \mathbf{X}, N) \mathbf{J}_{N}
    \\
    w_t(\mathbf{a}, \mathbf{X}, N)
    = &
    \frac{1}{N}
    \Big(
        \mathbbm{1}(a_{1} = t) Q(\mathbf{a}_{(-1)}| \mathbf{X}, N), 
        \dots, 
        \mathbbm{1}(a_{N} = t) Q(\mathbf{a}_{(-N)}| \mathbf{X}, N) 
    \Big)^\top
\end{align*}
from Lemma 1,
where 
$Q(\mathbf{a}_{(-j)} | \mathbf{X}, N) 
=  
Q(1, \mathbf{a}_{(-j)} | \mathbf{X}, N) 
+  
Q(0, \mathbf{a}_{(-j)} | \mathbf{X}, N)$. 
Similarly, let
$\phi(\mathbf{A}, \mathbf{X}, N; \mathbf{a})$ and
$\phi_t(\mathbf{A}, \mathbf{X}, N; \mathbf{a})$
denote the EIF of
$w(\mathbf{a}, \mathbf{x}, n)$ and
$w_t(\mathbf{a}, \mathbf{x}, n)$, respectively,
which are given by
\begin{align*}
    \phi(\mathbf{A}, \mathbf{X}, N; \mathbf{a})
    = &
    \frac{1}{N} \phi_{Q}(\mathbf{A}, \mathbf{X}, N; \mathbf{a}) \mathbf{J}_{N}
    \\
    \phi_t(\mathbf{A}, \mathbf{X}, N; \mathbf{a})
    = &
    \frac{1}{N}
    \Big(
        \mathbbm{1}(a_{1} = t) 
        \phi_{Q}(\mathbf{A}, \mathbf{X}, N; \mathbf{a}_{(-1)}),
        \dots,
        \mathbbm{1}(a_{N} = t) 
        \phi_{Q}(\mathbf{A}, \mathbf{X}, N; \mathbf{a}_{(-N)})
    \Big)^\top
\end{align*}
where
$\phi_{Q}(\mathbf{A}, \mathbf{X}, N; \mathbf{a}_{(-j)})
=
\phi_{Q}(\mathbf{A}, \mathbf{X}, N; (1,\mathbf{a}_{(-j)})) 
+
\phi_{Q}(\mathbf{A}, \mathbf{X}, N; (0,\mathbf{a}_{(-j)}))$
is the EIF of $Q(\mathbf{a}_{(-j)} | \mathbf{X}, N)$.
Under this specification, the uncentered EIFs of $\mu(Q)$ and $\mu_t(Q)$, namely $\varphi_{\mu(Q)}(\mathbf{O}; \boldsymbol{\eta})$ and $\varphi_{\mu_t(Q)}(\mathbf{O}; \boldsymbol{\eta})$, are given by
\begin{align*}
    \varphi_{\mu(Q)}(\mathbf{O}; \boldsymbol{\eta})
    = &
    \frac{1}{N}
    \sum_{j=1}^{N}
    \begin{aligned}[t]
        \Bigg[  
        &
        \sum_{\mathbf{a} \in \mathcal{A}(N)}
            \big\{ 
                    Q(\mathbf{a} | \mathbf{X}, N) 
                    + 
                    \phi_Q(\mathbf{A}, \mathbf{X}, N;\mathbf{a}) 
                \big\}
            G_j \big(
                \mathbf{a}, \mathbf{X}, N 
            \big)
        \\
        & + 
        \frac
            {Q(\mathbf{A} | \mathbf{X}, N) }
            {H(\mathbf{A},\mathbf{X}, N)}
        \left\{ 
            Y_{j} 
            - 
            G_j \big( 
                \mathbf{A}, \mathbf{X}, N 
            \big) 
        \right\}
        \Bigg]
    \end{aligned}
    \\
    \varphi_{\mu_t(Q)}(\mathbf{O}; \boldsymbol{\eta})
    = &
    \frac{1}{N}
    \sum_{j=1}^{N}
    \begin{aligned}[t]
        \Bigg[  
        &
        \sum_{\mathbf{a} \in \mathcal{A}(N)}
            \mathbbm{1}(a_{j} = t) 
            \big\{ 
                    Q(\mathbf{a}_{(-j)} | \mathbf{X}, N) 
                    + 
                    \phi_Q(\mathbf{A}, \mathbf{X}, N;\mathbf{a}_{(-j)}) 
                \big\}
            G_j \big(
                \mathbf{a}, \mathbf{X}, N 
            \big)
        \\
        & + 
        \frac
            {\mathbbm{1}(A_{j} = t) 
            Q(\mathbf{A}_{(-j)} | \mathbf{X}, N) }
            {H(\mathbf{A},\mathbf{X}, N)}
        \left\{ 
            Y_{j} 
            - 
            G_j \big( 
                \mathbf{A}, \mathbf{X}, N 
            \big) 
        \right\}
        \Bigg]
        ,
    \end{aligned}
\end{align*}
where $G_j$ is $j$ th component of $G$.
Based on these, the inference procedure for $\mu(Q)$, $\mu_t(Q)$, and causal effects can be described as follows:

\begin{enumerate}

    \item[\textbf{Step 1}] 
    Determine the policy distribution $Q$ of interest
    
    \item[\textbf{Step 2}] 
    Compute $\phi_Q$
    
    \item[\textbf{Step 3}] 
    Construct $w(\mathbf{a}, \mathbf{x}, n)$ and $w_t(\mathbf{a}, \mathbf{x}, n)$ for $t \in \{0,1\}$ from $Q$
    
    \item[\textbf{Step 4}] 
    Construct $\phi(\mathbf{A}, \mathbf{X}, N; \mathbf{a})$ and $\phi_t(\mathbf{A}, \mathbf{X}, N; \mathbf{a})$ from $\phi_Q$

    \item[\textbf{Step 5}] 
    Obtain the uncentered EIF of $\mu(Q)$ and $\mu_t(Q)$, namely, $\varphi_{\mu(Q)}(\mathbf{O}; \boldsymbol{\eta})$ and $\varphi_{\mu_t(Q)}(\mathbf{O}; \boldsymbol{\eta})$
    
    \item[\textbf{Step 6}] 
    Compute the proposed estimator
    $\widehat{\mu}(Q) 
    =
    K^{-1} 
    \sum_{k=1}^{K} 
    \mathbb{P}_m^k 
    \big\{ 
        \varphi_{\mu(Q)}(\mathbf{O}; \widehat{\boldsymbol{\eta}}^{(-k)}) 
    \big\}$
    and
    $\widehat{\mu}_t(Q) 
    =
    K^{-1} 
    \sum_{k=1}^{K} 
    \mathbb{P}_m^k 
    \big\{ 
        \varphi_{\mu_t(Q)}(\mathbf{O}; \widehat{\boldsymbol{\eta}}^{(-k)}) 
    \big\}$
    and their variance estimators. 
    Estimators of causal effects (e.g., direct effect) are obtained by the difference between the estimators $\widehat{\mu}(Q)$, $\widehat{\mu}_t(Q)$.

    \item[\textbf{Step 7}] 
    Check sufficient conditions for consistency and asymptotic normality of the proposed estimators according to Theorem 2 -- 4.

    \item[\textbf{Step 8}] 
    Perform inference on target causal estimands based on the large sample properties of the proposed estimators

\end{enumerate}

\noindent
Regarding the large sample conditions in step 7, 
it can be shown that 
\begin{align*}    
     \norm\Bigg{
        \sum_{\mathbf{a}\in \mathcal{A}(N)}
            \norm\bigg{
                \big(
                    \widehat{\phi}^{(-k)} - \phi
                \big)
                (\mathbf{A}, \mathbf{X}, N; \mathbf{a})
            }_2
    }_{L_2(\mathbb{P})}
    =
    O \left(
        \norm\Bigg{
            \sum_{\mathbf{a}\in \mathcal{A}(N)}
                \bigg|
                    \big(
                        \widehat{\phi}_Q^{(-k)} - \phi_Q
                    \big)
                    (\mathbf{A}, \mathbf{X}, N; \mathbf{a})
                \bigg|
        }_{L_2(\mathbb{P})}
    \right)
    ,
\end{align*}
\begin{align*}
    &
    \norm\Bigg{
        \sum_{\mathbf{a}\in \mathcal{A}(N)}
            \norm\bigg{
                \big(
                \widehat{w}^{(-k)}
                -
                w
                \big)
                (\mathbf{a}, \mathbf{X}, N)
                +
                \sum_{\mathbf{a}' \in \mathcal{A}(N)}
                    \widehat{\phi}^{(-k)}(\mathbf{a}', \mathbf{X}, N; \mathbf{a})
                    H(\mathbf{a}', \mathbf{X}, N)
            }_2
    }_{L_2(\mathbb{P})}
    \\
    = &
    O \left(
        \norm\Bigg{
            \sum_{\mathbf{a}\in \mathcal{A}(N)}
                \bigg|
                    \big(
                    \widehat{Q}^{(-k)}
                    -
                    Q
                    \big)
                    (\mathbf{a} | \mathbf{X}, N)
                    +
                    \sum_{\mathbf{a}' \in \mathcal{A}(N)}
                        \widehat{\phi}_Q^{(-k)}(\mathbf{a}', \mathbf{X}, N; \mathbf{a})
                        H(\mathbf{a}', \mathbf{X}, N)
                \bigg|
        }_{L_2(\mathbb{P})}
    \right)
\end{align*}
and
\begin{align*}    
     \norm\Bigg{
        \sum_{\mathbf{a}\in \mathcal{A}(N)}
            \norm\bigg{
                \big(
                    \widehat{\phi}_t^{(-k)} - \phi_t
                \big)
                (\mathbf{A}, \mathbf{X}, N; \mathbf{a})
            }_2
    }_{L_2(\mathbb{P})}
    =
    O \left(
        \norm\Bigg{
            \sum_{\mathbf{a}\in \mathcal{A}(N)}
                \bigg|
                    \big(
                        \widehat{\phi}_Q^{(-k)} - \phi_Q
                    \big)
                    (\mathbf{A}, \mathbf{X}, N; \mathbf{a})
                \bigg|
        }_{L_2(\mathbb{P})}
    \right)
    ,
\end{align*}
\begin{align*}
    &
    \norm\Bigg{
        \sum_{\mathbf{a}\in \mathcal{A}(N)}
            \norm\bigg{
                \big(
                \widehat{w}_t^{(-k)}
                -
                w_t
                \big)
                (\mathbf{a}, \mathbf{X}, N)
                +
                \sum_{\mathbf{a}' \in \mathcal{A}(N)}
                    \widehat{\phi}_t^{(-k)}(\mathbf{a}', \mathbf{X}, N; \mathbf{a})
                    H(\mathbf{a}', \mathbf{X}, N)
            }_2
    }_{L_2(\mathbb{P})}
    \\
    = &
    O \left(
        \norm\Bigg{
            \sum_{\mathbf{a}\in \mathcal{A}(N)}
                \bigg|
                    \big(
                    \widehat{Q}^{(-k)}
                    -
                    Q
                    \big)
                    (\mathbf{a} | \mathbf{X}, N)
                    +
                    \sum_{\mathbf{a}' \in \mathcal{A}(N)}
                        \widehat{\phi}_Q^{(-k)}(\mathbf{a}', \mathbf{X}, N; \mathbf{a})
                        H(\mathbf{a}', \mathbf{X}, N)
                \bigg|
        }_{L_2(\mathbb{P})}
    \right)
\end{align*}
where $\widehat{Q}$ and $\widehat{\phi}_Q$ are estimators of $Q$ and $\phi_Q$, respectively.
Thus, when investigating the rate conditions for $\widehat{w}$ and $\widehat{\phi}$ (or $\widehat{w}_t$ and $\widehat{\phi}_t$),
it suffices to assess the rate conditions for $\widehat{Q}$ and $\widehat{\phi}_Q$.

Below, steps 1, 2, 5, and 7 of this procedure are described for the four example policies discussed in the main text.

\subsection{Type B policy}
\label{example:TypeB}

\begin{enumerate}

    \item[\textbf{Step 1}]
    Policy distribution $Q$:  
    $$Q_{\scriptscriptstyle \textup{B}}(\mathbf{a}|\mathbf{X}, N; \alpha)
    = 
    \prod_{j=1}^{N} \alpha^{a_{j}} (1-\alpha)^{1-a_{j}}$$

    \item[\textbf{Step 2}]
    EIF $\phi_Q$: 
    $$\phi_{Q_{\textup{B}}}(\mathbf{A}, \mathbf{X}, N; \mathbf{a})
    = 
    0$$
    since $Q$ does not depend on the observed data distribution

    \item[\textbf{Step 5}] 
    Uncentered EIF of $\mu(Q)$ and $\mu_t(Q)$: 

    \begin{align*}
        \varphi_{\mu(Q)}(\mathbf{O}; \boldsymbol{\eta})
        = &
        \frac{1}{N}
        \sum_{j=1}^{N}
        \begin{aligned}[t]
            \Bigg[  
            &
            \sum_{\mathbf{a} \in \mathcal{A}(N)}
                G_j \big(
                    \mathbf{a}, \mathbf{X}, N 
                \big)
                \prod_{l=1}^{N} \alpha^{a_{l}} (1-\alpha)^{1-a_{l}}
            \\
            & + 
            \frac
                {\prod_{l=1}^{N} \alpha^{A_{l}} (1-\alpha)^{1-A_{l}}}
                {H(\mathbf{A},\mathbf{X}, N)}
            \left\{ 
                Y_{j} 
                - 
                G_j \big( 
                    \mathbf{A}, \mathbf{X}, N 
                \big) 
            \right\}
            \Bigg]
        \end{aligned}
        \\
        \varphi_{\mu_t(Q)}(\mathbf{O}; \boldsymbol{\eta})
        = &
        \frac{1}{N}
        \sum_{j=1}^{N}
        \begin{aligned}[t]
            \Bigg[  
            &
            \sum_{\mathbf{a} \in \mathcal{A}(N)}
                G_j \big(
                    \mathbf{a}, \mathbf{X}, N 
                \big)
                \mathbbm{1}(a_{j} = t) 
                \prod_{l \ne j} \alpha^{a_{l}} (1-\alpha)^{1-a_{l}} 
            \\
            & + 
            \frac
                {\mathbbm{1}(A_{j} = t) 
                \prod_{l \ne j} \alpha^{A_{l}} (1-\alpha)^{1-A_{l}}  }
                {H(\mathbf{A},\mathbf{X}, N)}
            \left\{ 
                Y_{j} 
                - 
                G_j \big( 
                    \mathbf{A}, \mathbf{X}, N 
                \big) 
            \right\}
            \Bigg]
        \end{aligned}
    \end{align*}
        

    \item[\textbf{Step 7}]
    Check large sample conditions:
    
    (B7) \textit{Convergence rate of} $\widehat{\phi}^{(-k)}$: 
    Since 
    $\phi \equiv \mathbf{0}$, 
    $\widehat{\phi}^{(-k)} =
    \widehat{\phi}_t^{(-k)} =
    \mathbf{0}$,
    and thus $r_\phi = 0$ 
    for both $\widehat{\mu}(Q)$ and $\widehat{\mu}_t(Q)$.

    (B8) \textit{Second order convergence rate of} $\widehat{w}^{(-k)}$: 
    Since $w$ is known and does not need to be estimated, 
    $\widehat{w}^{(-k)} = w$,
    and thus
    $\big(
        \widehat{w}^{(-k)}
        -
        w
    \big)
    (\mathbf{a}, \mathbf{X}, N)
    +
    \widehat{\phi}^{(-k)}(\mathbf{A}, \mathbf{X}, N; \mathbf{a})
    =
    0
    $,
    which implies $r_w = 0$
    for both $\widehat{\mu}(Q)$ and $\widehat{\mu}_t(Q)$.

    Thus, sufficient conditions for consistency and asymptotic normality are as follows:
    
    (i) Consistency: $r_G = o(1)$ or $r_H = o(1)$;
    
    (ii) Asymptotic normality: $r_G \cdot r_H = o(m^{-1/2})$;
    
    (iii) Consistent variance estimator: $r_G = r_H = o(1)$
    as $m \to \infty$. 

\end{enumerate}

\subsection{Cluster incremental propensity score policy}
\label{example:CIPS}

\begin{enumerate}

    \item[\textbf{Step 1}]
    Policy distribution $Q$:  
    $$Q_{\scriptscriptstyle \textup{CIPS}}(\mathbf{a} | \mathbf{X}, N; \delta) 
        =
        \prod_{j = 1}^{N} 
        \allowbreak
            (\pi_{j,\delta})^{a_{j}}
            \allowbreak
            (1 - \pi_{j,\delta})^{1-a_{j}} 
    ,
    $$
    where
    $\pi_{j, \delta} \allowbreak
    = \delta(\mathbf{X}, N)
    \pi_{j} / \allowbreak
    \{ \delta(\mathbf{X}, N) \pi_{j} + 1 - \allowbreak  \pi_{j} \}$
    denotes the shifted propensity score
    and
    $\delta$ is a known function of $\mathbf{X}$ and $N$

    \item[\textbf{Step 2}]
    EIF $\phi_Q$:
    \begin{align*}
        \phi_{Q_{\textup{CIPS}}}(\mathbf{A}, \mathbf{X}, N; \mathbf{a})
        =
        Q_{\scriptscriptstyle \textup{CIPS}}(\mathbf{a} | \mathbf{X}, N; \delta)
        \sum_{l=1}^{N}
            \frac
                {(2a_{l}-1) \delta(\mathbf{X}, N) (A_{l} - \pi_{l})}
                {(\pi_{l,\delta})^{a_{l}} (1 - \pi_{l,\delta})^{1-a_{l}}
                \{\delta(\mathbf{X}, N)\pi_{l} + 1 - \pi_{l}\}^2 }
    \end{align*}
    from the fact that the EIF of 
    $\pi_{j} \allowbreak
    = \mathbb{P}(A_{j} = 1 | \mathbf{X} = \mathbf{x}, N = n)$
    is
    \begin{align*}
        \textup{EIF}(\pi_{j})
        =
        \frac
            {\mathbbm{1}(\mathbf{X} = \mathbf{x}, N = n)}
            {d\mathbb{P}(\mathbf{x}, n)}
        \{A_j - \pi_j\}        
    \end{align*}
    and the EIF of 
    $\pi_{j, \delta}$
    is
    \begin{align*}
        \textup{EIF}(\pi_{j, \delta})
        =
        \textup{EIF}(\pi_{j})
        \frac
            {\delta(\mathbf{X}, N)}
            {\{\delta(\mathbf{X}, N)\pi_{j} + 1 - \pi_{j}\}^2}
        ,
    \end{align*}
    and thus
    \begin{align*}
        \textup{EIF}(Q_{\scriptscriptstyle \textup{CIPS}}(\mathbf{a} | \mathbf{x}, n; \delta))
        =
        \sum_{l=1}^{N}
        \left\{
        \prod_{j \ne l} 
            (\pi_{j,\delta})^{a_{j}}
            (1 - \pi_{j,\delta})^{1-a_{j}} 
        \right\}
        (2a_{l}-1)\textup{EIF}(\pi_{l,\delta})
        .
    \end{align*}

    \item[\textbf{Step 5}]
    Uncentered EIF of $\mu(Q)$ and $\mu_t(Q)$: 
    \begin{align*}
        \varphi_{\mu(Q)}(\mathbf{O}; \boldsymbol{\eta})
        = &
        \frac{1}{N}
        \sum_{j=1}^{N}
        \begin{aligned}[t]
            \Bigg[  
            &
            \sum_{\mathbf{a} \in \mathcal{A}(N)}
                G_j \big(
                    \mathbf{a}, \mathbf{X}, N 
                \big)
                \times
                Q_{\scriptscriptstyle \textup{CIPS}}(\mathbf{a} | \mathbf{X}, N; \delta)
                \\
                & \qquad \quad
                \times
                \left\{
                    1 
                    +
                    \sum_{l=1}^{N}
                    \frac
                        {(2a_{l}-1) \delta(\mathbf{X}, N) (A_{l} - \pi_{l})}
                        {(\pi_{l,\delta})^{a_{l}} (1 - \pi_{l,\delta})^{1-a_{l}}
                        \{\delta(\mathbf{X}, N)\pi_{l} + 1 - \pi_{l}\}^2}
                \right\}
            \\
            & + 
            \frac
                {Q_{\scriptscriptstyle \textup{CIPS}}(\mathbf{A} | \mathbf{X}, N; \delta)}
                {H(\mathbf{A},\mathbf{X}, N)}
            \left\{ 
                Y_{j} 
                - 
                G_j \big( 
                    \mathbf{A}, \mathbf{X}, N 
                \big) 
            \right\}
            \Bigg]
        \end{aligned}
        \\
        \varphi_{\mu_t(Q)}(\mathbf{O}; \boldsymbol{\eta})
        = &
        \frac{1}{N}
        \sum_{j=1}^{N}
        \begin{aligned}[t]
            \Bigg[  
            &
            \sum_{\mathbf{a} \in \mathcal{A}(N)}
                G_j \big(
                    \mathbf{a}, \mathbf{X}, N 
                \big)
                \frac
                    {\mathbbm{1}(a_{j} = t)}
                    {\mathbb{P}_{\delta}(a_{j}| \mathbf{X}, N)}
                \times
                Q_{\scriptscriptstyle \textup{CIPS}}(\mathbf{a} | \mathbf{X}, N; \delta)
                \\
                & \qquad \quad
                \times
                \left\{
                    1 
                    +
                    \sum_{l \ne j}
                    \frac
                        {(2a_{l}-1) \delta(\mathbf{X}, N) (A_{l} - \pi_{l})}
                        {(\pi_{l,\delta})^{a_{l}} (1 - \pi_{l,\delta})^{1-a_{l}}
                        \{\delta(\mathbf{X}, N)\pi_{l} + 1 - \pi_{l}\}^2}
                \right\}
            \\
            & + 
            \frac
                {\mathbbm{1}(A_{j} = t)}
                {\mathbb{P}_{\delta}(A_{j}| \mathbf{X}, N)}
            \times
            \frac
                {Q_{\scriptscriptstyle \textup{CIPS}}(\mathbf{A} | \mathbf{X}, N; \delta)}
                {H(\mathbf{A},\mathbf{X}, N)}
            \left\{ 
                Y_{j} 
                - 
                G_j \big( 
                    \mathbf{A}, \mathbf{X}, N 
                \big) 
            \right\}
            \Bigg]
        \end{aligned}
    \end{align*}
        
    \item[\textbf{Step 7}] 
    Check large sample conditions:

    Under CIPS policy, we assume conditional independence of $Y_{ij}$'s as well as that of $A_{ij}$'s.
    Therefore, individual-level nuisance functions $g$ and $\pi$ may be estimated instead of $G$ and $H$ and used to construct estimators of $G$, $H$, $w$, and $\phi$.
    
    Assume there exist $r_{\pi}, r_g > 0$ such that
    $ 
    \normp{
        \sum_{j=1}^{N}
            \lvert
                (\widehat{\pi}^{(-k)} - \pi) (j, \mathbf{X}, N)
            \rvert
    }
    =
    O_\mathbb{P}(r_{\pi})
    $
    and
    $
    \normp{
        \sum_{\mathbf{a} \in \mathcal{A}(N)}
        \sum_{j=1}^{N}
            \lvert
                (\widehat{g}^{(-k)} - g) (j, \mathbf{a}, \mathbf{X}, N)
            \rvert
    }
    =
    O_\mathbb{P}(r_g)$.
    We will assess the conditions using the inequalities that if
    $\overline{d}_i, d_i \in [0,u], i = 1, \dots, n$, 
    then 
    $|\prod_{i=1}^n \overline{d}_i - \prod_{i=1}^n d_i| 
    <
    2^n u^{n-1} \sum_{i=1}^n |\overline{d}_i - d_i|$
    and
    $|\prod_{i=1}^n \overline{d}_i 
    - 
    \prod_{i=1}^n d_i 
    - 
    \sum_{l=1}^n (\prod_{i \ne l} \overline{d}_i) \allowbreak (\overline{d}_l - d_l)  | 
    <
    2^n/u^{n-1} \sum_{i=1}^n |\overline{d}_i - d_i|^2$.    

    (B5) \textit{Convergence rate of} $\widehat{H}^{(-k)}$: 
    $r_H = O(r_{\pi})$
    since
    \begin{align*}
        \lvert
            \big(\widehat{H}-H\big)(\mathbf{a}, \mathbf{X}, N)
        \rvert
        =
        \Big|
            \prod_{j=1}^N
                \widehat{\mathbb{P}}(a_{j}| \mathbf{X}, N)
            -
            \prod_{j=1}^N
                \mathbb{P}(a_{j}| \mathbf{X}, N)
        \Big|
        \lesssim
        \sum_{j=1}^{N}
        \lvert
            (\widehat{\pi}_j - \pi_j)
            (\mathbf{X}, N)
        \rvert
    \end{align*}

    (B6) \textit{Convergence rate of} $\widehat{G}^{(-k)}$: 
    $r_G = O(r_g)$ since
    \begin{align*}
        \normt{
            \big(\widehat{G}-G\big)(\mathbf{a}, \mathbf{X}, N)
        }
        =
        \sqrt{
            \sum_{j=1}^N
            (\widehat{g} - g) (j, \mathbf{a}, \mathbf{X}, N)^2
        }
        =
        O\left(
            \sum_{j=1}^{N}
            \lvert
                (\widehat{g} - g) (j, \mathbf{a}, \mathbf{X}, N)
            \rvert
        \right)
    \end{align*}
    
    (B7) \textit{Convergence rate of} $\widehat{\phi}^{(-k)}$:
    $r_{\phi} = O(r_{\pi})$ since
    \begin{align*}
        &
        \widehat{\phi}_{Q_{\textup{CIPS}}}(\mathbf{A}, \mathbf{X}, N; \mathbf{a})
        -
        \phi_{Q_{\textup{CIPS}}}(\mathbf{A}, \mathbf{X}, N; \mathbf{a})
        \\
        = &
        \left(
            \widehat{Q}_{\scriptscriptstyle \textup{CIPS}}
            -
            Q_{\scriptscriptstyle \textup{CIPS}}
        \right)
        (\mathbf{a} | \mathbf{X}, N; \delta)
        \sum_{l=1}^{N}
            \frac
                {(2a_{l}-1) \delta(\mathbf{X}, N) (A_{l} - \widehat{\pi}_{l})}
                {\widehat{\mathbb{P}}_{\delta}(a_{l}| \mathbf{X}, N)
                \{\delta(\mathbf{X}, N)\widehat{\pi}_{l} + 1 - \widehat{\pi}_{l}\}^2}
        \\
        & +
        Q_{\scriptscriptstyle \textup{CIPS}}
        (\mathbf{a} | \mathbf{X}, N; \delta)
        \sum_{l=1}^{N}
            \left\{
                \frac
                    {(2a_{l}-1) \delta(\mathbf{X}, N) (A_{l} - \widehat{\pi}_{l})}
                    {\widehat{\mathbb{P}}_{\delta}(a_{l}| \mathbf{X}, N)
                    \{\delta(\mathbf{X}, N)\widehat{\pi}_{l} + 1 - \widehat{\pi}_{l}\}^2}
                -
                \frac
                    {(2a_{l}-1) \delta(\mathbf{X}, N) (A_{l} - \pi_{l})}
                    {\mathbb{P}_{\delta}(a_{l}| \mathbf{X}, N)
                    \{\delta(\mathbf{X}, N)\pi_{l} + 1 - \pi_{l}\}^2}
            \right\}
        \\
        \lesssim &
        \sum_{j=1}^{N}
        \lvert
            (\widehat{\pi}_j - \pi_j)
            (\mathbf{X}, N)
        \rvert
    \end{align*}
    which follows from
    \begin{align*}
        \left(
            \widehat{Q}_{\scriptscriptstyle \textup{CIPS}}
            -
            Q_{\scriptscriptstyle \textup{CIPS}}
        \right)
        (\mathbf{a} | \mathbf{X}, N; \delta)
        \lesssim &
        \sum_{j=1}^{N}
        \lvert
            \widehat{\mathbb{P}}_{\delta}(a_{j}| \mathbf{X}, N)
            -
            \mathbb{P}_{\delta}(a_{j}| \mathbf{X}, N)
        \rvert
        \\
        = &
        \sum_{j=1}^{N}
        \frac
            {\delta(\mathbf{X}, N)|\widehat{\pi}_{j} -\pi_{j}|}
            {\{\delta(\mathbf{X}, N)\widehat{\pi}_{j} + 1 - \widehat{\pi}_{j}\}
            \{\delta(\mathbf{X}, N)\pi_{j} + 1 - \pi_{j}\}}
    \end{align*}
    and
    \begin{align*}
        &
        \frac
            {A_{l} - \widehat{\pi}_{l}}
            {\widehat{\mathbb{P}}_{\delta}(a_{l}| \mathbf{X}, N)
            \{\delta(\mathbf{X}, N)\widehat{\pi}_{l} + 1 - \widehat{\pi}_{l}\}^2}
        -
        \frac
            {A_{l} - \pi_{l}}
            {\mathbb{P}_{\delta}(a_{l}| \mathbf{X}, N)
            \{\delta(\mathbf{X}, N)\pi_{l} + 1 - \pi_{l}\}^2}
        \\
        = &
        \frac
            {\pi_l - \widehat{\pi}_l}
            {\widehat{\mathbb{P}}_{\delta}(a_{l}| \mathbf{X}, N)
            \{\delta(\mathbf{X}, N)\widehat{\pi}_{l} + 1 - \widehat{\pi}_{l}\}^2}
        \\
        & +
        (A_l - \pi_l)
        \frac
            {\mathbb{P}_{\delta}(a_{l}| \mathbf{X}, N)
            \{\delta(\mathbf{X}, N)\pi_{l} + 1 - \pi_{l}\}^2
            -
            \widehat{\mathbb{P}}_{\delta}(a_{l}| \mathbf{X}, N)
            \{\delta(\mathbf{X}, N)\widehat{\pi}_{l} + 1 - \widehat{\pi}_{l}\}^2}
            {\widehat{\mathbb{P}}_{\delta}(a_{l}| \mathbf{X}, N)
            \{\delta(\mathbf{X}, N)\widehat{\pi}_{l} + 1 - \widehat{\pi}_{l}\}^2
            \mathbb{P}_{\delta}(a_{l}| \mathbf{X}, N)
            \{\delta(\mathbf{X}, N)\pi_{l} + 1 - \pi_{l}\}^2}
    \end{align*}
    with
    \begin{align*}
        &
        \mathbb{P}_{\delta}(a_{l}| \mathbf{X}, N)
        \{\delta(\mathbf{X}, N)\pi_{l} + 1 - \pi_{l}\}^2
        -
        \widehat{\mathbb{P}}_{\delta}(a_{l}| \mathbf{X}, N)
        \{\delta(\mathbf{X}, N)\widehat{\pi}_{l} + 1 - \widehat{\pi}_{l}\}^2
        \\
        = &
        \begin{cases}
            (\pi_l - \widehat{\pi}_l)
            \delta(\mathbf{X}, N)
            [\{\delta(\mathbf{X}, N)-1\}(\pi_l + \widehat{\pi}_l) + 1],
            & a_j = 1
            \\
            (\pi_l - \widehat{\pi}_l)
            [\delta(\mathbf{X}, N)-2 
            -\{\delta(\mathbf{X}, N)-1\}(\pi_l + \widehat{\pi}_l)],
            & a_j = 0
        \end{cases}
        \\
        \lesssim &
        | \pi_l - \widehat{\pi}_l |
        .
    \end{align*}

    (B8) \textit{Second order convergence rate of} $\widehat{w}^{(-k)}$: 
    $r_{w} = O(r_{\pi})$ since
    \begin{align*}
        &
        \widehat{Q}_{\scriptscriptstyle \textup{CIPS}}
        (\mathbf{a} | \mathbf{X}, N; \delta)
        -
        Q_{\scriptscriptstyle \textup{CIPS}}
        (\mathbf{a} | \mathbf{X}, N; \delta)
        +
        \sum_{\mathbf{a}' \in \mathcal{A}(N)}
            \widehat{\phi}_{Q_{\textup{CIPS}}}(\mathbf{a}', \mathbf{X}, N; \mathbf{a})
            H(\mathbf{a}', \mathbf{X}, N)
        \\
        = &
        \prod_{j=1}^N
            \widehat{\mathbb{P}}_{\delta}(a_{j}| \mathbf{X}, N)
        -
        \prod_{j=1}^N
            \mathbb{P}_{\delta}(a_{j}| \mathbf{X}, N)
        +
        \sum_{j=1}^N
            \left\{
            \prod_{l \ne j}
                \widehat{\mathbb{P}}_{\delta}(a_{l}| \mathbf{X}, N)
            \right\}
            \frac
                {(2a_{j}-1) \delta(\mathbf{X}, N) (\pi_{j} - \widehat{\pi}_{j})}
                {\{\delta(\mathbf{X}, N)\widehat{\pi}_{j} + 1 - \widehat{\pi}_{j}\}^2}
        \\
        =&
        \prod_{j=1}^N
            \widehat{\mathbb{P}}_{\delta}(a_{j}| \mathbf{X}, N)
        -
        \prod_{j=1}^N
            \mathbb{P}_{\delta}(a_{j}| \mathbf{X}, N)
        -
        \sum_{j=1}^N
            \left\{
            \prod_{l \ne j}
                \widehat{\mathbb{P}}_{\delta}(a_{l}| \mathbf{X}, N)
            \right\}
            (\widehat{\mathbb{P}}_{\delta} - \mathbb{P}_{\delta})
            (a_{j}| \mathbf{X}, N)
        \\
        & +
        \sum_{j=1}^N
            \left\{
                \prod_{l \ne j}
                    \widehat{\mathbb{P}}_{\delta}(a_{l}| \mathbf{X}, N)
            \right\}
            \left\{
                (\widehat{\mathbb{P}}_{\delta} - \mathbb{P}_{\delta})
                (a_{j}| \mathbf{X}, N)
                +
                \frac
                    {(2a_{j}-1) \delta(\mathbf{X}, N) (\pi_{j} - \widehat{\pi}_{j})}
                    {\{\delta(\mathbf{X}, N)\widehat{\pi}_{j} + 1 - \widehat{\pi}_{j}\}^2}
            \right\}
        \\
        \lesssim &
        \sum_{j=1}^{N}
        \lvert
            \widehat{\mathbb{P}}_{\delta}(a_{j}| \mathbf{X}, N)
            -
            \mathbb{P}_{\delta}(a_{j}| \mathbf{X}, N)
        \rvert^2
        +
        \lvert
            (\widehat{\pi}_j - \pi_j)
            (\mathbf{X}, N)
        \rvert^2
        \\
        \lesssim &
        \sum_{j=1}^{N}
        \lvert
            (\widehat{\pi}_j - \pi_j)
            (\mathbf{X}, N)
        \rvert^2
    \end{align*}
    from
    \begin{align*}
        \sum_{\mathbf{a}' \in \mathcal{A}(N)}
            \widehat{\phi}_{Q_{\textup{CIPS}}}(\mathbf{a}', \mathbf{X}, N; \mathbf{a})
            H(\mathbf{a}', \mathbf{X}, N)
        = &
        \E{\{}{\}}{
            \widehat{\phi}_{Q_{\textup{CIPS}}}(\mathbf{A}, \mathbf{X}, N; \mathbf{a})
            \middle| D, \mathbf{X}, N
        }
        \\
        = &
        \widehat{Q}_{\scriptscriptstyle \textup{CIPS}}
        (\mathbf{a} | \mathbf{X}, N; \delta)
        \sum_{j=1}^{N}
            \frac
                {(2a_{j}-1) \delta(\mathbf{X}, N) (\pi_{j} - \widehat{\pi}_{j})}
                {\widehat{\mathbb{P}}_{\delta}(a_{j}| \mathbf{X}, N)
                \{\delta(\mathbf{X}, N)\widehat{\pi}_{j} + 1 - \widehat{\pi}_{j}\}^2}
    \end{align*}
    and
    \begin{align*}
        (\widehat{\mathbb{P}}_{\delta} - \mathbb{P}_{\delta})
        (a_{j}| \mathbf{X}, N)
        +
        \frac
            {(2a_{j}-1) \delta(\mathbf{X}, N) (\pi_{j} - \widehat{\pi}_{j})}
            {\{\delta(\mathbf{X}, N)\widehat{\pi}_{j} + 1 - \widehat{\pi}_{j}\}^2}
        =
        \frac
            {(2a_{j}-1) \delta(\mathbf{X}, N) \{\delta(\mathbf{X}, N)-1\} (\pi_{j} - \widehat{\pi}_{j})^2}
            {\{\delta(\mathbf{X}, N)\widehat{\pi}_{j} + 1 - \widehat{\pi}_{j}\}^2
            \{\delta(\mathbf{X}, N)\pi_{j} + 1 - \pi_{j}\}^2}
        .
    \end{align*}

    Thus, sufficient conditions for consistency and asymptotic normality are as follows:
    
    (i) Consistency: $r_{\pi} = o(1)$;
    
    (ii) Asymptotic normality: $r_{\pi} = o(m^{-1/4})$ and $r_{\pi} \cdot r_g = o(m^{-1/2})$;
    
    (iii) Consistent variance estimator: $r_{\pi} = r_g = o(1)$ as $m \to \infty$.

\end{enumerate}

\subsection{Cluster multiplicative shift policy}
\label{example:CMS}

\begin{enumerate}

    \item[\textbf{Step 1}]
    Policy distribution $Q$:  
    $$Q_{\scriptscriptstyle \textup{CMS}}(\mathbf{a} | \mathbf{X}, N; \lambda) 
    =
    \prod_{j = 1}^{N} 
        (\pi_{j,\lambda})^{a_{j}}
        (1 - \pi_{j,\lambda})^{1-a_{j}} 
    ,
    $$
    where
    $\pi_{j, \lambda} \allowbreak
    = \mathbb{P}_{\lambda}(A_{j} = 1 | \mathbf{X}, N) \allowbreak
    = (1-\lambda) X^* + \pi_{j} (X^* \lambda + 1 - X^*)$
    is the shifted propensity score

    \item[\textbf{Step 2}]
    EIF $\phi_Q$:
    \begin{align*}
        \phi_{Q_{\textup{CMS}}}(\mathbf{A}, \mathbf{X}, N; \mathbf{a})
        =
        Q_{\scriptscriptstyle \textup{CMS}}(\mathbf{a} | \mathbf{X}, N; \lambda)
        \sum_{l=1}^{N}
            \frac
                {(2a_{l}-1) (A_{l} - \pi_{l}) (X^* \lambda + 1 - X^*)}
                {(\pi_{l,\lambda})^{a_{l}} (1 - \pi_{l,\lambda})^{1-a_{l}}}
    \end{align*}
    from the fact that the EIF of 
    $\pi_{j} \allowbreak
    = \mathbb{P}(A_{j} = 1 | \mathbf{X} = \mathbf{x}, N = n)$
    is
    \begin{align*}
        \textup{EIF}(\pi_{j})
        =
        \frac
            {\mathbbm{1}(\mathbf{X} = \mathbf{x}, N = n)}
            {d\mathbb{P}(\mathbf{x}, n)}
        \{A_j - \pi_j\}        
    \end{align*}
    and the EIF of 
    $\pi_{j, \lambda}$
    is
    \begin{align*}
        \textup{EIF}(\pi_{j, \lambda})
        =
        \textup{EIF}(\pi_{j})(X^* \lambda + 1 - X^*)
        ,
    \end{align*}
    and thus
    \begin{align*}
        \textup{EIF}(Q_{\scriptscriptstyle \textup{CMS}}(\mathbf{a} | \mathbf{x}, n; \lambda))
        =
        \sum_{l=1}^{N}
        \left\{
        \prod_{j \ne l} 
            (\pi_{j,\lambda})^{a_{j}}
            (1 - \pi_{j,\lambda})^{1-a_{j}} 
        \right\}
        (2a_{l}-1)\textup{EIF}(\pi_{l,\lambda})
        .
    \end{align*}

    \item[\textbf{Step 5}]
    Uncentered EIF of $\mu(Q)$ and $\mu_t(Q)$: 
    \begin{align*}
        \varphi_{\mu(Q)}(\mathbf{O}; \boldsymbol{\eta})
        = &
        \frac{1}{N}
        \sum_{j=1}^{N}
        \begin{aligned}[t]
            \Bigg[  
            &
            \sum_{\mathbf{a} \in \mathcal{A}(N)}
                G_j \big(
                    \mathbf{a}, \mathbf{X}, N 
                \big)
                \times
                Q_{\scriptscriptstyle \textup{CMS}}(\mathbf{a} | \mathbf{X}, N; \lambda)
                \\
                & \qquad \quad
                \times
                \left\{
                    1 
                    +
                    \sum_{l=1}^{N}
                    \frac
                        {(2a_{l}-1) (A_{l} - \pi_{l}) (X^* \lambda + 1 - X^*)}
                        {(\pi_{l,\lambda})^{a_{l}} (1 - \pi_{l,\lambda})^{1-a_{l}}}
                \right\}
            \\
            & + 
            \frac
                {Q_{\scriptscriptstyle \textup{CMS}}(\mathbf{A} | \mathbf{X}, N; \lambda)}
                {H(\mathbf{A},\mathbf{X}, N)}
            \left\{ 
                Y_{j} 
                - 
                G_j \big( 
                    \mathbf{A}, \mathbf{X}, N 
                \big) 
            \right\}
            \Bigg]
        \end{aligned}
        \\
        \varphi_{\mu_t(Q)}(\mathbf{O}; \boldsymbol{\eta})
        = &
        \frac{1}{N}
        \sum_{j=1}^{N}
        \begin{aligned}[t]
            \Bigg[  
            &
            \sum_{\mathbf{a} \in \mathcal{A}(N)}
                G_j \big(
                    \mathbf{a}, \mathbf{X}, N 
                \big)
                \frac
                    {\mathbbm{1}(a_{j} = t)}
                    {\mathbb{P}_{\lambda}(a_{j}| \mathbf{X}, N)}
                \times
                Q_{\scriptscriptstyle \textup{CMS}}(\mathbf{a} | \mathbf{X}, N; \lambda)
                \\
                & \qquad \quad
                \times
                \left\{
                    1 
                    +
                    \sum_{l \ne j}
                    \frac
                        {(2a_{l}-1) (A_{l} - \pi_{l}) (X^* \lambda + 1 - X^*)}
                        {(\pi_{l,\lambda})^{a_{l}} (1 - \pi_{l,\lambda})^{1-a_{l}}}
                \right\}
            \\
            & + 
            \frac
                {\mathbbm{1}(A_{j} = t)}
                {\mathbb{P}_{\lambda}(A_{j}| \mathbf{X}, N)}
            \times
            \frac
                {Q_{\scriptscriptstyle \textup{CMS}}(\mathbf{A} | \mathbf{X}, N; \lambda)}
                {H(\mathbf{A},\mathbf{X}, N)}
            \left\{ 
                Y_{j} 
                - 
                G_j \big( 
                    \mathbf{A}, \mathbf{X}, N 
                \big) 
            \right\}
            \Bigg]
        \end{aligned}
    \end{align*}
        
    \item[\textbf{Step 7}] 
    Check large sample conditions:

    Similar to CIPS policy, conditional independence of $Y_{ij}$'s as well as that of $A_{ij}$'s are assumed under CMS policy.
    Therefore, individual-level nuisance functions $g$ and $\pi$ may be estimated instead of $G$ and $H$ and used to construct estimators of $G$, $H$, $w$, and $\phi$.
    First, (B5) $r_H = O(r_{\pi})$ and (B6) $r_G = O(r_g)$ because of the same reasoning under CIPS policy.
    Other conditions are given as follows:

    (B7) \textit{Convergence rate of} $\widehat{\phi}^{(-k)}$:
    $r_{\phi} = O(r_{\pi})$ since
    \begin{align*}
        &
        \widehat{\phi}_{Q_{\textup{CMS}}}(\mathbf{A}, \mathbf{X}, N; \mathbf{a})
        -
        \phi_{Q_{\textup{CMS}}}(\mathbf{A}, \mathbf{X}, N; \mathbf{a})
        \\
        = &
        \left(
            \widehat{Q}_{\scriptscriptstyle \textup{CMS}}
            -
            Q_{\scriptscriptstyle \textup{CMS}}
        \right)
        (\mathbf{a} | \mathbf{X}, N; \lambda)
        \sum_{l=1}^{N}
            \frac
                {(2a_{l}-1) (A_{l} - \widehat{\pi}_{l}) (X_l^* \lambda + 1 - X_l^*)}
                {(\widehat{\pi}_{l,\lambda})^{a_{l}} (1 - \widehat{\pi}_{l,\lambda})^{1-a_{l}}}
        \\
        & +
        Q_{\scriptscriptstyle \textup{CMS}}
        (\mathbf{a} | \mathbf{X}, N; \lambda)
        \sum_{l=1}^{N}
            (2a_{l}-1)(X_l^* \lambda + 1 - X_l^*)
            \left\{
                \frac
                    {A_{l} - \widehat{\pi}_{l}}
                    {(\widehat{\pi}_{l,\lambda})^{a_{l}} (1 - \widehat{\pi}_{l,\lambda})^{1-a_{l}}}
                -
                \frac
                    {A_{l} - \pi_{l}}
                    {(\pi_{l,\lambda})^{a_{l}} (1 - \pi_{l,\lambda})^{1-a_{l}}}
            \right\}
        \\
        \lesssim &
        \sum_{j=1}^{N}
        \lvert
            (\widehat{\pi}_j - \pi_j)
            (\mathbf{X}, N)
        \rvert
    \end{align*}
    which follows from
    \begin{align*}
        \left(
            \widehat{Q}_{\scriptscriptstyle \textup{CMS}}
            -
            Q_{\scriptscriptstyle \textup{CMS}}
        \right)
        (\mathbf{a} | \mathbf{X}, N; \lambda)
        \lesssim
        \sum_{j=1}^{N}
        \lvert
            \widehat{\mathbb{P}}_{\lambda}(a_{j}| \mathbf{X}, N)
            -
            \mathbb{P}_{\lambda}(a_{j}| \mathbf{X}, N)
        \rvert
        =
        \sum_{j=1}^{N}
        \lvert
            \widehat{\pi}_{j}
            -
            \pi_{j}
        \rvert
        \times (X_j^* \lambda + 1 - X_j^*)
    \end{align*}
    and
    \begin{align*}
        &
        \frac
            {A_{l} - \widehat{\pi}_{l}}
            {(\widehat{\pi}_{l,\lambda})^{a_{l}} (1 - \widehat{\pi}_{l,\lambda})^{1-a_{l}}}
        -
        \frac
            {A_{l} - \pi_{l}}
            {(\pi_{l,\lambda})^{a_{l}} (1 - \pi_{l,\lambda})^{1-a_{l}}}
        \\
        = &
        \frac
            {\pi_l - \widehat{\pi}_l}
            {\widehat{\mathbb{P}}_{\lambda}(a_{l}| \mathbf{X}, N)}
        +
        (A_l - \pi_l)
        \frac
            {(2a_l-1)(\pi_{l}-\widehat{\pi}_{l})(X_l^* \lambda + 1 - X_l^*)}
            {\mathbb{P}_{\lambda}(a_{l}| \mathbf{X}, N) \widehat{\mathbb{P}}_{\lambda}(a_{l}| \mathbf{X}, N)}
        .
    \end{align*}

    (B8) \textit{Second order convergence rate of} $\widehat{w}^{(-k)}$: 
    $r_{w} = O(r_{\pi})$ since
    \begin{align*}
        &
        \widehat{Q}_{\scriptscriptstyle \textup{CMS}}
        (\mathbf{a} | \mathbf{X}, N; \lambda)
        -
        Q_{\scriptscriptstyle \textup{CMS}}
        (\mathbf{a} | \mathbf{X}, N; \lambda)
        +
        \sum_{\mathbf{a}' \in \mathcal{A}(N)}
            \widehat{\phi}_{Q_{\textup{CMS}}}(\mathbf{a}', \mathbf{X}, N; \mathbf{a})
            H(\mathbf{a}', \mathbf{X}, N)
        \\
        = &
        \prod_{j=1}^N
            \widehat{\mathbb{P}}_{\lambda}(a_{j}| \mathbf{X}, N)
        -
        \prod_{j=1}^N
            \mathbb{P}_{\lambda}(a_{j}| \mathbf{X}, N)
        +
        \sum_{j=1}^N
            \left\{
            \prod_{l \ne j}
                \widehat{\mathbb{P}}_{\lambda}(a_{l}| \mathbf{X}, N)
            \right\}
            (2a_{j}-1) (\pi_{j} - \widehat{\pi}_{j})
            (X_j^* \lambda + 1 - X_j^*)
        \\
        =&
        \prod_{j=1}^N
            \widehat{\mathbb{P}}_{\lambda}(a_{j}| \mathbf{X}, N)
        -
        \prod_{j=1}^N
            \mathbb{P}_{\lambda}(a_{j}| \mathbf{X}, N)
        -
        \sum_{j=1}^N
            \left\{
            \prod_{l \ne j}
                \widehat{\mathbb{P}}_{\lambda}(a_{l}| \mathbf{X}, N)
            \right\}
            (\widehat{\mathbb{P}}_{\lambda} - \mathbb{P}_{\lambda})
            (a_{j}| \mathbf{X}, N)
        \\
        & +
        \sum_{j=1}^N
            \left\{
                \prod_{l \ne j}
                    \widehat{\mathbb{P}}_{\lambda}(a_{l}| \mathbf{X}, N)
            \right\}
            \left\{
                (\widehat{\mathbb{P}}_{\lambda} - \mathbb{P}_{\lambda})
                (a_{j}| \mathbf{X}, N)
                +
                (2a_{j}-1) (\pi_{j} - \widehat{\pi}_{j})
                (X_j^* \lambda + 1 - X_j^*)
            \right\}
        \\
        \lesssim &
        \sum_{j=1}^{N}
        \lvert
            \widehat{\mathbb{P}}_{\lambda}(a_{j}| \mathbf{X}, N)
            -
            \mathbb{P}_{\lambda}(a_{j}| \mathbf{X}, N)
        \rvert^2
        \\
        \lesssim &
        \sum_{j=1}^{N}
        \lvert
            (\widehat{\pi}_j - \pi_j)
            (\mathbf{X}, N)
        \rvert^2
    \end{align*}
    from
    \begin{align*}
        \sum_{\mathbf{a}' \in \mathcal{A}(N)}
            \widehat{\phi}_{Q_{\textup{CMS}}}(\mathbf{a}', \mathbf{X}, N; \mathbf{a})
            H(\mathbf{a}', \mathbf{X}, N)
        = &
        \E{\{}{\}}{
            \widehat{\phi}_{Q_{\textup{CMS}}}(\mathbf{A}, \mathbf{X}, N; \mathbf{a})
            \middle| D, \mathbf{X}, N
        }
        \\
        = &
        \widehat{Q}_{\scriptscriptstyle \textup{CMS}}
        (\mathbf{a} | \mathbf{X}, N; \lambda)
        \sum_{j=1}^{N}
            \frac
                {(2a_{j}-1) (\pi_{j} - \widehat{\pi}_{j}) (X_j^* \lambda + 1 - X_j^*)}
                {(\widehat{\pi}_{j,\lambda})^{a_{j}} (1 - \widehat{\pi}_{j,\lambda})^{1-a_{j}}}
    \end{align*}
    and
    $
        (\widehat{\mathbb{P}}_{\lambda} - \mathbb{P}_{\lambda})
        (a_{j}| \mathbf{X}, N)
        +
        (2a_{j}-1) (\pi_{j} - \widehat{\pi}_{j})
        (X_j^* \lambda + 1 - X_j^*)
        = 0
    $.
    
    Thus, sufficient conditions for consistency and asymptotic normality are as follows:
    
    (i) Consistency: $r_{\pi} = o(1)$;
    
    (ii) Asymptotic normality: $r_{\pi} = o(m^{-1/4})$ and $r_{\pi} \cdot r_g = o(m^{-1/2})$;
    
    (iii) Consistent variance estimator: $r_{\pi} = r_g = o(1)$ as $m \to \infty$.
    
    
\end{enumerate}

\subsection{Treated proportion bound policy}

\begin{enumerate}

    \item[\textbf{Step 1}]
    Policy distribution $Q$:  
    $$
    Q_{\scriptscriptstyle \textup{TPB}}(\mathbf{a} | \mathbf{X}, N; \rho) 
    =
    \mathbbm{1}(\overline{\mathbf{a}} \ge \rho)
    \frac
        {H(\mathbf{a}, \mathbf{X}, N)}
        {\mathbb{P}(\overline{\mathbf{A}} \ge \rho | \mathbf{X}, N)}
    ,
    $$
    where
    $\mathbb{P}(\overline{\mathbf{A}} \ge \rho | \mathbf{X}, N)
    =
    \sum_{\overline{\mathbf{a}}' \ge \rho} 
        H(\mathbf{a}', \mathbf{X}, N)
    $
    is the observed probability of the proportion of treatment unit in a cluster to be at least $\rho$

    \item[\textbf{Step 2}]
    EIF $\phi_Q$:
    \begin{align*}
        \phi_{Q_{\textup{TPB}}}(\mathbf{A}, \mathbf{X}, N; \mathbf{a})
        =
        \frac
            {\mathbbm{1}(\overline{\mathbf{a}} \ge \rho)}
            {\mathbb{P}(\overline{\mathbf{A}} \ge \rho | \mathbf{X}, N)^2}
        \left\{
            \mathbbm{1}(\mathbf{A} = \mathbf{a})
            \mathbb{P}(\overline{\mathbf{A}} \ge \rho | \mathbf{X}, N)
            -
            \mathbbm{1}(\overline{\mathbf{A}} \ge \rho)
            H(\mathbf{a}, \mathbf{X}, N)
        \right\}
    \end{align*}
    from the fact that the EIF of 
    $H(\mathbf{a}, \mathbf{x}, n)$ is
    \begin{align*}
        \textup{EIF}(H(\mathbf{a}, \mathbf{x}, n))
        =
        \frac
            {\mathbbm{1}(\mathbf{X} = \mathbf{x}, N = n)}
            {d\mathbb{P}(\mathbf{x}, n)}
        \{
            \mathbbm{1}(\mathbf{A} = \mathbf{a}) 
            - 
            H(\mathbf{a}, \mathbf{X}, N)
        \}        
        ,
    \end{align*}
    and the EIF of 
    $\mathbb{P}(\overline{\mathbf{A}} \ge \rho | \mathbf{x}, n)
    =
    \sum_{\overline{\mathbf{a}}' \ge \rho} 
        H(\mathbf{a}', \mathbf{x}, n)
    $
    is
    \begin{align*}
        \textup{EIF}(
            \mathbb{P}(\overline{\mathbf{A}} \ge \rho | \mathbf{x}, n)
        )
        = &
        \sum_{\overline{\mathbf{a}}' \ge \rho} 
            \frac
                {\mathbbm{1}(\mathbf{X} = \mathbf{x}, N = n)}
                {d\mathbb{P}(\mathbf{x}, n)}
            \{
                \mathbbm{1}(\mathbf{A} = \mathbf{a}') 
                - 
                H(\mathbf{a}', \mathbf{X}, N)
            \}    
        \\
        = &
        \frac
            {\mathbbm{1}(\mathbf{X} = \mathbf{x}, N = n)}
            {d\mathbb{P}(\mathbf{x}, n)}
        \{
            \mathbbm{1}(\overline{\mathbf{A}} \ge \rho)
            -
            \mathbb{P}(\overline{\mathbf{A}} \ge \rho | \mathbf{x}, n)
        \}
        .
    \end{align*}

    \item[\textbf{Step 5}]
    Uncentered EIF of $\mu(Q)$ and $\mu_t(Q)$: 
    \begin{align*}
        \varphi_{\mu(Q)}(\mathbf{O}; \boldsymbol{\eta})
        = &
        \frac{1}{N}
        \sum_{j=1}^{N}
        \begin{aligned}[t]
            \Bigg[  
            &
            \sum_{\mathbf{a} \in \mathcal{A}(N)}
                G_j \big(
                    \mathbf{a}, \mathbf{X}, N 
                \big)
                \times
                Q_{\scriptscriptstyle \textup{TPB}}(\mathbf{a} | \mathbf{X}, N; \rho)
                \times
                \left\{
                    1 
                    +
                    \frac
                        {\mathbbm{1}(\mathbf{A} = \mathbf{a})}
                        {H(\mathbf{a}, \mathbf{X}, N)}
                    -
                    \frac
                        {\mathbbm{1}(\overline{\mathbf{A}} \ge \rho)}
                        {\mathbb{P}(\overline{\mathbf{A}} \ge \rho | \mathbf{X}, N)}
                \right\}
            \\
            & + 
            \frac
                {\mathbbm{1}(\overline{\mathbf{A}} \ge \rho)}
                {\mathbb{P}(\overline{\mathbf{A}} \ge \rho | \mathbf{X}, N)}
            \left\{ 
                Y_{j} 
                - 
                G_j \big( 
                    \mathbf{A}, \mathbf{X}, N 
                \big) 
            \right\}
            \Bigg]
        \end{aligned}
        \\
        \varphi_{\mu_t(Q)}(\mathbf{O}; \boldsymbol{\eta})
        = &
        \frac{1}{N}
        \sum_{j=1}^{N}
        \begin{aligned}[t]
            \Bigg[  
            &
            \sum_{\mathbf{a} \in \mathcal{A}(N)}
                G_j \big(
                    \mathbf{a}, \mathbf{X}, N 
                \big)
                \mathbbm{1}(a_{j} = t)
                \times
                \begin{aligned}[t]
                    \Big\{
                        &
                        Q_{\scriptscriptstyle \textup{TPB}}(\mathbf{a} | \mathbf{X}, N; \rho)
                        \left(
                            1 
                            -
                            \frac
                                {\mathbbm{1}(\overline{\mathbf{A}} \ge \rho)}
                                {\mathbb{P}(\overline{\mathbf{A}} \ge \rho | \mathbf{X}, N)}
                        \right)
                        \\
                        &
                        +
                        \frac
                            {\mathbbm{1}(\overline{\mathbf{A}} \ge \rho)
                            \mathbbm{1}(\mathbf{A}_{(-j)} = \mathbf{a}_{(-j)})}
                            {\mathbb{P}(\overline{\mathbf{A}} \ge \rho | \mathbf{X}, N)}
                    \Big\}
                \end{aligned}
            \\
            & + 
            \frac
                {\mathbbm{1}(A_{j} = t)
                Q_{\scriptscriptstyle \textup{TPB}}(\mathbf{A}_{(-j)} | \mathbf{X}, N; \rho)}
                {H(\mathbf{A},\mathbf{X}, N)}
            \left\{ 
                Y_{j} 
                - 
                G_j \big( 
                    \mathbf{A}, \mathbf{X}, N 
                \big) 
            \right\}
            \Bigg]
        \end{aligned}
    \end{align*}
        
    \item[\textbf{Step 7}] 
    Check large sample conditions:
    
    (B7) \textit{Convergence rate of} $\widehat{\phi}^{(-k)}$:
    $r_{\phi} = O(r_{H})$ since
    \begin{align*}
        &
        \widehat{\phi}_{Q_{\textup{TPB}}}(\mathbf{A}, \mathbf{X}, N; \mathbf{a})
        -
        \phi_{Q_{\textup{TPB}}}(\mathbf{A}, \mathbf{X}, N; \mathbf{a})
        \\
        = &
        \mathbbm{1}(\overline{\mathbf{a}} \ge \rho)
        \begin{aligned}[t]
            \Bigg[
                &
                \mathbbm{1}(\mathbf{A} = \mathbf{a})
                \left\{
                    \frac
                        {1}
                        {\widehat{\mathbb{P}}(\overline{\mathbf{A}} \ge \rho | \mathbf{X}, N)}
                    -
                    \frac
                        {1}
                        {\mathbb{P}(\overline{\mathbf{A}} \ge \rho | \mathbf{X}, N)}
                \right\}
                \\
                & -
                \mathbbm{1}(\overline{\mathbf{A}} \ge \rho)
                \left\{
                    \frac
                        {\widehat{H}(\mathbf{a}, \mathbf{X}, N)}
                        {\widehat{\mathbb{P}}(\overline{\mathbf{A}} \ge \rho | \mathbf{X}, N)^2}
                    -
                    \frac
                        {H(\mathbf{a}, \mathbf{X}, N)}
                        {\mathbb{P}(\overline{\mathbf{A}} \ge \rho | \mathbf{X}, N)^2}
                \right\}
            \Bigg]
        \end{aligned}
        \\
        \lesssim &
        \lvert
            (\widehat{\mathbb{P}}-\mathbb{P})(\overline{\mathbf{A}} \ge \rho | \mathbf{X}, N)
        \rvert
        +
        \lvert
            (\widehat{H}-H)(\mathbf{a}, \mathbf{X}, N)
        \rvert
        +
        \lvert
            \widehat{\mathbb{P}}(\overline{\mathbf{A}} \ge \rho | \mathbf{X}, N)^2
            -
            \mathbb{P}(\overline{\mathbf{A}} \ge \rho | \mathbf{X}, N)^2
        \rvert
        \\
        \lesssim &
        \suma
            \lvert
                (\widehat{H}-H)(\mathbf{a}, \mathbf{X}, N)
            \rvert
    \end{align*}
    which follows from
    \begin{align*}
        \lvert
            (\widehat{\mathbb{P}}-\mathbb{P})(\overline{\mathbf{A}} \ge \rho | \mathbf{X}, N)
        \rvert
        =
        \sum_{\overline{\mathbf{a}}' \ge \rho} 
            \lvert
                (\widehat{H}-H)(\mathbf{a}', \mathbf{X}, N)
            \rvert
    \end{align*}
    and the inequality 
    $
    \lvert
        x_1/y_1 - x_2/y_2
    \rvert
    \lesssim
    \lvert
        x_1 - x_2
    \rvert
    +
    \lvert
        y_1 - y_2
    \rvert
    $
    if $y_1, y_2$ are bounded.

    (B8) \textit{Second order convergence rate of} $\widehat{w}^{(-k)}$: 
    $r_{w} = O(r_{H})$ since
    \begin{align*}
        &
        \widehat{Q}_{\scriptscriptstyle \textup{TPB}}
        (\mathbf{a} | \mathbf{X}, N; \rho)
        -
        Q_{\scriptscriptstyle \textup{TPB}}
        (\mathbf{a} | \mathbf{X}, N; \rho)
        +
        \sum_{\mathbf{a}' \in \mathcal{A}(N)}
            \widehat{\phi}_{Q_{\textup{TPB}}}(\mathbf{a}', \mathbf{X}, N; \mathbf{a})
            H(\mathbf{a}', \mathbf{X}, N)
        \\
        = &
        \mathbbm{1}(\overline{\mathbf{a}} \ge \rho)
        \left\{
            \frac
                {\widehat{H}(\mathbf{a}, \mathbf{X}, N)}
                {\widehat{\mathbb{P}}(\overline{\mathbf{A}} \ge \rho | \mathbf{X}, N)}
            -
            \frac
                {H(\mathbf{a}, \mathbf{X}, N)}
                {\mathbb{P}(\overline{\mathbf{A}} \ge \rho | \mathbf{X}, N)}
            +
            \frac
                {H(\mathbf{a}, \mathbf{X}, N)}
                {\widehat{\mathbb{P}}(\overline{\mathbf{A}} \ge \rho | \mathbf{X}, N)}
            -
            \frac
                {\mathbb{P}(\overline{\mathbf{A}} \ge \rho | \mathbf{X}, N)
                \widehat{H}(\mathbf{a}, \mathbf{X}, N)}
                {\widehat{\mathbb{P}}(\overline{\mathbf{A}} \ge \rho | \mathbf{X}, N)^2}
        \right\}
        \\
        = &
        \mathbbm{1}(\overline{\mathbf{a}} \ge \rho)
        \left\{
            \frac
                {\widehat{H}(\mathbf{a}, \mathbf{X}, N)}
                {\widehat{\mathbb{P}}(\overline{\mathbf{A}} \ge \rho | \mathbf{X}, N)}
            -
            \frac
                {H(\mathbf{a}, \mathbf{X}, N)}
                {\mathbb{P}(\overline{\mathbf{A}} \ge \rho | \mathbf{X}, N)}
        \right\}
        \left\{
            1
            -
            \frac
                {\mathbb{P}(\overline{\mathbf{A}} \ge \rho | \mathbf{X}, N)}
                {\widehat{\mathbb{P}}(\overline{\mathbf{A}} \ge \rho | \mathbf{X}, N)}
        \right\}
        \\
        \lesssim &
        \left\{
            \suma
                \lvert
                    (\widehat{H}-H)(\mathbf{a}, \mathbf{X}, N)
                \rvert
        \right\}^2
    \end{align*}
    from
    \begin{align*}
        &
        \sum_{\mathbf{a}' \in \mathcal{A}(N)}
            \widehat{\phi}_{Q_{\textup{TPB}}}(\mathbf{a}', \mathbf{X}, N; \mathbf{a})
            H(\mathbf{a}', \mathbf{X}, N)
        \\
        = &
        \E{\{}{\}}{
            \widehat{\phi}_{Q_{\textup{TPB}}}(\mathbf{A}, \mathbf{X}, N; \mathbf{a})
            \middle| D, \mathbf{X}, N
        }
        \\
        =&
        \mathbbm{1}(\overline{\mathbf{a}} \ge \rho)
        \E{\{}{\}}{
            \frac
                {\mathbbm{1}(\mathbf{A} = \mathbf{a})}
                {\widehat{\mathbb{P}}(\overline{\mathbf{A}} \ge \rho | \mathbf{X}, N)}
            -
            \mathbbm{1}(\overline{\mathbf{A}} \ge \rho)
            \frac
                {\widehat{H}(\mathbf{a}, \mathbf{X}, N)}
                {\widehat{\mathbb{P}}(\overline{\mathbf{A}} \ge \rho | \mathbf{X}, N)^2}
            \middle| D, \mathbf{X}, N
        }
        \\
        = &
        \mathbbm{1}(\overline{\mathbf{a}} \ge \rho)
        \left\{
            \frac
                {H(\mathbf{a}, \mathbf{X}, N)}
                {\widehat{\mathbb{P}}(\overline{\mathbf{A}} \ge \rho | \mathbf{X}, N)}
            -
            \frac
                {\mathbb{P}(\overline{\mathbf{A}} \ge \rho | \mathbf{X}, N)
                \widehat{H}(\mathbf{a}, \mathbf{X}, N)}
                {\widehat{\mathbb{P}}(\overline{\mathbf{A}} \ge \rho | \mathbf{X}, N)^2}
        \right\}
    .
    \end{align*}
    
    Thus, sufficient conditions for consistency and asymptotic normality are as follows:
    
    (i) Consistency: $r_{H} = o(1)$;
    
    (ii) Asymptotic normality: $r_{H} = o(m^{-1/4})$ and $r_{H} \cdot r_G = o(m^{-1/2})$;
    
    (iii) Consistent variance estimator: $r_H = r_G = o(1)$ as $m \to \infty$.
    
\end{enumerate}

\newpage

\section{Additional Simulation results}

In this section, additional simulation results are presented which are not included in the main text.

\subsection{CIPS policy with varying $\delta(\mathbf{X}_i, N_i) = \delta_0 (1+1/N_i)$}

The simulation results for CIPS policy with varying 
$\delta(\mathbf{X}_i, N_i) = \delta_0 (1+1/N_i)$,
$\delta_0 \in \{0.5, 1, 2\}$ 
are given in Table~\ref{tab:simulCIPSvary}.
The simulation setting is the same as the main text.
As discussed in the main text,
the nonparametric estimators performed well, 
while the parametric estimator performed poorly. 

\begin{table}[H]
\caption{\small Simulation results for nonparametric and parametric sample splitting estimators for CIPS policy with varying $\delta$}
\label{tab:simulCIPSvary}
\resizebox{\textwidth}{!}{%
\begin{tabular}{cccccccccccccccc}
\hline
 &  &  & \multicolumn{5}{c}{Nonparametric} &  & \multicolumn{5}{c}{Parametric} &  & RMSE \\ \cline{4-8} \cline{10-14}
Estimand & Truth &  & Bias & RMSE & ASE & ESE & Cov \% &  & Bias & RMSE & ASE & ESE & Cov \% &  & Ratio \\ \hline
$\mu_{\scriptscriptstyle \textup{CIPS}}(0.5)$ & 0.426 &  & 0.003 & 0.016 & 0.016 & 0.016 & 95.2\% &  & 0.011 & 0.021 & 0.017 & 0.018 & 90.3\% &  & 0.76 \\
$\mu_{\scriptscriptstyle \textup{CIPS}, \scriptstyle 1}(0.5)$ & 0.261 &  & -0.001 & 0.014 & 0.014 & 0.014 & 95.0\% &  & -0.010 & 0.018 & 0.015 & 0.015 & 88.5\% &  & 0.78 \\
$\mu_{\scriptscriptstyle \textup{CIPS}, \scriptstyle 0}(0.5)$ & 0.551 &  & 0.006 & 0.022 & 0.021 & 0.021 & 94.6\% &  & 0.034 & 0.042 & 0.023 & 0.025 & 65.6\% &  & 0.52 \\
$DE_{\scriptscriptstyle \textup{CIPS}}(0.5)$ & -0.290 &  & -0.007 & 0.022 & 0.021 & 0.021 & 93.7\% &  & -0.044 & 0.050 & 0.023 & 0.024 & 49.0\% &  & 0.44 \\
$SE_{\scriptscriptstyle \textup{CIPS}, \scriptstyle 1}(0.5, 1)$ & 0.021 &  & 0.001 & 0.011 & 0.011 & 0.011 & 95.2\% &  & 0.004 & 0.013 & 0.013 & 0.012 & 95.5\% &  & 0.85 \\
$SE_{\scriptscriptstyle \textup{CIPS}, \scriptstyle 0}(0.5, 1)$ & 0.025 &  & 0.003 & 0.019 & 0.017 & 0.018 & 93.7\% &  & 0.006 & 0.022 & 0.020 & 0.021 & 93.2\% &  & 0.84 \\
$OE_{\scriptscriptstyle \textup{CIPS}}(0.5, 1)$ & 0.072 &  & 0.003 & 0.013 & 0.012 & 0.013 & 92.9\% &  & 0.013 & 0.020 & 0.014 & 0.015 & 83.3\% &  & 0.68 \\
$TE_{\scriptscriptstyle \textup{CIPS}}(0.5, 1)$ & -0.265 &  & -0.005 & 0.016 & 0.016 & 0.016 & 94.1\% &  & -0.038 & 0.042 & 0.018 & 0.017 & 43.3\% &  & 0.39 \\ \hline
$\mu_{\scriptscriptstyle \textup{CIPS}}(1)$ & 0.354 &  & 0.000 & 0.012 & 0.012 & 0.012 & 94.8\% &  & -0.002 & 0.011 & 0.012 & 0.011 & 95.3\% &  & 1.03 \\
$\mu_{\scriptscriptstyle \textup{CIPS}, \scriptstyle 1}(1)$ & 0.240 &  & -0.002 & 0.011 & 0.011 & 0.011 & 94.6\% &  & -0.014 & 0.017 & 0.011 & 0.011 & 73.3\% &  & 0.65 \\
$\mu_{\scriptscriptstyle \textup{CIPS}, \scriptstyle 0}(1)$ & 0.526 &  & 0.004 & 0.015 & 0.015 & 0.015 & 94.8\% &  & 0.028 & 0.032 & 0.016 & 0.015 & 56.1\% &  & 0.48 \\
$DE_{\scriptscriptstyle \textup{CIPS}}(1)$ & -0.287 &  & -0.006 & 0.014 & 0.013 & 0.012 & 93.1\% &  & -0.042 & 0.044 & 0.014 & 0.013 & 13.5\% &  & 0.31 \\ \hline
$\mu_{\scriptscriptstyle \textup{CIPS}}(2)$ & 0.293 &  & -0.003 & 0.023 & 0.022 & 0.023 & 94.7\% &  & -0.011 & 0.025 & 0.021 & 0.022 & 90.1\% &  & 0.95 \\
$\mu_{\scriptscriptstyle \textup{CIPS}, \scriptstyle 1}(2)$ & 0.222 &  & -0.004 & 0.024 & 0.023 & 0.024 & 94.2\% &  & -0.017 & 0.028 & 0.022 & 0.022 & 83.7\% &  & 0.86 \\
$\mu_{\scriptscriptstyle \textup{CIPS}, \scriptstyle 0}(2)$ & 0.504 &  & 0.001 & 0.025 & 0.025 & 0.025 & 95.0\% &  & 0.022 & 0.034 & 0.027 & 0.026 & 88.4\% &  & 0.74 \\
$DE_{\scriptscriptstyle \textup{CIPS}}(2)$ & -0.283 &  & -0.005 & 0.025 & 0.024 & 0.025 & 94.2\% &  & -0.039 & 0.047 & 0.025 & 0.025 & 65.4\% &  & 0.54 \\
$SE_{\scriptscriptstyle \textup{CIPS}, \scriptstyle 1}(2, 1)$ & -0.018 &  & -0.001 & 0.020 & 0.018 & 0.019 & 93.6\% &  & -0.004 & 0.019 & 0.017 & 0.018 & 93.3\% &  & 1.04 \\
$SE_{\scriptscriptstyle \textup{CIPS}, \scriptstyle 0}(2, 1)$ & -0.022 &  & -0.002 & 0.020 & 0.019 & 0.019 & 94.4\% &  & -0.006 & 0.022 & 0.021 & 0.021 & 93.3\% &  & 0.90 \\
$OE_{\scriptscriptstyle \textup{CIPS}}(2, 1)$ & -0.061 &  & -0.002 & 0.019 & 0.018 & 0.019 & 93.3\% &  & -0.010 & 0.021 & 0.017 & 0.018 & 88.6\% &  & 0.92 \\
$TE_{\scriptscriptstyle \textup{CIPS}}(2, 1)$ & -0.304 &  & -0.007 & 0.026 & 0.024 & 0.025 & 92.6\% &  & -0.045 & 0.052 & 0.024 & 0.024 & 48.2\% &  & 0.50 \\ \hline
\end{tabular}
}
\caption*{\footnotesize RMSE: root mean squared error, ASE: average standard error estimates, ESE: standard deviation of estimates, Cov \%: 95\% CI coverage, RMSE Ratio: RMSE ratio of nonparametric and parametric estimators}
\end{table}

\subsection{Comparison of NSS and IPW estimators}

\begin{figure}[]
 \centerline{\includegraphics[width = 0.9\textwidth]{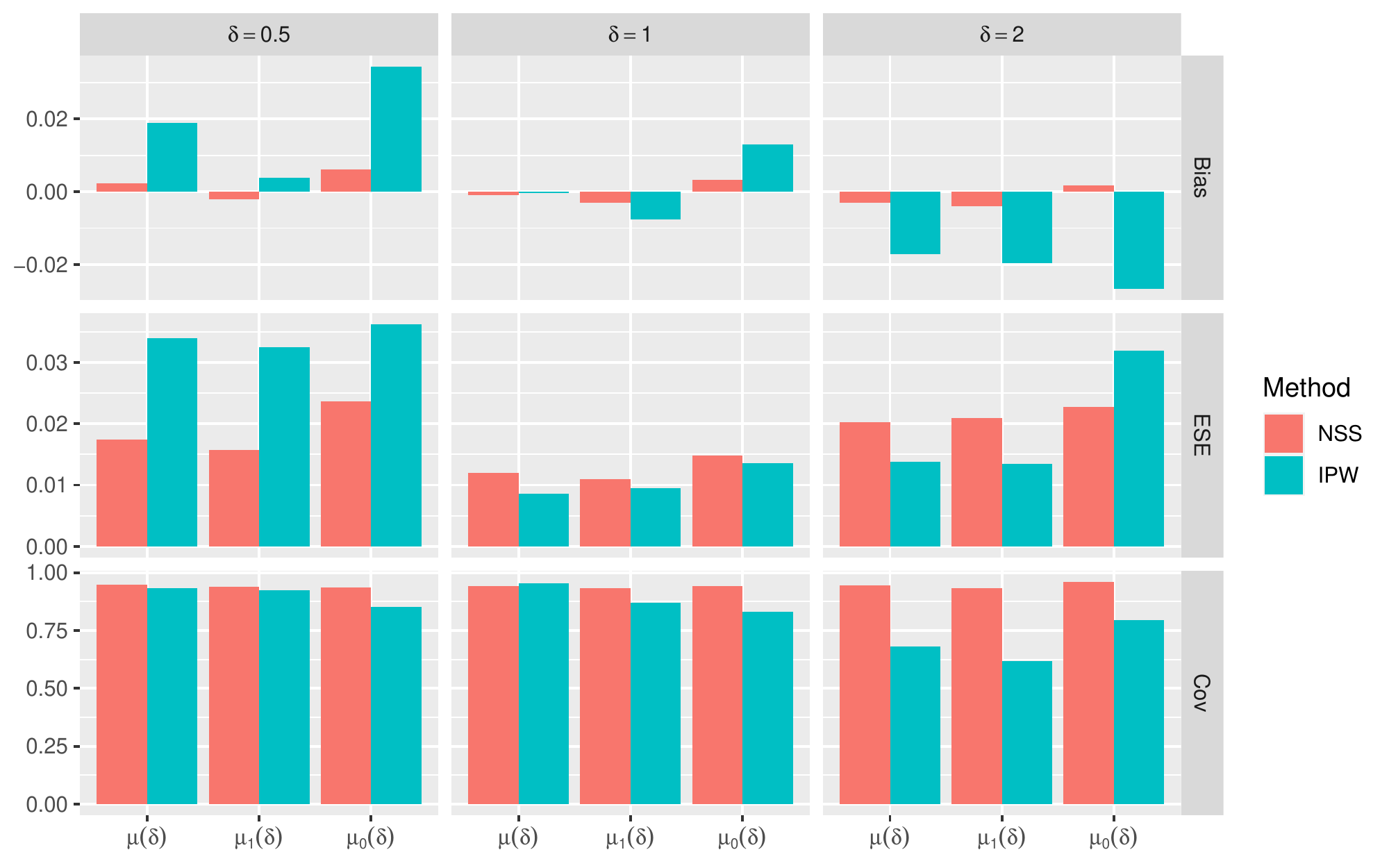}}
    \caption{\footnotesize
    Finite sample performance of the proposed nonparametric sample splitting (NSS) and Barkley estimators for CIPS policy.
    }
    \label{fig:CIPSproIPW}
\end{figure}

The CIPS policy with constant $\delta$ can be viewed as the nonparametric counterpart of the policy in \citet{Papadogeorgou19} and \citet{barkley20}.
\citet{Papadogeorgou19} and \citet{barkley20} proposed parametric inverse propensity weighted (IPW) estimators, given by
$$
\widehat{\Psi}^{\textup{IPW}}(w) 
=
\frac{1}{m}
\sum_{i = 1}^{m}
    \frac
        {\widehat{w}(\mathbf{A}_i, \mathbf{X}_i, N_i)^\top
        \mathbf{Y}_i }
        {\widehat{H}(\mathbf{A}_i,\mathbf{X}_i, N_i)}
,
$$
where $\widehat{H}$ and $\widehat{w}$ denote estimators of $H$ and $w$ based on generalized linear mixed models.
The finite sample performance of the proposed NSS estimators and the IPW estimator was compared using the same simulation setting as described in the main text. 
For both estimators the bias, empirical SE (ESE) and corresponding 95\% point-wise Wald confidence interval coverage (Cov) were computed for $\mu(\delta), \mu_1(\delta), \mu_0(\delta)$.
The results presented in Figure~\ref{fig:CIPSproIPW} show the proposed NSS estimator had smaller bias, smaller ESE, and 95\% CI coverage closer to the nominal level compared to the IPW estimator. 
These results are not surprising given mis-specification of the nuisance functions by the IPW estimator. 

\subsection{Finite sample performance over $r$}
\label{simul:r}

\begin{figure}[H]
 \centerline{\includegraphics[width = \textwidth]{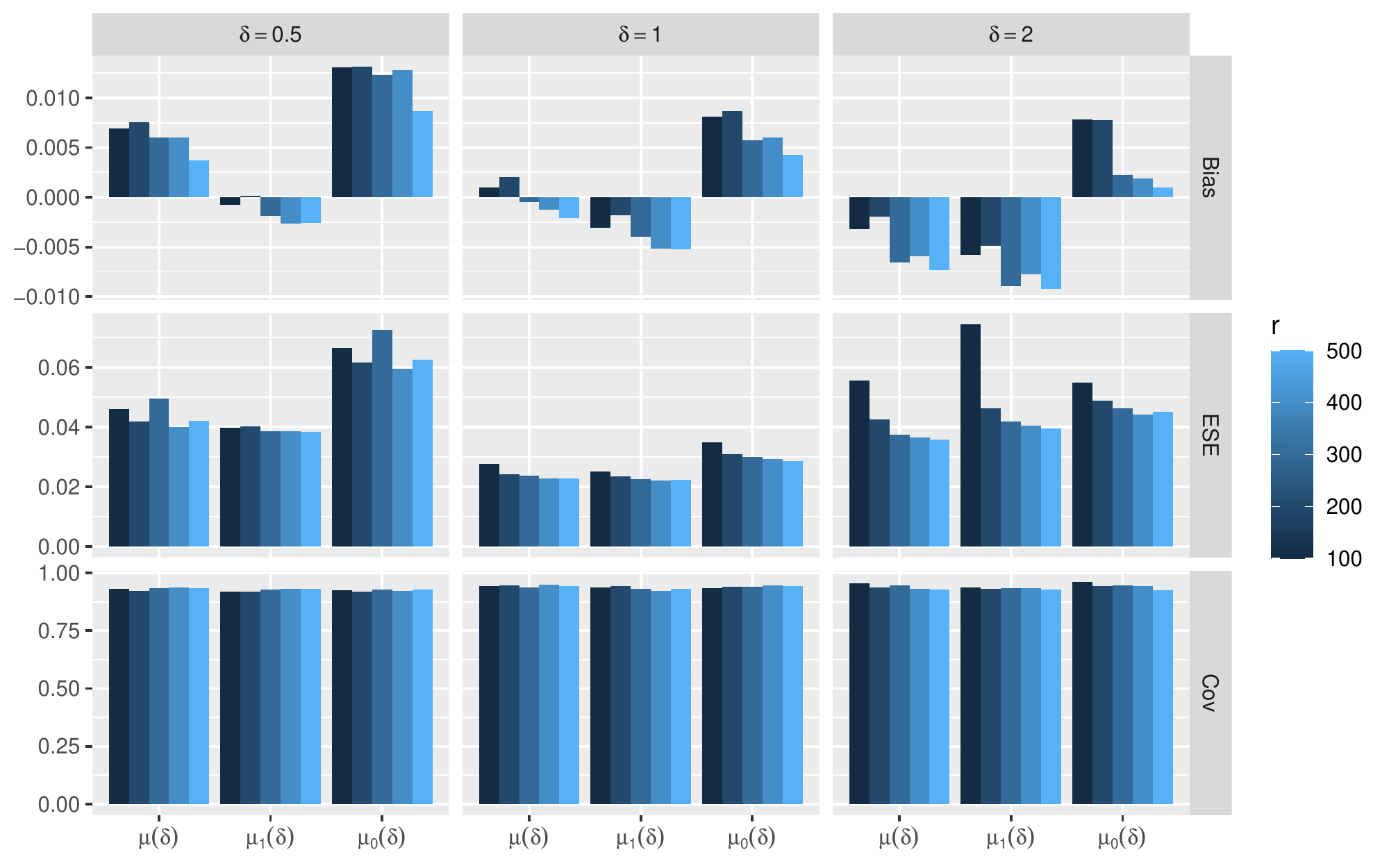}}
 \caption{\footnotesize
    Finite sample performance of the proposed NSS estimator over $r = 100, 200, 300, 400, 500$
    }
    \label{fig:Simulr}
\end{figure}

The finite sample performance of the proposed NSS estimators under the subsampling approximation was investigated, and the results are presented in Figure~\ref{fig:Simulr}.
Bias, empirical SE, and 95\% CI coverage were computed for different CIPS policy estimands with constant $\delta \in \{0.5, 1, 2\}$
over $r = 100, 200, 300, 400, 500$ when $m = 500$ (number of clusters).
The bias was insensitive to $r$, 
but the empirical SE of the estimators tended to decrease in $r$, while the 95\% CI coverage achieved the nominal level regardless of $r$.

\subsection{Finite sample performance over distribution of $N_i$}

\begin{figure}[H]
 \centerline{\includegraphics[width = \textwidth]{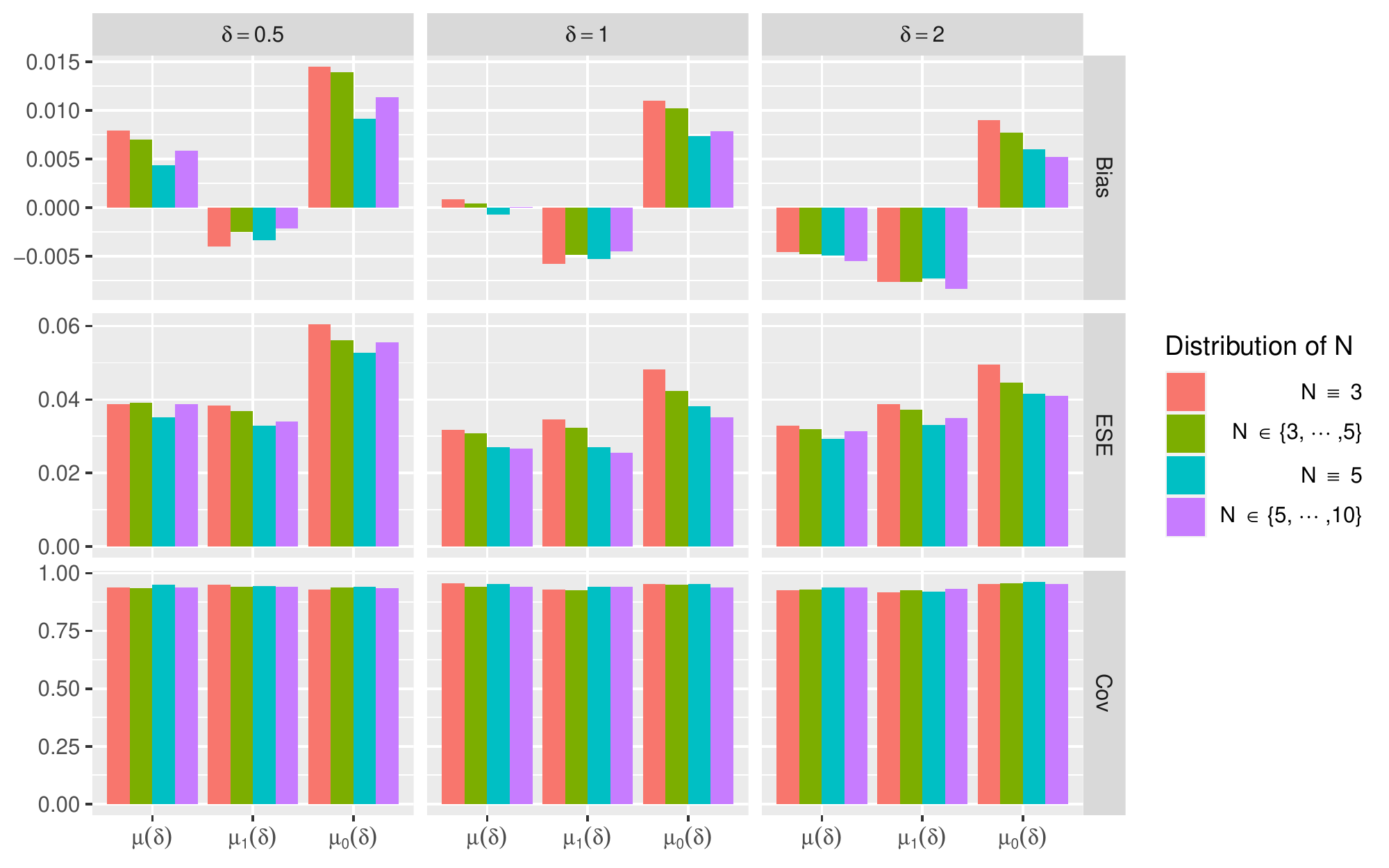}}
 \caption{\footnotesize
    Finite sample performance of the proposed NSS estimator over various distribution of $N_i$ (cluster size)
    }
\label{fig:SimulNdist}
\end{figure}

The finite sample performance of the NSS estimators under various $N_i$ distributions was investigated.
Bias, empirical SE, and 95\% CI coverage were computed for different CIPS policy estimands with constant $\delta \in \{0.5, 1, 2\}$ and $m=500$
over the following $N_i$ distributions:
(i) $N_i \equiv 3$; 
(ii) $\mathbb{P}(N_i = n) = 1/3$ for $n = 3,4,5$;
(iii) $N_i \equiv 5$; 
(iv) $\mathbb{P}(N_i = n) = 1/6$ for $n = 5,\dots,10$.
The results are presented in Figure~\ref{fig:SimulNdist}.
The bias of the NSS estimator tended to be small and the 95\% CI coverage achieved the nominal level for all scenarios,
demonstrating that the proposed inference procedure is robust to the distribution of $N_i$.

\newpage
\section{Details of Senegal DHS data analysis}

\subsection{Data distribution}

\begin{figure}[H]
 \centerline{\includegraphics[width = \textwidth]{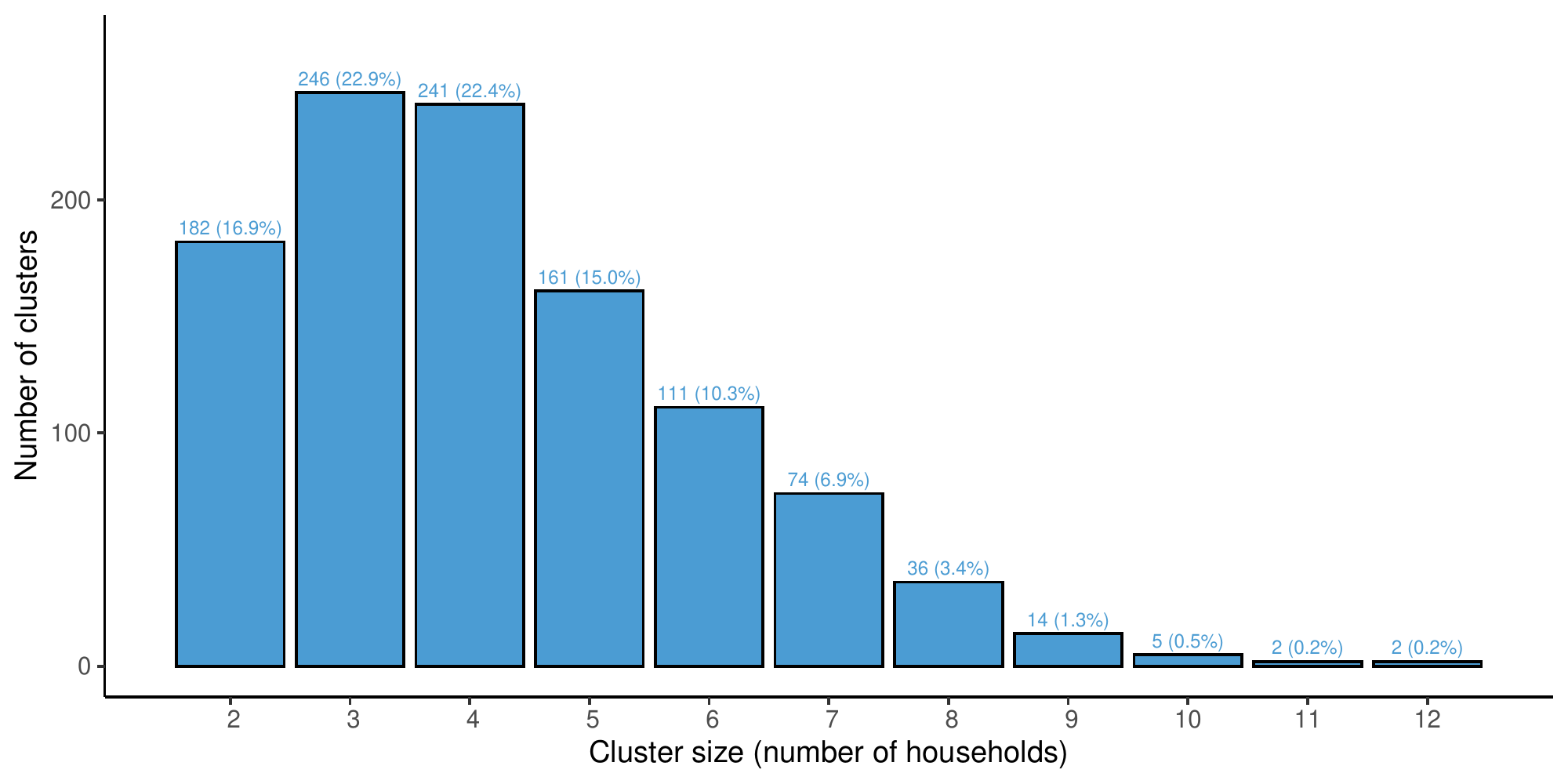}}
 \caption{\footnotesize
    Distribution of cluster size (number of households) from the analysis of the Senegal DHS data
    }
\label{fig:SenegalDHS_Ndist}
\end{figure}

Restricting the sample to households composed of complete data resulted in 1,074 clusters with 4,565 households (mean cluster size: 4.25).
Figure \ref{fig:SenegalDHS_Ndist} shows the distribution of the cluster size (number of households in a census block).
Most clusters include at most 5 households (77.2\%).

\begin{figure}[H]
 \centerline{\includegraphics[width = \textwidth]{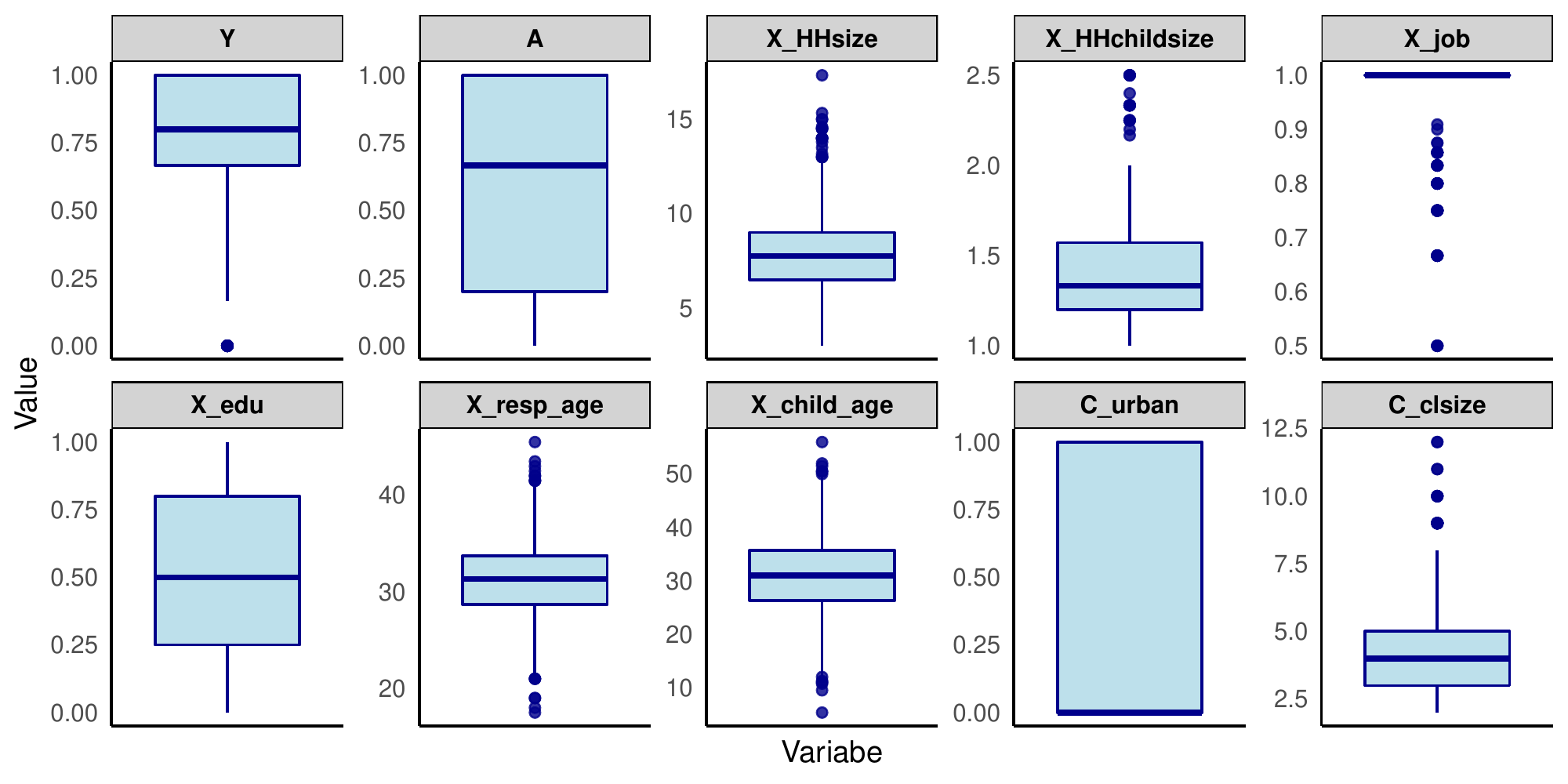}}
 \caption{\footnotesize
    Distribution of cluster-level averaged data values
    from the analysis of the Senegal DHS data.
    Y: diarrhea-free status,
    A: WASH facility,
    X\_HHsize: household size,
    X\_HHchildsize: number of children,
    X\_job: parents had a job,
    X\_edu: parents ever attended school,
    X\_resp\_age: mother's age,
    X\_child\_age: average age of children,
    C\_urban: cluster is at urban,
    C\_clsize: cluster size
    }
\label{fig:SenegalDHS_datadist}
\end{figure}

For each cluster, 
outcome (diarrhea-free status), 
treatment (WASH facility), 
and covariates (household size, number of children, whether parents had a job, whether parents ever attended school, mother’s age, average age of children, cluster size, whether the cluster was in an urban area) 
were averaged across households within the cluster and the distribution of cluster-level averages are shown in Figure \ref{fig:SenegalDHS_datadist}.
77.6\% of households were diarrhea-free, and 58.9\% of households had WASH facilities.
The average household size (number of family members) was 7.8, while the average number of children in a household was 1.3.
Only 1.1\% of parents did not have a job, and 53\% of parents ever attended to school.
Mother's average age was 31.2 years old, and children's average age was 30.9 months old.
Finally, 45\% of clusters were located in urban areas, 
and average cluster size was 4.25.

\subsection{Weights of super learner ensemble estimator}

\begin{figure}[H]
 \centerline{\includegraphics[width = \textwidth]{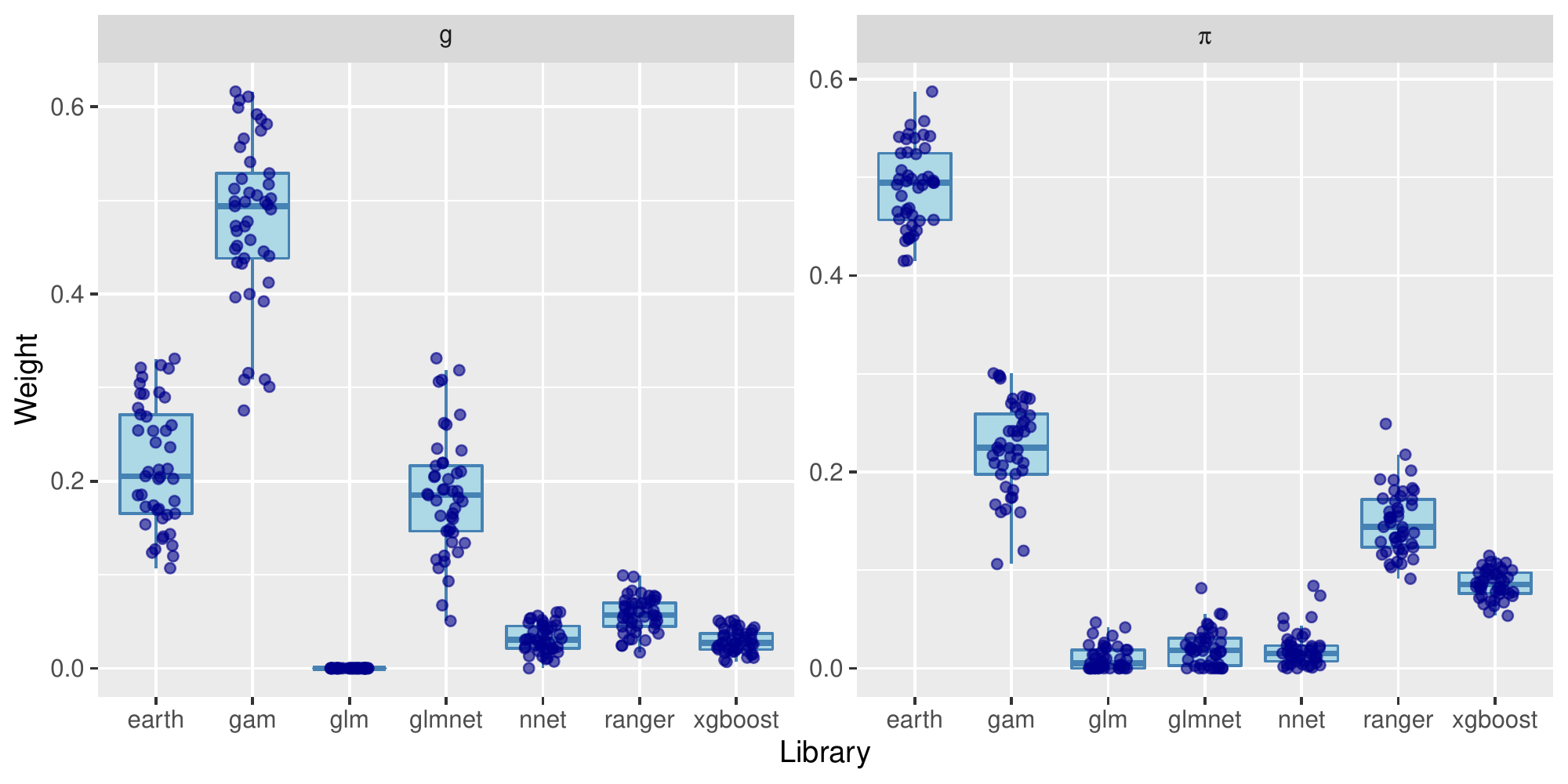}}
 \caption{\footnotesize
    Weights of estimators in Super Learner ensemble library for individual-level outcome regression model $g$ and individual-level propensity score model $\pi$ from the analysis of the Senegal DHS with $S = 30$.
    }
\label{fig:SenegalDHS_SLWeight}
\end{figure}

For nuisance function estimation, individual-level nuisance functions $g: Y_{ij} \sim A_{ij} + \overline{\mathbf{A}}_{i(-j)} + \mathbf{X}_{ij}$ 
and 
$\pi: A_{ij} \sim \mathbf{X}_{ij}$ were estimated,
where $j = 1, \dots, N_i$; $i = 1, \dots, 1074$,
$A_{ij}$ is the WASH facility status, 
and
$\mathbf{X}_{ij}$ is a covariates vector of $j$ th household in $i$ th cluster.
Both nuisance functions were fit using the ensemble of
logistic regression via \textbf{glm}, 
logistic lasso/elastic net via \textbf{glmnet}, 
spline regression via \textbf{earth}, 
generalized additive model via \textbf{gam}, 
gradient boosting machine via \textbf{xgboost}, 
random forest via \textbf{ranger}, 
and neural net via \textbf{nnet} 
R packages using the super learner algorithm in R \citep{superlearner}.
The weights of each method in our super learner library are shown in Figure \ref{fig:SenegalDHS_SLWeight}.
Since we constructed estimators $S=30$ times to generate split-robust estimators, 
there were 30 replicates of weights also.
For $g$, the generalize additive model (\textbf{gam}) estimator had the largest weight,
while the spline regression (\textbf{earth}) estimator had the greatest weight in the ensemble used to estimate $\pi$.
The minimal weights of generalized linear model (\textbf{glm}) suggest that the parametric modeling of nuisance functions might have suffered from mis-specification.

\subsection{Choice of $S$}

The asymptotic properties of the sample splitting estimator do not depend on a specific sample split,
but different choices of sample split may affect the finite sample performance of the estimator.
In practice, one can repeat splitting the sample to construct the estimator $S$ times and then take the median of $S$ estimators to get a split-robust estimator \citep{chernozhukov18}.

\begin{figure}[H]
 \centerline{\includegraphics[width = \textwidth]{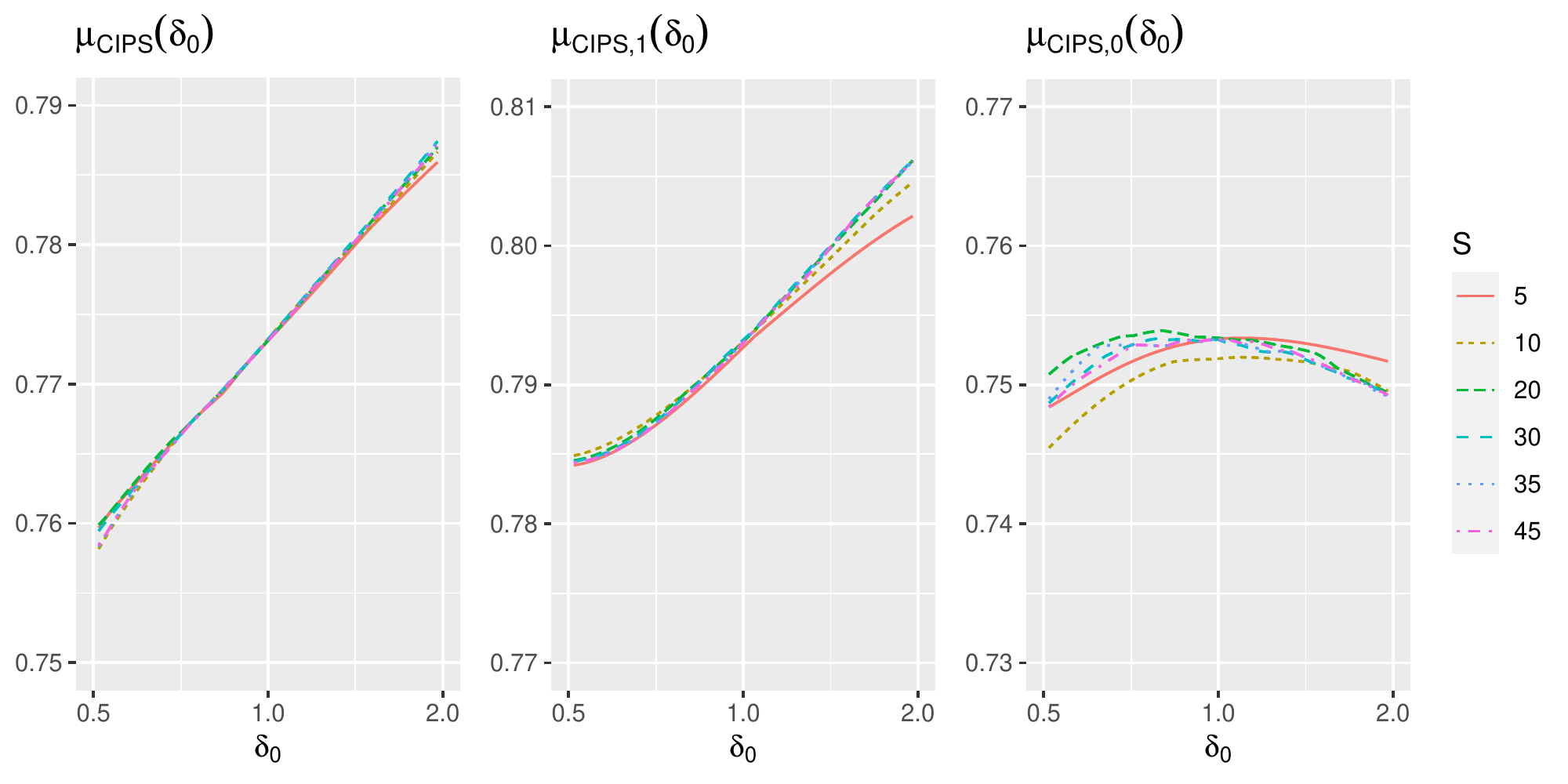}}
 \caption{\footnotesize
    Point estimates of CIPS estimands from the analysis of the Senegal DHS over $S = 5, 10, 20, 30, 35, 45$.
    }
\label{fig:ScompCIPS}
\end{figure}

\begin{figure}[H]
 \centerline{\includegraphics[width = \textwidth]{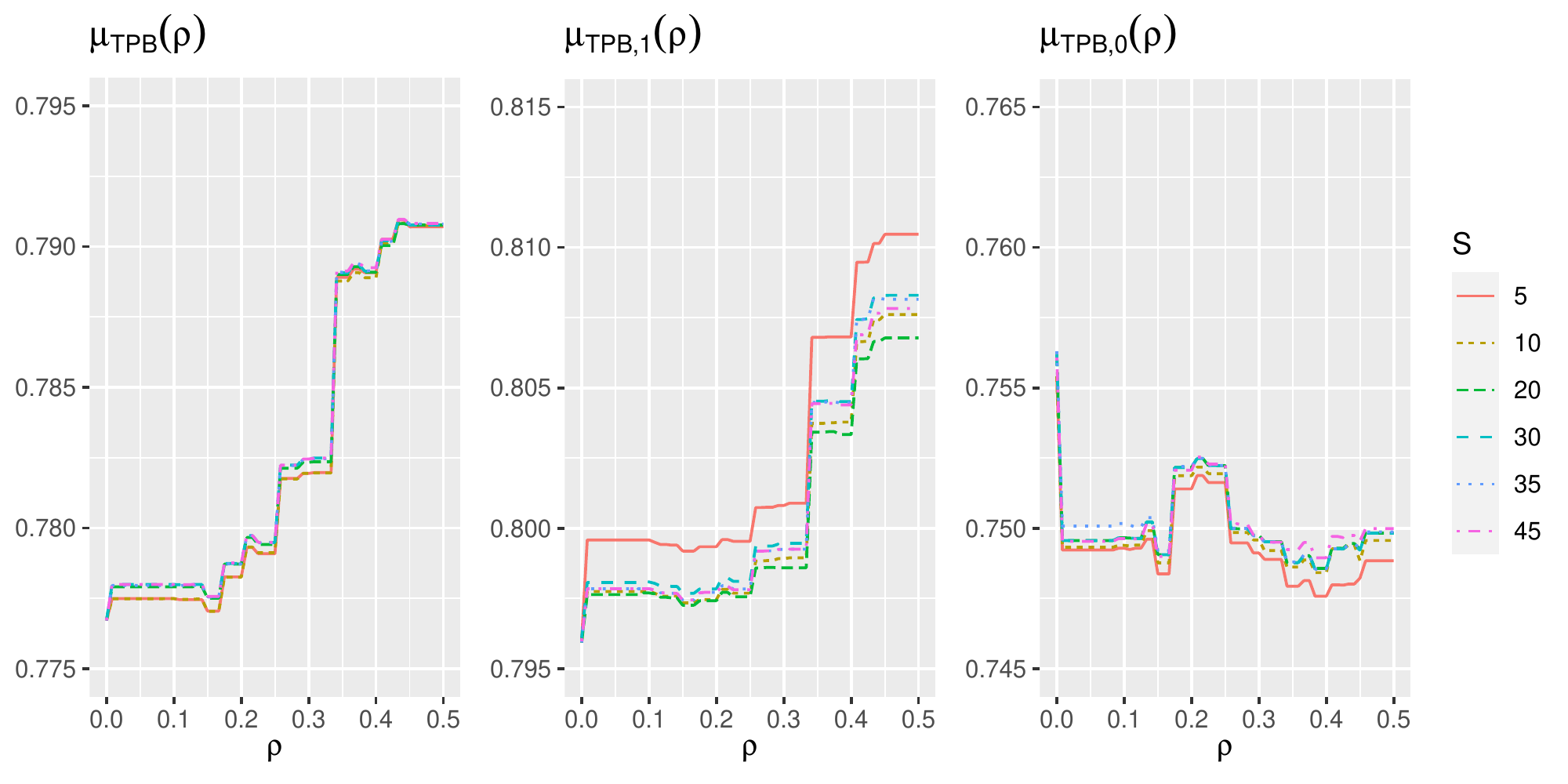}}
 \caption{\footnotesize
    Point estimates of TPB estimands from the analysis of the Senegal DHS over $S = 5, 10, 20, 30, 35, 45$.
    }
\label{fig:ScompTPB}
\end{figure}

In general, larger values of $S$ are recommended because the results will be less dependent on the arbitrary sample partitions used to construct the estimator.
In our simulation study, for the sample size of $m =$ 500, $S=1$ worked well.
For the Senegal DHS analysis ($m =$ 1,074), Figures \ref{fig:ScompCIPS} and \ref{fig:ScompTPB} show the point estimates under CIPS policy with constant $\delta  = \delta_0 \in [0.5, 2]$ and TPB policy with $\rho \in [0, 0.5]$, respectively, over $S = 5, 10, 20, 30, 35, 45$. 
These figures show very small differences in the DHS analysis results for $S \ge 30$. 



\subsection{Comparison with the estimation results in \citet{park21}}

\cite{park21} considered a variant of type B policy such that a cluster of size $N_i$ with cluster-level covariate $\mathbf{X}_i$ receives treatment $\mathbf{a}_i$ with probability 
$Q_{\scriptscriptstyle \textup{Park}}(\mathbf{a}_i|\mathbf{X}_i, N_i; \alpha) = \prod_{j=1}^{N_i} \alpha(\mathbf{X}_i)^{a_{ij}} \{1-\alpha(\mathbf{X}_i)\}^{1-a_{ij}}$,
where $\alpha(\mathbf{X}_i)$ is an unknown function of cluster-level covariate $\mathbf{X}_i$ which equals the probability of an individual being treated in cluster $i$.
Their goal was to find the minimum probability $\alpha^{\text{opt}}(\mathbf{X}_i)$ as a function of $\mathbf{X}_i$ such that
$
\alpha^{\text{opt}}(\mathbf{X}_i) = \inf_{\alpha \in [0,1]} 
\allowbreak
\Big\{
    \allowbreak
    \alpha \Big| \sum_{s = 0}^{N_i} 
    \allowbreak
    \binom{N_i}{s}
    \allowbreak
    g\left({s}/{N_i}, \mathbf{X}_i\right)
    \allowbreak
    \alpha^{s} (1-\alpha)^{N_i-s}
    \allowbreak
    \ge
    \allowbreak
    \mathcal{T}
    \allowbreak
\Big\}
$,
where:
$g\left({s}/{N_i}, \mathbf{X}_i\right) 
= 
\mathbb{E} \left(
    \overline{Y}_{i} | \overline{A}_{i} = {s}/{N_i}, \mathbf{X}_i, N_i 
\right) $;
$\overline{Y}_{i} 
= 
{N_i}^{-1} \sum_{j=1}^{N_i} Y_{ij}$;
$\overline{A}_{i} 
= 
{N_i}^{-1} \sum_{j=1}^{N_i} A_{ij}$;
$Y_{ij}$, $A_{ij}$, and $\mathbf{X}_i$ are defined as in Section 6 of the main text;
and $\mathcal{T}$ is the target diarrhea-free rate in each cluster, which is set to be in [0.64, 0.73].
The value of $\alpha^{\text{opt}}(\mathbf{X}_i)$ is the smallest proportion of households in census block $i$ to receive WASH facilities in order to ensure the expected average potential outcome $\mu_{\scriptscriptstyle \textup{Park}}(\alpha)$ 
under policy
$Q_{\scriptscriptstyle \textup{Park}}(\mathbf{a}_i|\mathbf{X}_i, N_i; \alpha)$
to be greater than the target level $\mathcal{T}$.

\begin{figure}[H]
 \centerline{\includegraphics[width = 0.7\textwidth]{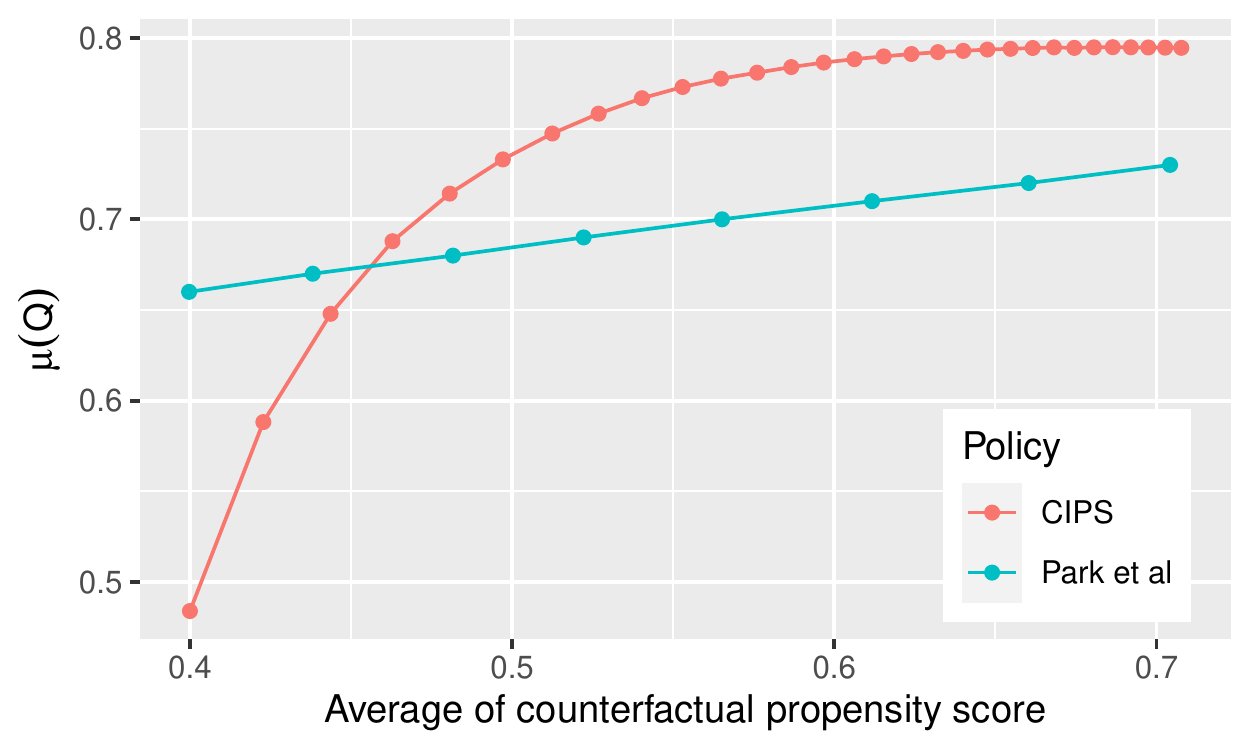}}
 \caption{\footnotesize
    Point estimates of $\mu(Q)$ for CIPS and \cite{park21} policies from the analysis of the Senegal DHS.
    }
\label{fig:CIPSvsOMAR}
\end{figure}

In contrast to the \cite{park21} analysis, our goal was to estimate average outcomes under different treatment allocation policies, in particular the CIPS and TPB policies. 
Thus, direct comparison of the results from the two analyses is not straightforward. 
Nevertheless, Figure \ref{fig:CIPSvsOMAR} presents a comparison of point estimates of the expected average potential outcome $\mu(Q)$ for the CIPS and \cite{park21} policies applied to the Senegal DHS data.


In this plot, the horizontal axis is the average of the estimated counterfactual propensity score under each policy, i.e., the average of $\widehat{\pi}_{ij, \delta}$ over $\delta \in [0,2]$ for CIPS policy and the average of $\widehat{\alpha}^{\text{opt}}(\mathbf{X}_i)$ over $\mathcal{T} \in [0.64, 0.73]$,
and the vertical axis is the estimated expected average potential outcome $\mu(Q)$.
For both policies, the $\mu(Q)$ estimates increase as the average counterfactual propensity score increases, 
implying in both cases that increasing the prevalence of WASH facilities should lead to more diarrhea-free households.

\end{document}